\documentclass[pre,11pt,preprint,superscriptaddress, amsmath, amssymb, floatfix]{revtex4-1} 
\usepackage{graphicx}  
\usepackage{graphics}
\usepackage{dcolumn}   
\usepackage{bm}        
\usepackage{amssymb}   
\usepackage{color}
\usepackage{amsmath}
\usepackage{threeparttable}
\usepackage[font=scriptsize]{caption}
\usepackage{url}
\usepackage{makecell}
\usepackage{diagbox}
\usepackage{booktabs}
\usepackage{subfigure}
\usepackage{amsfonts}
\usepackage{rotating}
\usepackage{fontenc}
\usepackage{cancel}
\usepackage{hyperref}
\usepackage[normalem]{ulem}
\usepackage{mathtools}

\hypersetup{CJKbookmarks,
       bookmarksnumbered,
              colorlinks,
               linkcolor=black,
               citecolor=black,
              plainpages=false,
            pdfstartview=FitH}

\definecolor{darkred}{RGB}{139,0,0}
\definecolor{chartreuse}{RGB}{127,255,0}
\definecolor{goldenrod}{RGB}{218,165,32}
\definecolor{gray}{RGB}{127,127,127}

\definecolor{Magenta}{RGB}{255, 0,255}
\definecolor{Orange}{RGB}{255,165, 0}
\definecolor{Gray}{RGB}{127,127,127}

\begin{document}

\title{The maximum capability of a topological feature in link prediction}
\author{Yijun Ran}
\affiliation{College of Computer and Information Science, Southwest University, Chongqing, 400715, P. R. China}
\affiliation{Center for Computational Communication Research, Beijing Normal University, Zhuhai, 519087, P. R. China}
\affiliation{School of Journalism and Communication, Beijing Normal University, Beijing, 100875, P. R. China}
\author{Xiao-Ke Xu}
\affiliation{Center for Computational Communication Research, Beijing Normal University, Zhuhai, 519087, P. R. China}
\affiliation{School of Journalism and Communication, Beijing Normal University, Beijing, 100875, P. R. China}
\author{Tao Jia}
\email{tjia@swu.edu.cn}
\affiliation{College of Computer and Information Science, Southwest University, Chongqing, 400715, P. R. China}

\begin{abstract}
Networks offer a powerful approach to modeling complex systems by representing the underlying set of pairwise interactions. Link prediction is the task that predicts links of a network that are not directly visible, with profound applications in biological, social, and other complex systems. Despite intensive utilization of the topological feature in this task, it is unclear to what extent a feature can be leveraged to infer missing links. Here, we aim to unveil the capability of a topological feature in link prediction by identifying its prediction performance upper bound. We introduce a theoretical framework that is compatible with different indexes to gauge the feature, different prediction approaches to utilize the feature, and different metrics to quantify the prediction performance. The maximum capability of a topological feature follows a simple yet theoretically validated expression, which only depends on the extent to which the feature is held in missing and nonexistent links. Because a family of indexes based on the same feature shares the same upper bound, the potential of all others can be estimated from one single index. Furthermore, a feature's capability is lifted in the supervised prediction, which can be mathematically quantified, allowing us to estimate the benefit of applying machine learning algorithms. The universality of the pattern uncovered is empirically verified by 550 structurally diverse networks. The findings have applications in feature and method selection, and shed light on network characteristics that make a topological feature effective in link prediction.
\end{abstract}

\maketitle 

Complex systems can be described by networks, in which nodes are the components of the system and links are the interactions between the components \cite{barabasi2016network,newman2018networks}. Link prediction is a task to infer missing connections that should exist but are not directly visible due to the incomplete information of the system \cite{clauset2008hierarchical,guimera2009missing,guimera2020one,wang2022full}, which has wide applications in predicting molecular interactions \cite{Ravasz2002Hierarchical,barzel2013network}, drug targets \cite{ryu2018deep,gysi2021network}, protein-protein interactions \cite{yu2008high,kovacs2019network}, recommendations on online social platforms \cite{santos2021link} and online shopping \cite{godoy2016accurate}. Because the topology information is usually directly available for a given network, significant effort has been devoted to utilizing topological features to predict missing links. Indeed, despite recent developments in computational tools \cite{xu2016representing,xue2022quantifying}, especially the network embedding by deep learning techniques \cite{cao2019network,xie2019sim2vec}, the topological feature is still widely used in link prediction due to its simplicity, interpretability, and overall good performance \cite{martinez2016survey,kumar2020link,zhou2021progresses,ji2023signal,jusup2022social}.

Here, we ask a simple question: to what extent could a topological feature be leveraged in link prediction? This question is of great importance given the vast link prediction methods developed in the past and current research, and its direct applications in the optimization of the feature and method selection. But seemingly, this question can only be answered on a case-by-case basis, as a myriad of factors could play a role. There are multiple indexes to quantify the same topological feature, each of which generates different index values. The obtained index value can be directly used for prediction \cite{Liben2007The,lu2011link,lee2021collaborative}, or further processed by a machine learning algorithm for a supervised prediction \cite{benson2018simplicial,ghasemian2020stacking,ghorbanzadeh2021hybrid,zhang2021semi,kumar2022link,jalili2017link}. The accuracy of the prediction can be gauged by either AUC or precision  \cite{ghasemian2020stacking,ghorbanzadeh2021hybrid,zhang2021semi,kumar2022link,lu2015toward,sun2020revealing}. Given all these variables, a theoretical approach to a unified answer is not expected. Indeed, the most relevant study in this direction is the quantification of the link predictability \cite{ghasemian2020stacking,lu2015toward,sun2020revealing,tang2020predictability}, corresponding to the extreme that any prediction method can ever reach in a given network. However, if we narrow down and focus on a particular topological feature, its potential in inferring missing links remains unclear.

In this work, we demonstrate that this question can be addressed in surprisingly general, yet, mathematically precise manners. By identifying patterns underlying the link prediction performance, we find that the maximum capability of a topological feature follows a simple and precise mathematical expression. The capability depends on the percentage of missing and nonexistent links that hold the feature, but is independent of the way that an index quantifies the feature. Hence, a family of indexes based on the same topological feature shares the same upper bound of prediction accuracy. The potential of all other indexes can be readily estimated through one single measurement. We also demonstrate that the supervised prediction in principle gives a more accurate result compared with the unsupervised one. But this improvement is not obtained merely by pushing the performance to the same upper bound. Instead, the maximum capability of the topological feature is lifted by utilizing the supervised method. The results derived can be applied to optimize the feature and the method selection, and also advance our understanding of network characteristics associated with the utilization of a topological feature in link prediction. In the following, we first give a brief introduction to the link prediction problem, the features and different indexes usually considered, and the evaluation metrics. We then derive the mathematical expressions of the prediction upper bound for a given feature in unsupervised and supervised link prediction tasks. We show how to utilize these expressions in deciding when to switch to a different feature or apply a different method. Finally, we take the common neighbor feature as an example, and quantitatively explain how the motifs of open and closed triangles determine the prediction outcome, which is not fully captured by a network's clustering coefficient. Our work benefits from a recently announced large corpus of 550 structurally diverse networks \cite{ghasemian2020stacking,broido2019scale}, allowing us to empirically verify the universality of the pattern uncovered. In short, our work contributes to link prediction by offering a tool to theoretically estimate the best performance of a topological feature, which can be applied to optimize the feature and method section, and to probe the crucial topological characteristics in link prediction.

{\bf Results}

{\bf Problem definition.}
We select 21 indexes commonly used in link prediction, ranging from the traditional common neighbor index (CN) in social networks \cite{Liben2007The} to the recently proposed paths of length three index (L3) for protein-protein interaction networks \cite{kovacs2019network}. According to the associated topological feature, these indexes can be classified into 4 families: common neighbor \cite{Ravasz2002Hierarchical,Liben2007The,Adamic2003Friends,zhou2009predicting,leicht2006vertex}, path \cite{lu2009similarity,papadimitriou2011friendlink,chen2012discovering,ran2021novel,ran2022predicting}, heterogeneity \cite{shang2019link}, and path of length three \cite{kovacs2019network,muscoloni2018local,muscoloni2022adaptive,muscoloni2023stealing}.
A given feature can be quantified differently by indexes in the same family. For example, the feature common neighbor can be expressed as the number of common neighbors \cite{Liben2007The} or the percentage of the neighborhood overlap \cite{santos2021link}. The path feature can be quantified by different combinations of paths \cite{kumar2020link,zhou2021experimental}. We list the 21 indexes of the 4 features in Table. \ref{table:features} with detailed descriptions presented in Supplementary Section \ref{section:s1}. To balance the number of indexes in each family, some indexes \cite{li2023link} associated with the common neighbor feature are tested but not explicitly shown in the paper. In existing studies, some features are quantified by only one index. We choose the index preferential attachment \cite{lu2011link} as an example and explicitly show in Supplementary Section \ref{section:s2} that our finding also works for this feature with one single index.

Essentially, the link prediction is to assign a score $S_{ab}$ to two nodes $a$ and $b$, which is proportional to the chance that nodes $a$ and $b$ should be truly connected. An index value for a topological feature can be utilized in two ways. One is to input the value to a machine learning based classifier, which finds a mapping function $y = f(x)$ to transfer the index value $x$ to the score value $y$. The other is to directly use the index value as the score, corresponding to a simple mapping function $y=x$. The two approaches, with the same input, differ only in the choice of the mapping function. They are named differently depending on the taxonomy applied \cite{martinez2016survey,kumar2020link,zhou2021progresses}. The former is sometimes called an algorithm-based or learning-based approach, whereas the latter is named a feature-based, similarity-based, or heuristic approach. To unify the name, we call the former  {\it supervised} approach and the latter {\it unsupervised} approach in this paper.

We adopt the most common settings of the link prediction problem \cite{zhou2021progresses,lu2011link}. Assume an undirected simple network composed of $N$ nodes and $L$ links, in which a node can not connect to itself (no self-loops) nor share more than one link with another node (no repeated links). Because missing links are actually unknown, a prediction can hardly be tested. To make the problem technically testable, a small portion of existing links are removed from the original network. They are considered as missing links to be inferred, which gives rise to the positive testing set $L^P$. As the control group of $L^P$, the negative testing set $L^N$ is constructed by randomly picking node pairs that are not connected in the original network, forming a set of nonexistent links. More details are discussed in Materials and Methods and Supplementary Section \ref{section:s3}. 

Despite some different opinions \cite{lichtnwalter2012link,muscoloni2022early,zhou2023discriminating}, link prediction is often treated as a binary classification problem, whose performance is gauged by the extent to which $L^P$ outscores $L^N$. One measure most commonly applied is AUC \cite{ghasemian2020stacking,ghorbanzadeh2021hybrid,zhang2021semi,kumar2022link}. The AUC can be calculated by a sampling method \cite{lu2011link,zhou2009predicting,ran2021novel,shang2019link}. In each comparison, we randomly pick one node pair from $L^P$ and one from $L^N$, and compare their scores. If out of $n$ comparisons, there are $n'$ times that samples from $L^P$ have a higher score than that from $L^N$, and $n''$ times that they have the same score, the AUC can be calculated as 
\begin{equation}
\text{AUC} = \frac{n' + 0.5n''}{n}.
\label{equation:auc}
\end{equation}
An example of the link prediction task and the quantification by AUC is illustrated in Supplementary Fig. \ref{fig:aucexample}.
 
Another metric in link prediction is precision \cite{ghorbanzadeh2021hybrid,lu2015toward,sun2020revealing}, which quantifies the percentage of true missing links in the top-k node pairs with the highest score. Recently, to deal with the sample imbalance issue, the area under the magnified ROC (AUC-mROC) is proposed \cite{muscoloni2022early}. Mathematical definitions of precision and AUC-mROC are presented in Materials and Methods. To avoid switching between different measures and to make the paper's flow more consistent, we choose to present results based on AUC in the main text. Results based on the other two metrics are reported in Supplementary Sections \ref{section:s4} and \ref{section:s5}. The framework proposed in this study applies to prediction capability measured by all these three metrics.

\begin{figure*}[ht]
\centering
\includegraphics[width=.5\linewidth]{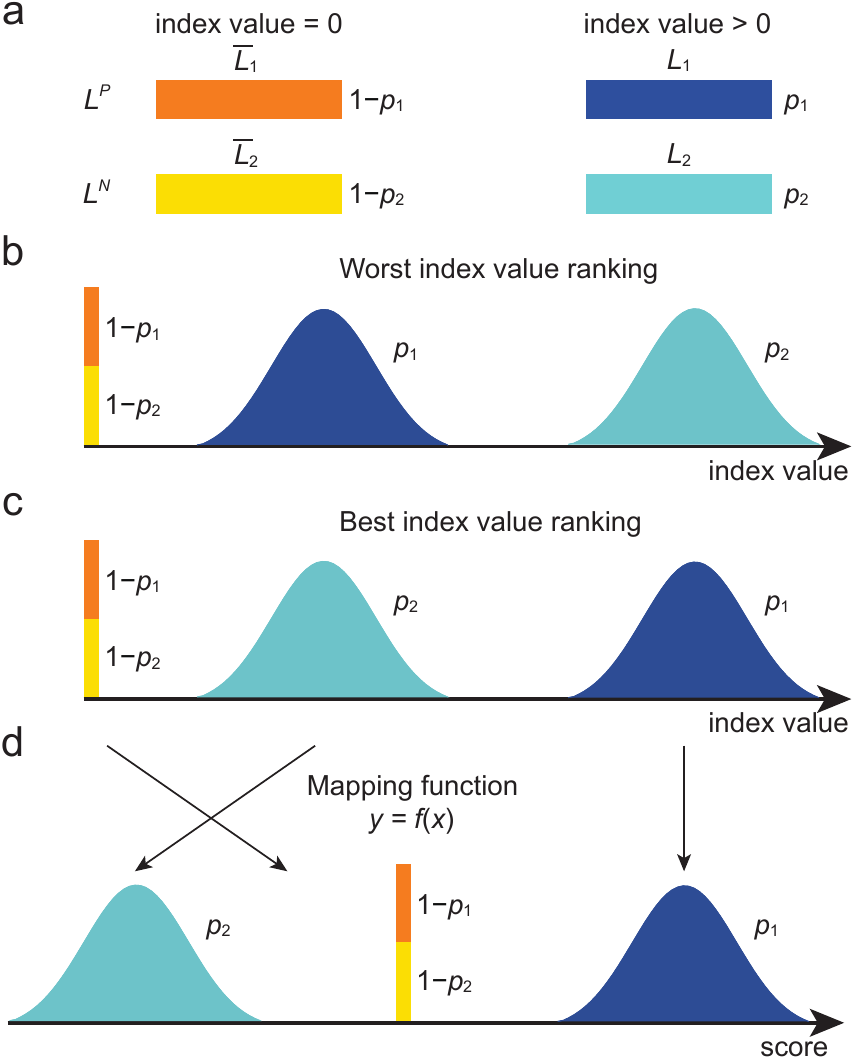}
\caption{{\bf An illustration of different link prediction performance.}
({\bf a}) Samples in the positive set $L^P$ can be divided into two subsets based on whether the feature is held or not. $L_1$ is the subset of $L^P$ in which node pairs hold the feature, whereas the complement set $\overline{L}_1$ is composed of node pairs that do not hold the feature. As the index is designed to quantify the feature, it should assign non-zero values to samples in $L_1$ and value 0 to samples in $\overline{L}_1$. Similarly, the negative set $L^N$ can also be divided into two subsets $L_2$ and $\overline{L}_2$. Assume that $L_1$ takes a fraction $p_1$ of $L^P$ and $L_2$ takes a fraction $p_2$ of $L^N$. Because samples in $\overline{L}_1$ and $\overline{L}_2$ have the same index value 0, the prediction performance mainly relies on the ranking of $L_1$ and $L_2$. 
({\bf b}) The worst index value ranking is when $L_2$ is systematically ranked ahead of $L_1$. ({\bf c}) The best index value ranking is just the opposite when $L_1$ is systematically ranked ahead of $L_2$. Note that in both cases, $\overline{L}_1$ and $\overline{L}_2$ are always ranked behind $L_1$ and $L_2$ in the unsupervised approach.
({\bf d}) In supervised prediction, the machine learning based classifier can find a mapping function $y=f(x)$ to transfer the index value to the score for prediction. Hence, the relative position among $L_1$, $L_2$ and $\overline{L}_1 \cup \overline{L}_2$ can be further optimized. Because samples in $\overline{L}_1 \cup \overline{L}_2$ have the same index value, they should have the same score. The optimal score ranking is to assign a score to $\overline{L}_1 \cup \overline{L}_2$ that makes it lie between $L_1$ and $L2$. In this case, no negative samples have a higher score than positive samples.
Note that different scenarios described here can also be used to explain different precision values obtained (see Supplementary Section \ref{section:s4}). 
}
\label{fig:unsupervised}
\end{figure*}

{\bf The capability of a feature in the unsupervised prediction.}
The index value is directly used to predict missing links in the unsupervised approach. Hence, how the index values are distributed in $L^P$ and $L^N$ determines the prediction performance. Different indexes gauge the same feature differently. But since they are proposed to quantify the feature, they should follow one rule in common: entities that do not hold the feature have the same and the lowest value. Without loss of generality, these entities are often assigned by the value 0 (see further discussion in Supplementary Section \ref{section:s6}). For instance, a pair of nodes without any overlap of neighborhoods has the value 0 for any indexes based on the common neighbor feature. Likewise, all indexes using network distance assign the value 0 to a pair of nodes not connected by a path. Denote $L_1$ and $L_2$ by the subset of node pairs that hold the topological feature in $L^P$ and $L^N$ (index values greater than 0), respectively (Fig. \ref{fig:unsupervised}a). Consequently, $\overline{L}_1 = L^P \backslash L_1$ and $\overline{L}_2 = L^N \backslash L_2$ are the subsets of node pairs with index value 0. The prediction performance mainly relies on the value distribution in $L_1$ and $L_2$, as node pairs in $\overline{L}_1$ and $\overline{L}_2$ all have the same value 0 and are indistinguishable by their index values.

Let us first consider the worst scenario, when the index values are assigned such that the highest value in $L_1$ is less than the lowest in $L_2$ (Fig. \ref{fig:unsupervised}b). A positive sample outscores a negative sample only when one is from $L_1$ and the other is from $\overline{L}_2$. A positive sample and a negative sample have the same score only when they are from $\overline{L}_1 \cup \overline{L}_2$. Denoting $p_1 = |L_1|/|L^P|$ and $p_2 = |L_2|/|L^N|$, the AUC in the worst scenario can be calculated as 
\begin{align}
\text{AUC}_\text{lower} &= \frac{n'}{n} + \frac{1}{2}\frac{n''}{n} \nonumber \\
                                      &= p_{1}(1-p_{2})+\frac{1}{2}(1-p_{1})(1-p_{2}) \nonumber \\ 
                                      &=  \frac{1}{2}+\frac{p_{1}-p_{2}-p_{1}p_{2}}{2}. \label{equation:lower}
\end{align}
On the contrary, the best scenario is when the lowest value in $L_1$ is greater than the highest in $L_2$ (Fig. \ref{fig:unsupervised}c). Node pairs in $L_1$ outscore all negative samples, which gives the AUC upper bound
\begin{align}
\text{AUC}_\text{upper} &= \frac{n'}{n} + \frac{1}{2}\frac{n''}{n} \nonumber \\
                                      &= p_{1}+\frac{1}{2}(1-p_{1})(1-p_{2}) \nonumber \\
                                      &=  \frac{1}{2}+\frac{p_{1}-p_{2}+p_{1}p_{2}}{2}. \label{equation:upper}
\end{align} 
For general cases, the index values distributed in $\overline{L}_1$ and $\overline{L}_2$ have overlaps. Hence, the performance lies between $\text{AUC}_\text{lower}$ and $\text{AUC}_\text{upper}$.

Eq. (\ref{equation:lower}) and Eq. (\ref{equation:upper}) suggest that the performance of the unsupervised link prediction depends on how well an index is able to rank positive samples ahead of the negative samples. Two factors set the bounds of this process. First, there are $(1-p_{1}) + (1-p_{2})$ samples with the same index value that can not be effectively ranked. Second, for the rest $p_1+p_2$ samples, once the sets $L_1$ and $L_2$ are well separated and properly ranked by their index values, no further improvement can be achieved. Note that we start with the performance of an index, but Eq. (\ref{equation:lower}) and Eq. (\ref{equation:upper}) only depends on $p_1$ and $p_2$. The performance limit of an index only relies on the percentage of missing and nonexistent links that hold the feature, not on how the index quantitatively gauges the feature. Different indexes in the same family can give rise to different prediction results, with different distances to the upper bound set by Eq. (\ref{equation:upper}). But they all fall within the same range determined by the properties of the topological feature (Supplementary Fig. \ref{fig:unsupupper}). Hence, Eq. (\ref{equation:upper}) not only depicts the performance limit of one single index, but also effectively gives the maximum capability of a topological feature in unsupervised link prediction that can be ever reached by any index associated with it. 

Furthermore, Eq. (\ref{equation:lower}) and Eq. (\ref{equation:upper}) indicate that the gap between the upper and lower bound is $p_1 \times p_2$. For common neighbor feature and path feature whose $p_1 \times p_2$ values are small (Supplementary Table \ref{table:p1p2}), the performance of their indexes should not fluctuate significantly and is mainly determined by $p_{1}-p_{2}$. Hence, it is predicted that the AUC by different indexes in different networks, regardless of their types and sizes, scale as $p_{1}-p_{2}$, which is empirically verified in Fig. \ref{fig:p1-p2} (an extended discussion is presented in Supplementary Section \ref{section:s7}).

{\bf The capability of a feature in the supervised prediction.}
In the supervised approach, the index value of a sample is input to a machine learning based classifier. We use the Random Forest classifier \cite{ghasemian2020stacking,wang2011human} in the main text. Similar results by Gradient Boosting \cite{kumar2022link,mahapatra2021improved} and AdaBoost \cite{ghasemian2020stacking,shan2020supervised} are presented in Supplementary Section \ref{section:s8}. The classifier finds a mapping function $y=f(x)$ to transform the input $x$ to the score $y$ for prediction, with the aim to further improve the prediction performance. Therefore, a more accurate prediction is expected as the non-fixed mapping function provides more flexibility to properly rank samples in $L^P$ and $L^N$. Indeed, the best index value ranking is not the optimal score ranking. Because node pairs in $\overline{L}_1$ are ranked behind $L_2$, part of the negative samples still outscore positive samples. With the mapping function, the score ranking of $L_1$, $L_2$ and $\overline{L}_1 \cup \overline{L}_2$ can be rearranged. In Supplementary Section \ref{section:s9}, we compare the AUC value under different rankings of the three sets. The optimal score ranking is when samples in $L_1$ rank ahead of $L_2$ and samples in $\overline{L}_1$ and $\overline{L}_2$ (note that they have the same index value $x$, hence with the same score $y$) lie between $L_1$ and $L_2$ (Fig. \ref{fig:unsupervised}d). In this case, no negative sample has a higher score than the positive sample. Hence, if the classifier can find the right mapping function, the prediction's upper bound becomes
\begin{align}
\text{AUC}^{\prime}_\text{upper} &= \frac{n'}{n} + \frac{1}{2}\frac{n''}{n} \nonumber \\
                                                    &= p_{1} + (1-p_{1})p_{2} +\frac{1}{2}(1-p_{1})(1-p_{2}) \nonumber \\
                                                    &=  \frac{1}{2}+\frac{p_{1}+p_{2}-p_{1}p_{2}}{2}. \label{equation:upper2}                                   
\end{align}

\begin{figure*}[ht]
\centering
\includegraphics[width=.8\linewidth]{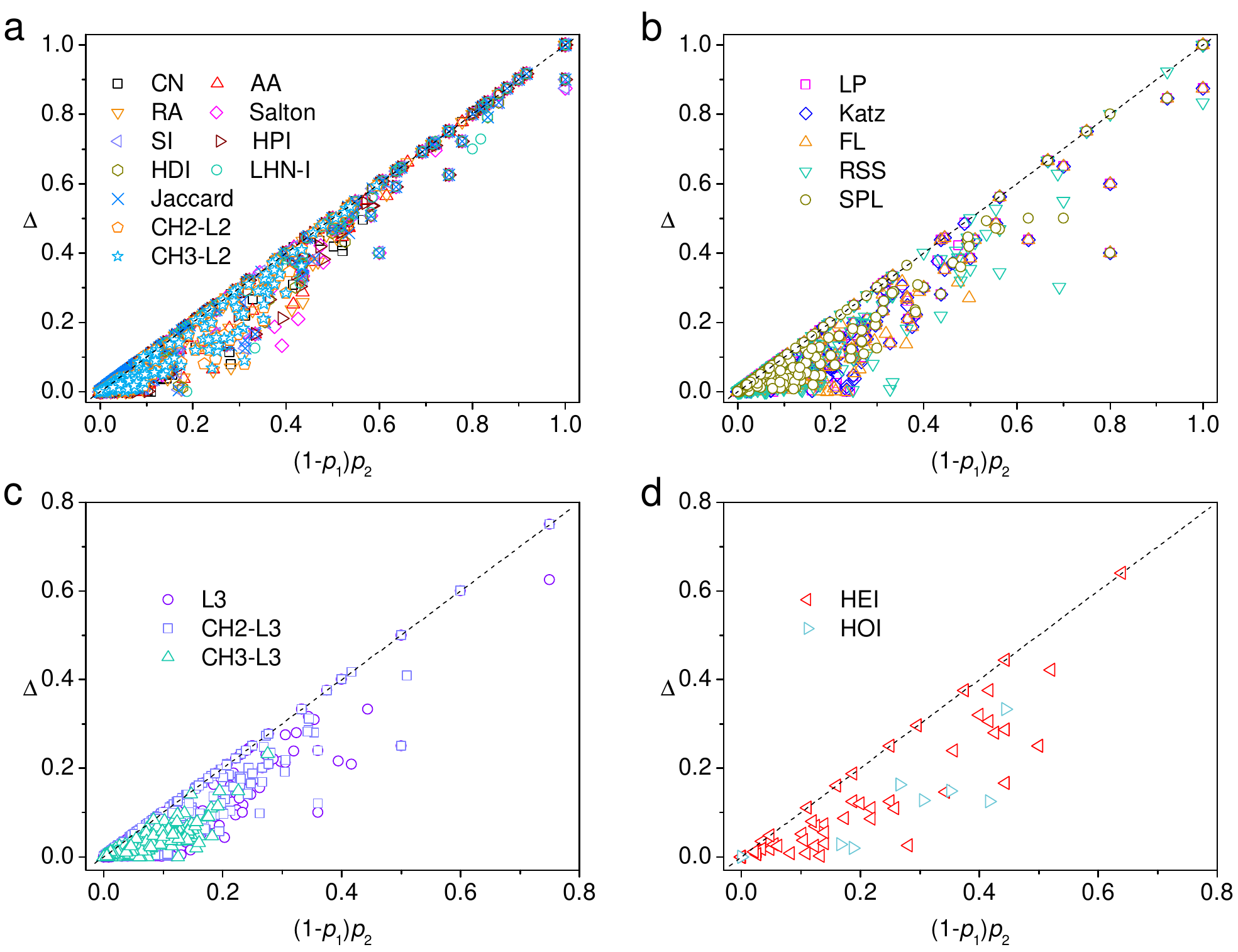}
\caption{{\bf The actual and predicted improvement by the supervised approach.}
Eq. (\ref{equation:upper2}) suggests that the supervised approach can lift the capability of a feature by $(1-p_{1})p_{2}$. To test it, we select networks in which the unsupervised prediction by an index is already close to its upper bound (measured AUC is more than 95\% of $\text{AUC}_\text{upper}$). For these networks, we input the same index values of $L^P$ and $L^N$ to the classifier to obtain the supervised prediction results. For each network, we randomly generate 200 realizations of networks with link removal, as well as 200 pairs of $L^P$ and $L^N$ sets (Materials and Methods). We pick the network realization that gives rise to the highest AUC in supervised prediction. $\Delta$ for a given network is measured as the AUC difference between the supervised and unsupervised prediction for that particular network realization. The empirically measured $\Delta$ for different networks and indexes are close to $(1-p_{1})p_{2}$, in line with the prediction by Eq. (\ref{equation:upper2}).
}
\label{fig:(1-p1)p2}
\end{figure*}

The capability of a topological feature in supervised link prediction ($\text{AUC}^{\prime}_\text{upper}$) is empirically verified in 550 networks (Supplementary Fig. \ref{fig:supupper}). In general, the performance of supervised prediction improves, in line with the expectation (Supplementary Fig. \ref{fig:simvsml}). More importantly, Eq. (\ref{equation:upper2}) suggests that the improvement is not merely by getting closer to the original upper bound. Instead, the maximum capability of a feature is lifted by an extent $\Delta = (1-p_{1})p_{2}$. Therefore, if the results about the maximum capability were correct, we would expect that when an unsupervised prediction is already close to the maximum capability $\text{AUC}_\text{upper}$, the prediction performance would be further improved in this network by $\Delta$ if a machine learning algorithm is applied. This hypothesis is verified in Fig. \ref{fig:(1-p1)p2}, supporting the proposed capability of a feature.

{\bf Applications.}
In link prediction, a critical choice we typically face is to decide whether to switch to another topological feature for a better prediction or to keep the same topological feature but try other indexes. Traditionally, such a decision can only be made by enumerating the performance of all indexes associated with a feature. With the mathematical expression of $\text{AUC}_\text{upper}$ and $\text{AUC}^{\prime}_\text{upper}$, however, the capability of a feature can be easily estimated from the measurement of one single index. Because $p_1$ and $p_2$ are only related to node pairs that hold the topological feature, their values are simultaneously known once an index is applied to $L^P$ and $L^N$. The capability can therefore serve as theoretical guidance for feature selection. Take the two networks in Table \ref{table:unsupervisedf} as an example. When using the index LHN-I to make an unsupervised link prediction, we obtain $\text{AUC}=0.497$ in both networks. It is obvious that LHN-I should not be considered as its prediction is worse than a random guess. But the next question is, should we try other indexes of the common neighbor feature or should we turn to a different topological feature? With the $p_1$ and $p_2$ values obtained through the calculation of LHN-I, Eq. (\ref{equation:upper}) gives the maximum capability of the common neighbor feature: $\text{AUC}_\text{upper} = 0.497$ for network A and $\text{AUC}_\text{upper} = 0.974$ for network B. Consequently, the strategy is to switch to a new topological feature in network A, as any indexes related to the common neighbor are doomed to fail. For network B, however, we can keep using the common neighbor feature and try other indexes. Indeed, for 8 indexes in the family of common neighbors, the prediction results halt at $\text{AUC}=0.497$ in network A, but increase to $\text{AUC}=0.77$ in network B, supporting the decision made through the estimated maximum capability (see another example in Supplementary Section \ref{section:s10}).

Likewise, the value $\Delta = (1-p_{1})p_{2}$ theoretically predicts the expected benefits of applying a supervised method, which can be helpful in method selection. When the performance of unsupervised prediction is similar in two networks (Table \ref{table:unsupervisedm}), can we decide and explain if an improvement is expected by adopting a machine learning based classifier? With the $p_1$ and $p_2$ values, first, we can tell that the unsupervised prediction in both networks is close to the upper bound $\text{AUC}_\text{upper}$. Furthermore, $\Delta$ predicts that the upper bound $\text{AUC}^{\prime}_\text{upper}$ can increase as much as 0.111 in network C, but could not go any further in network D. Hence, the machine learning algorithm can be beneficial only in Network C, which is verified by the performance of supervised prediction in Table \ref{table:unsupervisedm}.

\begin{figure*}[ht]
\centering
\includegraphics[width=.8\linewidth]{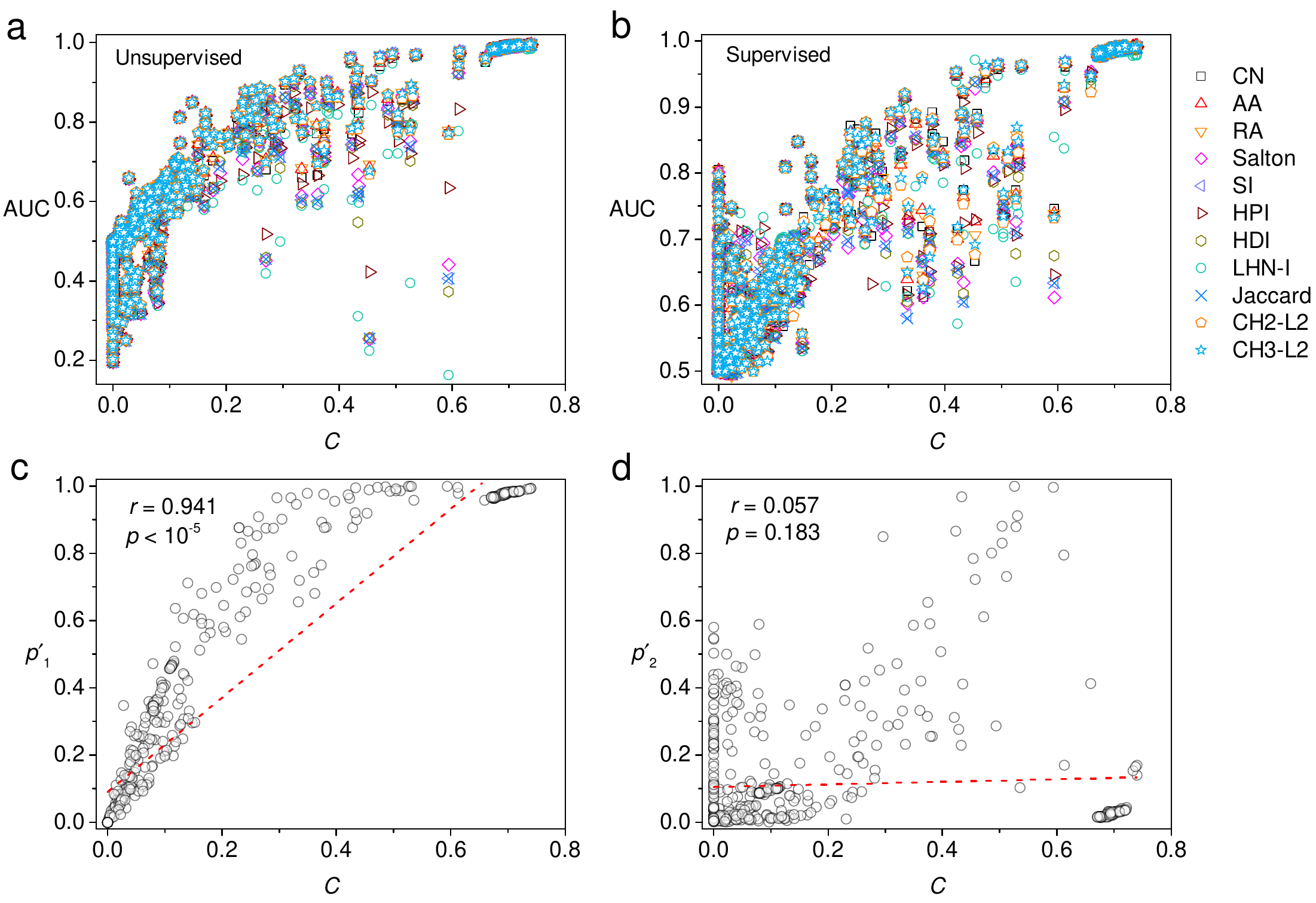}
\caption{{\bf The structural characteristics related to the common neighbor feature in link prediction.}
({\bf a, b}) It is intuitively expected that the clustering coefficient $C$ is directly related to the performance of indexes based on the common neighbor feature. But the unsupervised ({\bf a}) and supervised ({\bf b}) prediction results show that $C$ can not fully explain the effectiveness of the common neighbor feature. AUC demonstrates a significant variability for some $C$ values in both cases.
({\bf c, d}) According to Eq. (\ref{equation:p1}) and Eq. (\ref{equation:p2}), $p_{1}$ depends on the number of closed triangles, and $p_{2}$ depends on the number of open triangles. Therefore, $p_{1}$ should demonstrate a strong correlation with the clustering coefficient $C$, and $p_{2}$ should be independent of $C$, which are empirically confirmed. The $r$ is the Pearson correlation coefficient, and the p-value is from the Student's t-test.
}
\label{fig:tp1p2}
\end{figure*}

Moreover, once the theoretical expression of $p_{1}$ and $p_{2}$ are derived, we can quantitatively explore the structural characteristics that make a topological feature effective in link prediction. Take the common neighbor as an example. It is one of the most frequently used topological features. Yet, we still lack a quantitative assessment of the type of networks in which the common neighbor feature would work or fail. One may intuitively expect that the performance of the common neighbor is associated with the clustering coefficient $C$, as this feature is more prominent in clustered networks \cite{feng2012link,liu2016degree}. However, numerical results show that the clustering coefficient can not fully explain the prediction performance (Figs. \ref{fig:tp1p2}a, b). While the AUC reaches a high value and saturates when $C$ is sufficiently large, it fluctuates significantly for small values of $C$. This is particularly an issue when the supervised approach is applied: even when the network is not clustered at all ($C=0$), a supervised prediction can still give rise to an AUC value close to 0.8.

To explore the structural characteristics of the common neighbor feature, we derive the analytical expression of $p_1$ and $p_2$ as (see the deduction in Supplementary Section \ref{section:s11})
\begin{equation}
p^{\prime}_{1} =  \frac{3*N_{\triangle}-S_{\triangle}}{L}
\label{equation:p1}
\end{equation}
\begin{equation}
p^{\prime}_{2} =  \frac{N_{\wedge}-S_{\wedge}}{\frac{N(N-1)}{2}-L}.
\label{equation:p2}
\end{equation}
In Eq. (\ref{equation:p1}) and Eq. (\ref{equation:p2}), $N_{\triangle}$ and $N_{\wedge}$ are the number of closed and open triangles, respectively. $S_{\triangle}$ is the number of times that a link is shared by multiple triangles, and $S_{\wedge}$ is the number of times that an unconnected node pair is shared by other open triangles. The mathematical expressions by Eq. (\ref{equation:p1}) and Eq. (\ref{equation:p2}) quantitatively reveal structural characteristics as well as their interplay related to the common neighbor feature in link prediction. $p^{\prime}_{1}$ depends on the number of closed triangles, hence strongly correlated with the clustering coefficient $C$ (Fig. \ref{fig:tp1p2}c). But $p^{\prime}_{2}$ depends on the number of open triangles which is unrelated to $C$ (Fig. \ref{fig:tp1p2}d). This explains why the clustering coefficient alone is insufficient to characterize the capability of the common neighbor feature, as the prediction performance is determined by a combination of both $p^{\prime}_{1}$ and $p^{\prime}_{2}$. What is more, the denominator in the expression of $p^{\prime}_{2}$ scales with $N^2$. Therefore, it is expected that prediction results in large networks should demonstrate a stronger dependence with $C$, as the $p^{\prime}_{2}$ in general vanishes quickly with $N$ (Supplementary Fig. \ref{fig:tp2n}a). But networks with many ``leaves'' structures have the number of open triangles scaling non-linearly with $N$, giving rise to a non-zero $p^{\prime}_{2}$ value for large $N$ (Supplementary Fig. \ref{fig:tp2n}b). In such networks, even though the network is sufficiently large, the abundance of local clusters is still insufficient to quantify the capability of the common neighbor feature.

{\bf Discussion}

To summarize, we quantify the maximum capability of a topological feature in link prediction by analyzing the upper bound of prediction accuracy. Given a variety of indexes, different approaches to utilize the index, and different measures of prediction performance, there seems no simple answer to this question. 
Nevertheless, we identify regularities underlying the link prediction task, leading to the mathematical expression for the performance upper bound by a topological feature. The maximum capability of a feature only depends on the extent to which the feature is held in missing and nonexistent links but does not depend on how a related index gauges the feature. Hence, a family of indexes associated with one topological feature shares the same performance upper bound, which can be used to decide if a new feature or a new index is needed to advance the prediction. The capability of a feature is lifted by applying the supervised approach, whose magnitude can be theoretically derived, allowing us to estimate the benefit of applying machine learning algorithms in link prediction. Finally, using the common neighbor feature as an example, we quantitatively show how the interplay of different structural characteristics determines the prediction performance in different networks, which can not be fully explained by the clustering coefficient. Our theoretical results are verified by 550 empirical networks, demonstrating a strong universality.

To make the flow of the paper more consistent, we choose to present the performance by AUC measure in the main text, but the framework also works for precision and AUC-mROC, whose results are presented in Supplementary Material. It is noteworthy that different metrics reflect different aspects of prediction performance. This is already illustrated by different upper bound values for the same feature and network measured by different metrics. While AUC is commonly used in link prediction, it may give biased evaluation when the network size is large and samples are very imbalanced \cite{lichtnwalter2012link}. In addition, results presented in Supplementary Section \ref{section:s4} bring insights into the interpretation of results measured by the precision and AUC. For a network with a small $p_1$ value, link prediction measured by AUC can be poor. But the precision can be high if we only focus on the top-ranked node pairs such that $L_\text{k} < p_1|L^P|$ (see an example in Supplementary Section \ref{section:s12}). This is because AUC tends to provide the overall extent to which the positive samples outscore the negative samples \cite{ghasemian2020stacking,zhou2023discriminating}, while precision is mainly affected by how close the actual score ranking is to the optimal ranking. The precision upper bound is 100\% if $L_\text{k}$ is relatively small and only the top tires are considered (Supplementary Eq. (\ref{equation:preupper1})). Therefore, when comparing link prediction performance among different networks, more details, such as the sample imbalance ratio and the $p_1$ value, need to be considered for a more comprehensive interpretation of results. 

When making use of a new topological feature, the $\text{AUC}_\text{upper}$ and $\text{AUC}^{\prime}_\text{upper}$ derived in this work can serve as the pre-evaluation of its potential. It is interesting to note that even for unsupervised prediction, the average accuracy by indexes of common neighbor, path, and path of length three is close to the feature's maximum capability. This suggests that future development on indexes related to these features may only bring marginal advances. A promising direction is to merge multiple features for link prediction. One approach is to numerically combine (such as adding and multiplying) multiple index values associated with different features to form a new index \cite{yang2016predicting,rafiee2020link,ahmad2020missing,yuliansyah2023new}. Another approach is to use a new feature to sub-rank samples with zero index value \cite{muscoloni2022adaptive,muscoloni2023stealing}. When the performance is already close to the maximum capability and $p_1$ is relatively small, sub-ranking by a new feature would be effective, because the $p_1$ value is increased without changing the original ranking. On the contrary, when $p_1$ is already large, numerically combining multiple indexes would be a better choice, as it can further fine-tune the rank of samples. With the framework proposed in this paper, one can systematically evaluate and explain the effectiveness of different feature-merging approaches in different groups of networks. The relationship between the capability of a feature and the link predictability of the network is another intriguing question to tackle. Identifying the single or a set of topological features that have the capability closest to the link predictability may reveal the intrinsic characteristics of the network. 
Taken together, our work uncovers a regularity in the link prediction problem, which not only provides a theoretical upper bound for leveraging a topological feature but also sheds light on a range of related questions that can be analyzed in the future.

{\bf Materials and Methods}

{\bf Experiment setup.}
In link prediction, a small portion of existing links are removed from the original network. They are the positive samples that need to be distinguished from the negative samples, which are randomly selected from the non-connected node pairs. Because the topological features are measured from the network after the link removal, the portion of removed links tends to be as small to preserve the network topology \cite{muscoloni2018local,cannistraci2013link}. Compared with the unsupervised prediction, the supervised link prediction requires an additional training set to adjust the mapping function. Therefore, the portion of removed links is usually larger in supervised prediction than in unsupervised prediction \cite{benson2018simplicial,ghasemian2020stacking}. In this study, the performance of both unsupervised and supervised prediction are analyzed and compared. It is important to make the network topology the same for both prediction approaches. For this reason, we use the setup that is commonly applied in supervised link prediction \cite{benson2018simplicial,ghasemian2020stacking} in the main text. 

In particular, we randomly remove 20\% of $L$ links from the original network. Half of these removed links, or equivalently 10\% of $L$ links, are used as the positive testing set $L^P$. The other half forms the positive training set that is only used in supervised prediction. We adopt balanced positive and negative samples. Negative samples with the same size as the positive samples are formed by randomly selected nonexistent links. While 10\% of $L$ links are not used in unsupervised prediction, the setup makes sure that the network topology is the same for unsupervised and supervised prediction.

For each network, we run the above procedure 200 times. We get 200 realizations of networks with link removal, and 200 pairs of $L^P$ and $L^N$ sets. Unless otherwise specified, in unsupervised prediction, we average the capability derived and the prediction performance for the 200 realizations. In supervised prediction, to avoid the performance fluctuation caused by the choice of training and testing set, we report the best prediction performance and the corresponding capability for that particular realization of a network.

For robustness check, we adopt the common setup in unsupervised link prediction that removes only 10\% of original links \cite{muscoloni2018local,cannistraci2013link}. We repeat the analyses and the results are shown in Supplementary Section \ref{section:s13}.  In addition, we perform two alternative sampling methods whose results are reported in Supplementary Section \ref{section:s3}. Our results are not affected by different setups. Finally, for the supervised prediction, we take the grid search method \cite{bergstra2012random} to choose the best parameters that optimize the AUC.

{\bf Other evaluation metrics.} 
Besides AUC, we also consider precision and AUC-mROC to evaluate the link prediction performance. The precision measures the percentage of the correct prediction (node pairs indeed in $L^P$) among the top-k predicted candidates \cite{ghorbanzadeh2021hybrid,lu2015toward,sun2020revealing}. After ranking the node pairs in both $L^P$ and $L^N$ according to their scores in descending order, we select $L_\text{k}$ node pairs with the highest score. The precision is given as
\begin{equation}
\text{Precision} = \frac{L_\text{r}}{L_\text{k}},
\end{equation} 
where $L_\text{r}$ is the number of selected node pairs that are included in $L^P$. 

Link prediction usually faces imbalanced positive and negative samples because real-world networks are often sparse. To cope with this issue, the area under the magnified ROC (AUC-mROC) is recently proposed \cite{muscoloni2022early}. Denote $n_1 = |L^P|$ and $n_2 = |L^N|$ by the size of the positive and negative testing set. Given the index values of samples, we can count the number of true positive (\text{TP}@r) and false positive (\text{FP}@r) samples at a varying ranking threshold $r \in [1, n_1+n_2]$. The non-normalized magnified TPR (\text{nmTPR}@r) and non-normalized magnified FPR (\text{nmFPR}@r) at each ranking threshold $r$ are defined as
\begin{equation}
\text{nmTPR}@r = \frac{\ln(1+\text{TP}@r)}{\ln(1+n_1)},
\end{equation} 

\begin{equation}
\text{nmFPR}@r = \frac{\ln(1+\text{FP}@r)}{\ln(1+n_2)}.
\end{equation} 
To make the AUC-mROC of a random prediction equal to 0.5, the \text{nmTPR}@r is normalized to \text{mTPR}@r as 
\begin{align}
\text{mTPR}@r &=  \frac{\text{nmTPR}@r-\ln_{(1+n_1)}(1+\text{FP}@r \cdot \frac{n_1}{n_2})}{1-\ln_{(1+n_1)}(1+\text{FP}@r \cdot \frac{n_1}{n_2})} \nonumber \\
                        & \quad \times (1-\text{nmFPR}@r) +\text{nmFPR}@r.
\end{align} 
The mROC curve is composed of data points at coordinates (\text{nmFPR}@r, \text{mTPR}@r) for each $r \in [1, n_1+n_2]$. The AUC-mROC is then obtained by computing the area under the mROC curve using the trapezoidal rule.

{\bf Network data.} 
To verify the universality of the pattern uncovered, we consider the dataset from ``CommunityFitNet corpus'' \cite{ghasemian2020stacking,broido2019scale}. The dataset includes networks from 6 different domains: 32.55\% (179) biological, 22.55\% (124) social, 22.55\% (124) economic, 12.73\% (70) technological, 6.36\% (35) transportation, and 3.27\% (18) information networks. The basic statistics of networks in different domains are in Supplementary Section \ref{section:s14}.

{\bf Code availability}

The code used in this study is available at \href{https://github.com/YijunRan/Maximum-Capability-Link-Prediction}{https://github.com/YijunRan/MCLP}.

{\bf Acknowledgements}

This work is supported by the National Natural Science Foundation of China (No. 72374173 and No. 62173065), University Innovation Research Group of Chongqing (No. CXQT21005), the Fundamental Research Funds for the Central Universities (No. SWU-XDJH202303), and the Postdoctoral Fellowship Program of CPSF (No. GZC20230281).

{\bf Author contributions}

YJ. R., XK. X. and T. J. designed the research. YJ. R. performed numerical analyses. YJ. R. and T. J. derived the analytical results. YJ. R. and T. J. prepared the initial draft. XK. X. and T. J. revised the draft to the final version of the paper.

{\bf Competing financial interests} 

The authors declare no competing financial interests.

\clearpage

\bibliographystyle{plain}

\clearpage

\begin{table}[h!]
\caption{\bf The 21 indexes related to 4 topological features.}
\centering
\begin{tabular}{cc}
\toprule
Features & Indexes (Abbr.)\\
\hline
Common Neighbor& \makecell{Common Neighbor Index ({\bf CN})\cite{Liben2007The}, \\ Adamic-Adar Index ({\bf AA})\cite{Adamic2003Friends}, \\ Resource Allocation Index ({\bf RA})\cite{zhou2009predicting}, \\ Salton Index ({\bf Salton})\cite{sun2020revealing}, \\ S$\o$rensen Index ({\bf SI})\cite{ghorbanzadeh2021hybrid}, \\ Hub Promoted Index ({\bf HPI})\cite{Ravasz2002Hierarchical},\\ Hub Depressed Index ({\bf HDI})\cite{zhou2009predicting}, \\ Jaccard Index ({\bf Jaccard})\cite{Liben2007The}, \\ Leicht-Holme-Newman Index ({\bf LHN-I})\cite{leicht2006vertex},\\ CH2-L2 Index ({\bf CH2-L2})\cite{muscoloni2018local,muscoloni2022adaptive,muscoloni2023stealing}, \\ CH3-L2 Index ({\bf CH3-L2})\cite{muscoloni2018local,muscoloni2022adaptive,muscoloni2023stealing}} \\
\hspace*{\fill} \\
Path & \makecell{Local Path Index ({\bf LP})\cite{lu2009similarity}, \\ Katz Index ({\bf Katz})\cite{ran2021novel},  \\ FriendLink Index ({\bf FL})\cite{papadimitriou2011friendlink}, \\ Shortest Path Length Index ({\bf SPL})\cite{ran2021novel,ran2022predicting}, \\ Relation Strength Similarity Index ({\bf RSS})\cite{chen2012discovering}}\\
\hspace*{\fill} \\
Heterogeneity& \makecell{Heterogeneity Index ({\bf HEI})\cite{shang2019link}, \\ Homogeneity Index ({\bf HOI})\cite{shang2019link}}\\
\hspace*{\fill} \\
Path of Length Three & \makecell{Paths of Length Three Index ({\bf L3})\cite{kovacs2019network}, \\ CH2-L3 Index ({\bf CH2-L3})\cite{muscoloni2018local,muscoloni2022adaptive,muscoloni2023stealing}, \\ CH3-L3 Index ({\bf CH3-L3})\cite{muscoloni2018local,muscoloni2022adaptive,muscoloni2023stealing}}\\
\hline
\end{tabular}
\label{table:features}
\end{table}  

\begin{table*}[htbp]
	\centering
	\caption{{\bf An example that utilizes the maximum capability for feature selection.} The performance of the unsupervised prediction using LHN-I is the same in networks A and B. However, using the $p_1$ and $p_2$ values, the maximum capability of the common neighbor feature can be estimated. The common neighbor feature is not suitable for network A but has potential in network B. The estimation is confirmed by the prediction results of other indexes. The network A is ``Norwegian\_Board\_of\_Directors\_net2mode\_2010-09-01'', and the network B is ``577ee40d58d31bd664bac0ef'' in the dataset  \cite{ghasemian2020stacking,broido2019scale}.}
	\resizebox{17cm}{!}{
	\begin{tabular}{l|rrrrrrrrrrrrrrrr}
	\hline
		 \multicolumn{1}{c}{}& \multicolumn{1}{c}{LHN-I} & \multicolumn{1}{c}{$p_{1}$} & \multicolumn{1}{c}{$p_{2}$}& \multicolumn{1}{c}{$\text{AUC}_\text{upper}$}& \multicolumn{1}{c}{CN} & \multicolumn{1}{c}{AA} & \multicolumn{1}{c}{RA}& \multicolumn{1}{c}{Salton} & \multicolumn{1}{c}{SI}& \multicolumn{1}{c}{HPI}& \multicolumn{1}{c}{HDI}& \multicolumn{1}{c}{Jaccard}& \multicolumn{1}{c}{CH2-L2}& \multicolumn{1}{c}{CH3-L2}\\
		\hline
		\multicolumn{1}{c}{Network A} & \multicolumn{1}{c}{0.497} & \multicolumn{1}{c}{$\textsl{0.0}$}& \multicolumn{1}{c}{$\textsl{0.006}$}& \multicolumn{1}{c}{${\bf 0.497}$} & \multicolumn{1}{c}{0.497} & \multicolumn{1}{c}{0.497}& \multicolumn{1}{c}{0.497} & \multicolumn{1}{c}{0.497} & \multicolumn{1}{c}{0.497}& \multicolumn{1}{c}{0.497}& \multicolumn{1}{c}{0.497}& \multicolumn{1}{c}{0.497}& \multicolumn{1}{c}{0.497}& \multicolumn{1}{c}{0.497} \\
		\multicolumn{1}{c}{Network B} &\multicolumn{1}{c}{0.497} & \multicolumn{1}{c}{$\textsl{0.972}$}& \multicolumn{1}{c}{$\textsl{0.872}$}& \multicolumn{1}{c}{${\bf 0.974}$} &\multicolumn{1}{c}{0.766} & \multicolumn{1}{c}{0.768}& \multicolumn{1}{c}{0.755} & \multicolumn{1}{c}{0.725}& \multicolumn{1}{c}{0.705}& \multicolumn{1}{c}{0.717}& \multicolumn{1}{c}{0.679}& \multicolumn{1}{c}{0.705}& \multicolumn{1}{c}{0.776}& \multicolumn{1}{c}{0.773} \\
		\hline
	\end{tabular}%
	}
\label{table:unsupervisedf}
\end{table*}%

\begin{table*}[htbp]
	\centering
	\caption{{\bf An example that utilizes the maximum capability for method selection.} The unsupervised prediction using index CN gives rise to similar AUC values in networks C and D. Using the $p_1$ and $p_2$ values, the improvement by applying a supervised method can be estimated in both networks. The actual supervised prediction performance by different indexes associated with the common neighbor feature is listed, which is in line with the estimation. The network C is  ``Malaria\_var\_DBLa\_HVR\_networks\_HVR\_networks\_3'', and the network D is ``57574842bd3e93b53c695556'' in the dataset \cite{ghasemian2020stacking,broido2019scale}.}
	\resizebox{17cm}{!}{
	\begin{tabular}{l|rrrrrrrrrrrrrrrr}
	\hline
		  \multicolumn{1}{c}{}& \multicolumn{5}{c|}{Unsupervised}& \multicolumn{9}{c}{Supervised}\\
		\cline{2-17}
		   \multicolumn{1}{c}{}& \multicolumn{1}{c}{CN} & \multicolumn{1}{c}{$p_{1}$} & \multicolumn{1}{c}{$p_{2}$}& \multicolumn{1}{c}{$\text{AUC}_\text{upper}$}& \multicolumn{1}{c|}{$\Delta$}& \multicolumn{1}{c}{CN} & \multicolumn{1}{c}{AA} & \multicolumn{1}{c}{RA}& \multicolumn{1}{c}{Salton} & \multicolumn{1}{c}{SI}& \multicolumn{1}{c}{HPI}& \multicolumn{1}{c}{HDI}& \multicolumn{1}{c}{LHN-I}& \multicolumn{1}{c}{Jaccard}& \multicolumn{1}{c}{CH2-L2}& \multicolumn{1}{c}{CH3-L2}\\
		  \hline
		\multicolumn{1}{c}{Network C} & \multicolumn{1}{c}{0.778} & \multicolumn{1}{c}{$\textsl{0.667}$}& \multicolumn{1}{c}{$\textsl{0.333}$}& \multicolumn{1}{c}{0.778} & \multicolumn{1}{c|}{${\bf 0.111}$}& \multicolumn{1}{c}{0.889} & \multicolumn{1}{c}{0.889}& \multicolumn{1}{c}{0.889} & \multicolumn{1}{c}{0.889} & \multicolumn{1}{c}{0.889}& \multicolumn{1}{c}{0.889}& \multicolumn{1}{c}{0.889}& \multicolumn{1}{c}{0.889}& \multicolumn{1}{c}{0.889}& \multicolumn{1}{c}{0.889}& \multicolumn{1}{c}{0.889} \\
		\multicolumn{1}{c}{Network D} &\multicolumn{1}{c}{0.773} & \multicolumn{1}{c}{$\textsl{0.545}$}& \multicolumn{1}{c}{$\textsl{0.0}$}& \multicolumn{1}{c}{0.773} & \multicolumn{1}{c|}{${\bf 0.0}$}&\multicolumn{1}{c}{0.773} & \multicolumn{1}{c}{0.773}& \multicolumn{1}{c}{0.773} & \multicolumn{1}{c}{0.773}& \multicolumn{1}{c}{0.773}& \multicolumn{1}{c}{0.773}& \multicolumn{1}{c}{0.773}& \multicolumn{1}{c}{0.773}& \multicolumn{1}{c}{0.773}& \multicolumn{1}{c}{0.773}& \multicolumn{1}{c}{0.773} \\
		\hline
	\end{tabular}%
	}
	\label{table:unsupervisedm}
\end{table*}%

\clearpage

\begin{figure*}[ht]
\centering
\includegraphics[width=.8\linewidth]{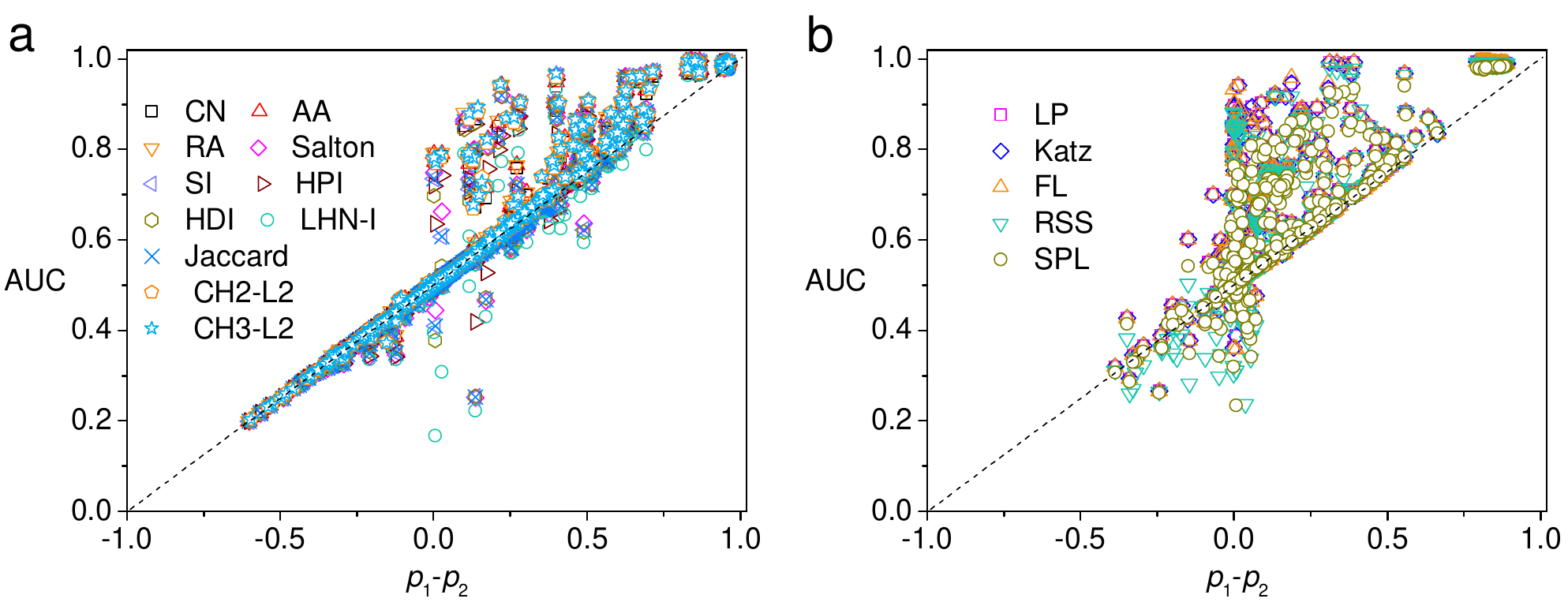}
\caption{{\bf The scaling of the AUC values.}
Eq. (\ref{equation:lower}) and Eq. (\ref{equation:upper}) suggest that the actual prediction by an index fluctuates within $p_1 \times p_2$. Therefore, for the common neighbor feature ($\bf{a}$) and the path feature ($\bf{b}$) whose $p_1 \times p_2$ values are small, the link prediction performance by different indexes roughly scales as $p_{1}-p_{2}$. For each network, we randomly generate 200 realizations of networks with link removal, as well as 200 pairs of $L^P$ and $L^N$ sets (Materials and Methods). The $p_{1}$, $p_{2}$, and the corresponding AUC obtained may vary slightly in different sampled $L^P$'s and $L^N$'s. In the figure, we use the average value.  
}
\label{fig:p1-p2}
\end{figure*}

\clearpage

\clearpage

\clearpage

\clearpage

\renewcommand{\figurename}{{\bf Supplementary Figure}}
\renewcommand{\thefigure}{{\bf S\arabic{figure}}}
\renewcommand{\thesection}{S\arabic{section}}
\renewcommand{\thesubsection}{S\arabic{section}.\arabic{subsection}}
\renewcommand{\tablename}{{\bf Supplementary Table}}
\renewcommand{\thetable}{{\bf S\arabic{table}}}
\renewcommand{\theequation}{S\arabic{equation}}
\def\msec#1{\bigskip\textbf{#1}}
\def\note#1{{\small\color{red}\textbf{[[#1]]}}}

\centerline{\bf Supplementary Material}

\hspace*{\fill}

\centerline{\bf The maximum capability of a topological feature in link prediction}

\hspace*{\fill}

\centerline{Yijun Ran, Xiao-Ke Xu, Tao Jia$^1$}

\hspace*{\fill}

\centerline{\it $^1$To whom correspondence should be addressed: tjia@swu.edu.cn}

\clearpage

\setcounter{figure}{0} 
\setcounter{table}{0} 
\setcounter{equation}{0}

\begin{figure}[ht]
\begin{center}
\resizebox{14cm}{!}{\includegraphics{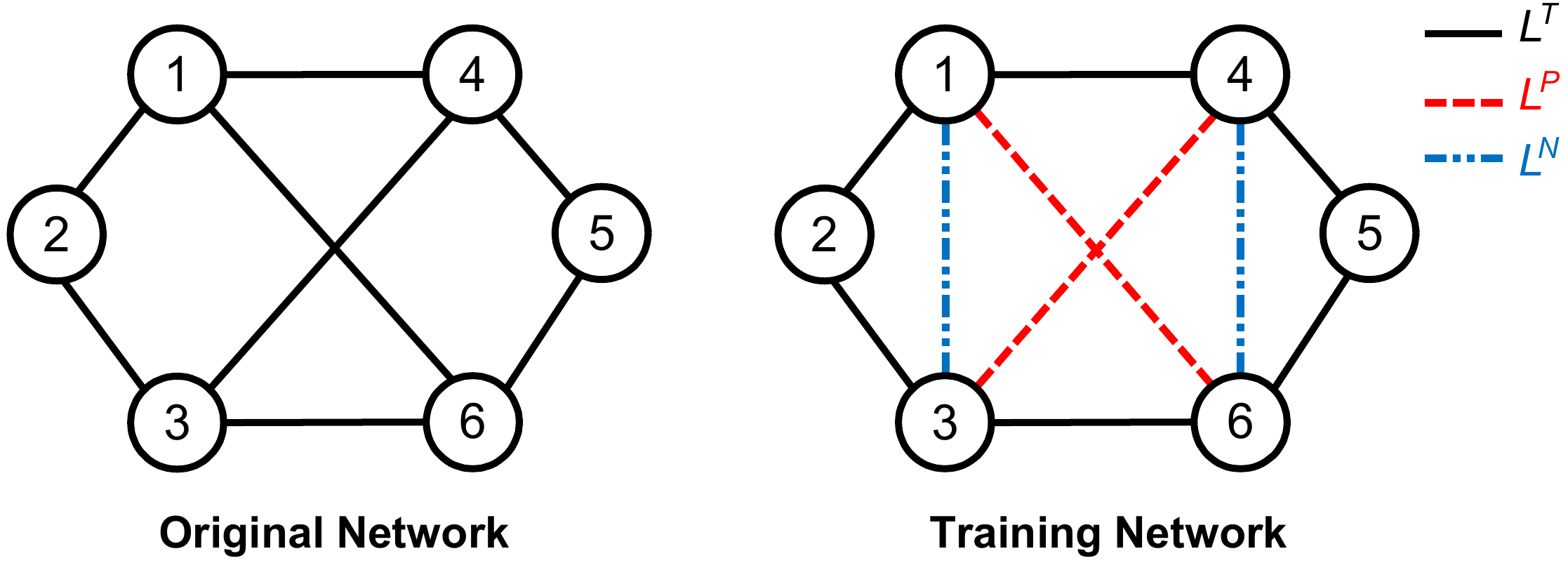}}
\caption{{\bf An example of the link prediction problem.}
The common process of link prediction is that a set of existing links is removed randomly from the original network, which is marked as the positive testing set $L^P$ (the red dash lines in the training network). As the control group, a random set of node pairs that are not connected in the original network is selected as the negative testing set $L^N$ (the blue dash lines in the training network). An index considers the topology based on the rest of the links $L^T$ (the black solid lines in the training network) and assigns a value to each node pair in $L^P$ and $L^N$. In unsupervised prediction, the index values are directly used as the score of samples in $L^P$ and $L^N$. Assume that according to the index value we have $S_{16} = 0.3, S_{34} = 0.58, S_{13} = 0.2$, and $S_{46} = 0.58$. 
The prediction quality is measured by how samples in $L^P$ are ranked ahead of those in $L^N$. 
When using AUC to measure the prediction performance, we usually apply the random sampling approach. In each comparison, we randomly draw a node pair from $L^P$ and a node pair from $L^N$, and compare their scores. Suppose 3 random comparisons are made. Node pairs 1-6 and 1-3, node pairs 1-6 and 4-6, and node pairs 3-4 and 4-6 are selected in each comparison. We have one case where node pair from $L^P$ outscores that from $L^N$, and one case where node pairs from $L^P$ and $L^N$ have an equal score. According to Eq. (\ref{equation:auc}) of the main text, the AUC can be estimated as $\frac{1+0.5}{3} = 0.5$.
When using precision to measure the prediction performance, we rank node pairs according to their scores in descending order. In the example shown, the rank is 3-4, 4-6, 1-6, 1-3. If we select the hyper-parameter $L_\text{k}=2$, the top-two node pairs (3-4 and 4-6) are considered. As node pair 3-4 is the true positive sample whereas node pair 4-6 is not, the precision is 0.5.
}
\label{fig:aucexample}
\end{center}
\end{figure}\noindent 

\clearpage

\begin{figure}[ht]
\begin{center}
\resizebox{14cm}{!}{\includegraphics{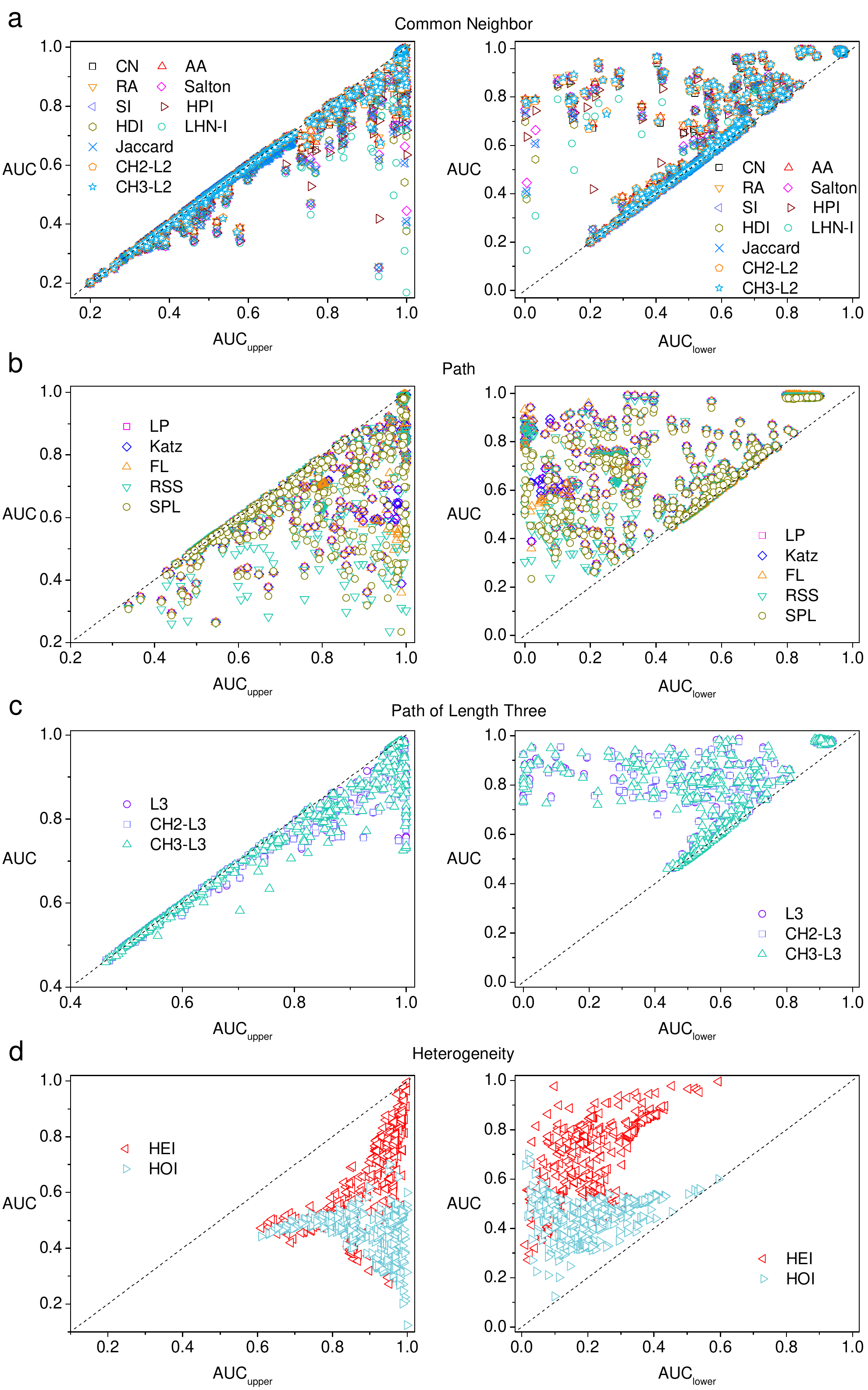}}
\end{center}
\end{figure}\noindent 

\clearpage

\begin{figure}[ht]
\begin{center}
\caption{{\bf The capability of the topological features measured by AUC in the unsupervised prediction.}
Eqs. (\ref{equation:lower}) and (\ref{equation:upper}) in the main text suggest that different indexes have different prediction performances, but all indexes associated with one topological feature share the same $\text{AUC}_\text{upper}$ and $\text{AUC}_\text{lower}$. This is confirmed by 21 indexes related to 4 topological features: common neighbor $\bf{(a)}$, path $\bf{(b)}$, path of length three $\bf{(c)}$, and heterogeneity $\bf{(d)}$. For each network, we randomly generate 200 realizations of networks with link removal, as well as 200 pairs of $L^P$ and $L^N$ sets. In the figure, we use the average value of 200 samples.
}
\label{fig:unsupupper}
\end{center}
\end{figure}\noindent 

\clearpage

\begin{figure}[ht]
\begin{center}
\resizebox{16cm}{!}{\includegraphics{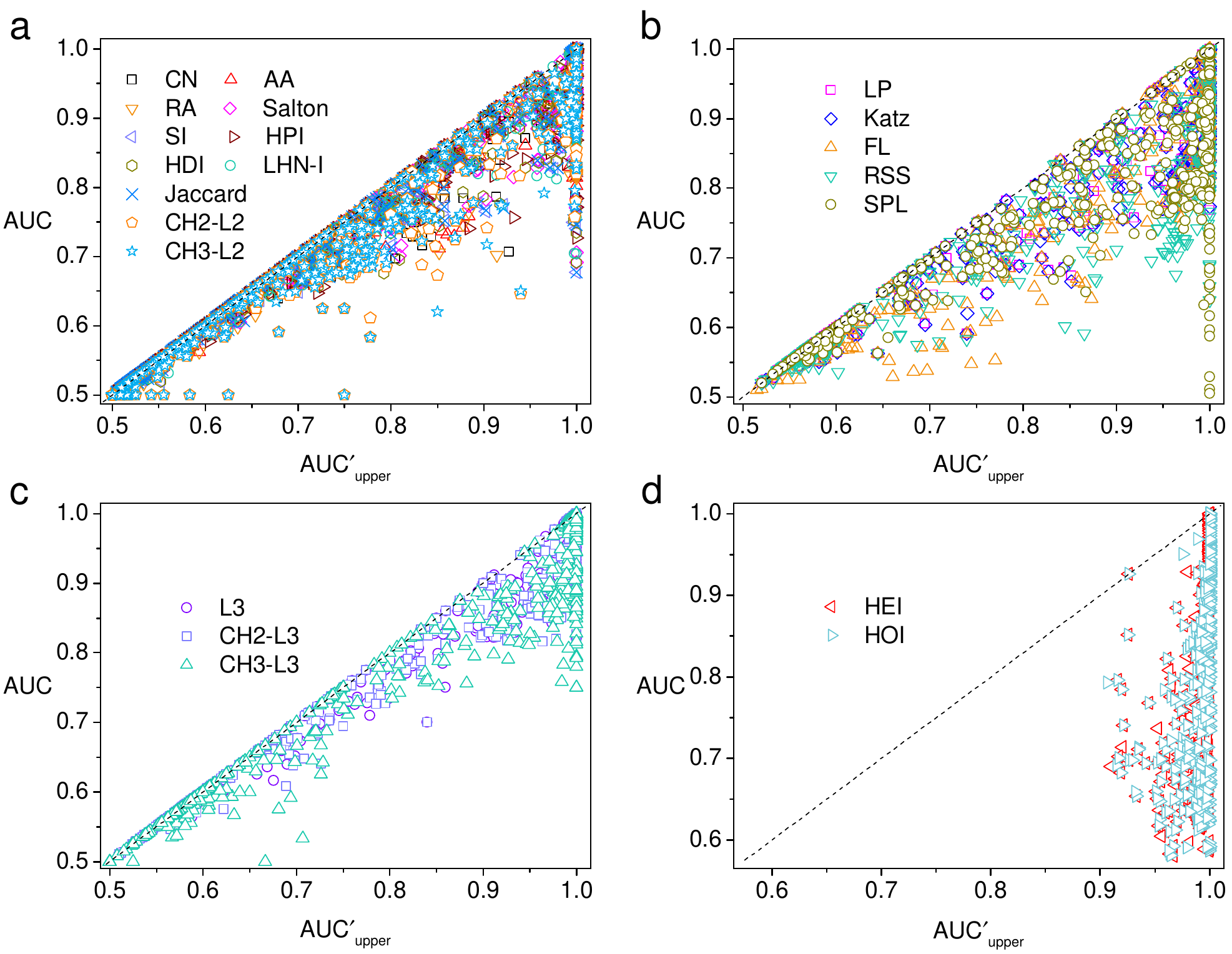}}
\caption{{\bf The maximum capability of the topological features measured by AUC in the supervised prediction.}
Eq. (\ref{equation:upper2}) in the main text suggests that $\text{AUC}^{\prime}_\text{upper}$ sets the upper bound of the supervised prediction. This is confirmed by 21 indexes related to 4 topological features: common neighbor $\bf{(a)}$, path $\bf{(b)}$, path of length three $\bf{(c)}$, and heterogeneity $\bf{(d)}$. For each network, we randomly generate 200 realizations of networks with link removal, as well as 200 pairs of $L^P$ and $L^N$ sets. In the figure, we choose the highest AUC from 200 samples as the performance of an index.}
\label{fig:supupper}
\end{center}
\end{figure}\noindent 

\clearpage

\begin{figure}[ht]
\begin{center}
\resizebox{16cm}{!}{\includegraphics{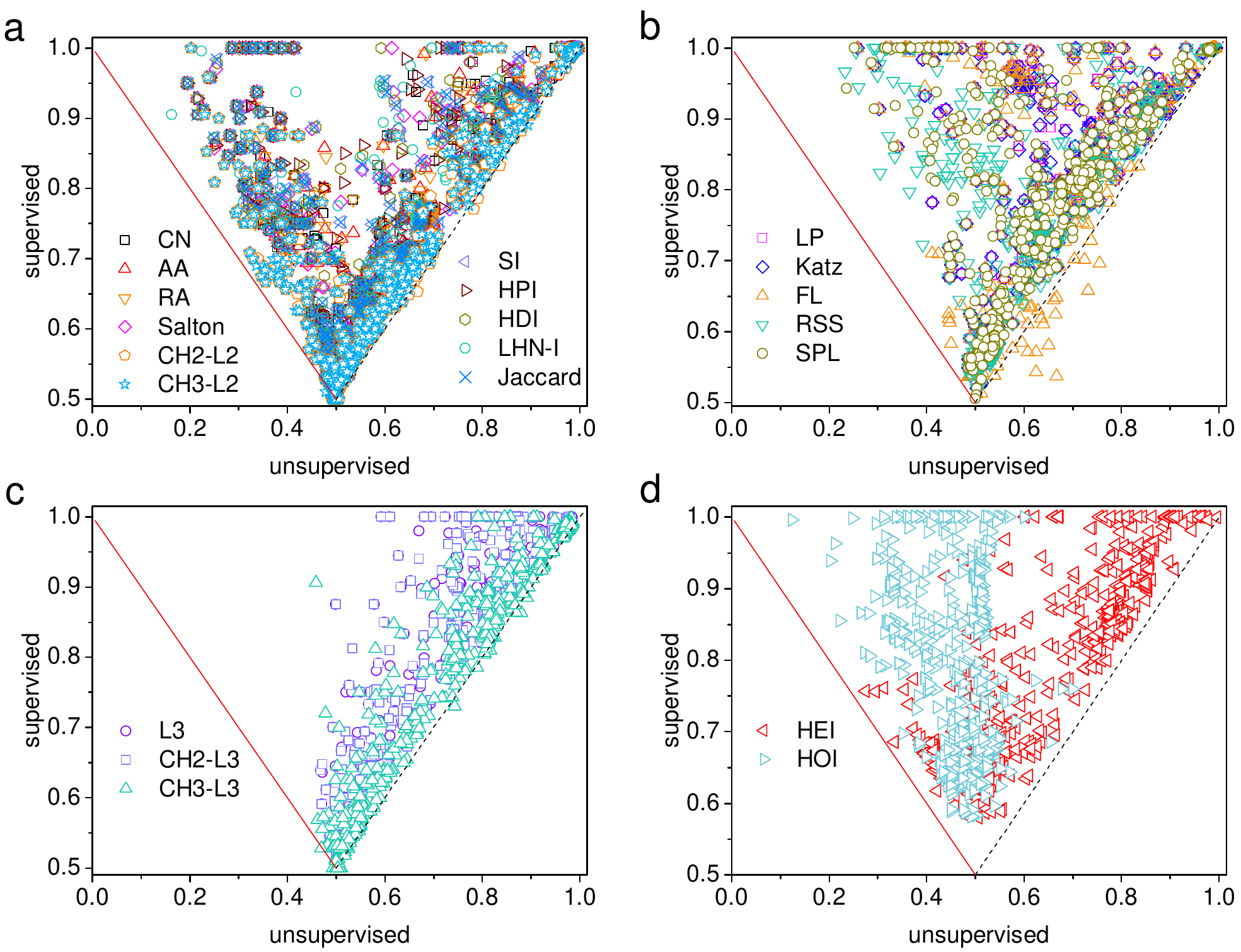}}
\caption{{\bf Performance comparison measured by AUC between the unsupervised and supervised prediction.}
The dashed line is $y=x$. Almost all data points are above the line $y=x$, indicating that the supervised prediction in general gives rise to a higher AUC value compared with the unsupervised prediction in the same network based on the same index. In unsupervised prediction, the AUC value can be less than 0.5  (Fig. \ref{fig:unsupupper}), suggesting that the feature is more prominent in negative samples. A simple fix in this circumstance is to consider the feature as the ``negative feature''. When the AUC is below 0.5, the predictor will consider that a smaller index value corresponds to a higher probability that two nodes are truly connected. After this modification is applied, predictions with $\text{AUC} < 0.5$ will be lifted. The red solid line is $y = 1-x$, corresponding to the performance after considering the ``negative feature''. All data points are above $y = 1-x$. Hence, the improvement by the supervised approach is not merely taking advantage of the ``negative feature''. We use 21 indexes related to 4 topological features: common neighbor $\bf{(a)}$, path $\bf{(b)}$, path of length three $\bf{(c)}$, and heterogeneity $\bf{(d)}$. For each network, we randomly generate 200 realizations of networks with link removal, as well as 200 pairs of $L^P$ and $L^N$ sets. In the figure, we use the average value of 200 samples for unsupervised results. For supervised results, we choose the highest AUC from 200 samples as the performance of an index.  
}
\label{fig:simvsml}
\end{center}
\end{figure}\noindent 

\clearpage

\begin{figure}[ht]
\begin{center}
\resizebox{16cm}{!}{\includegraphics{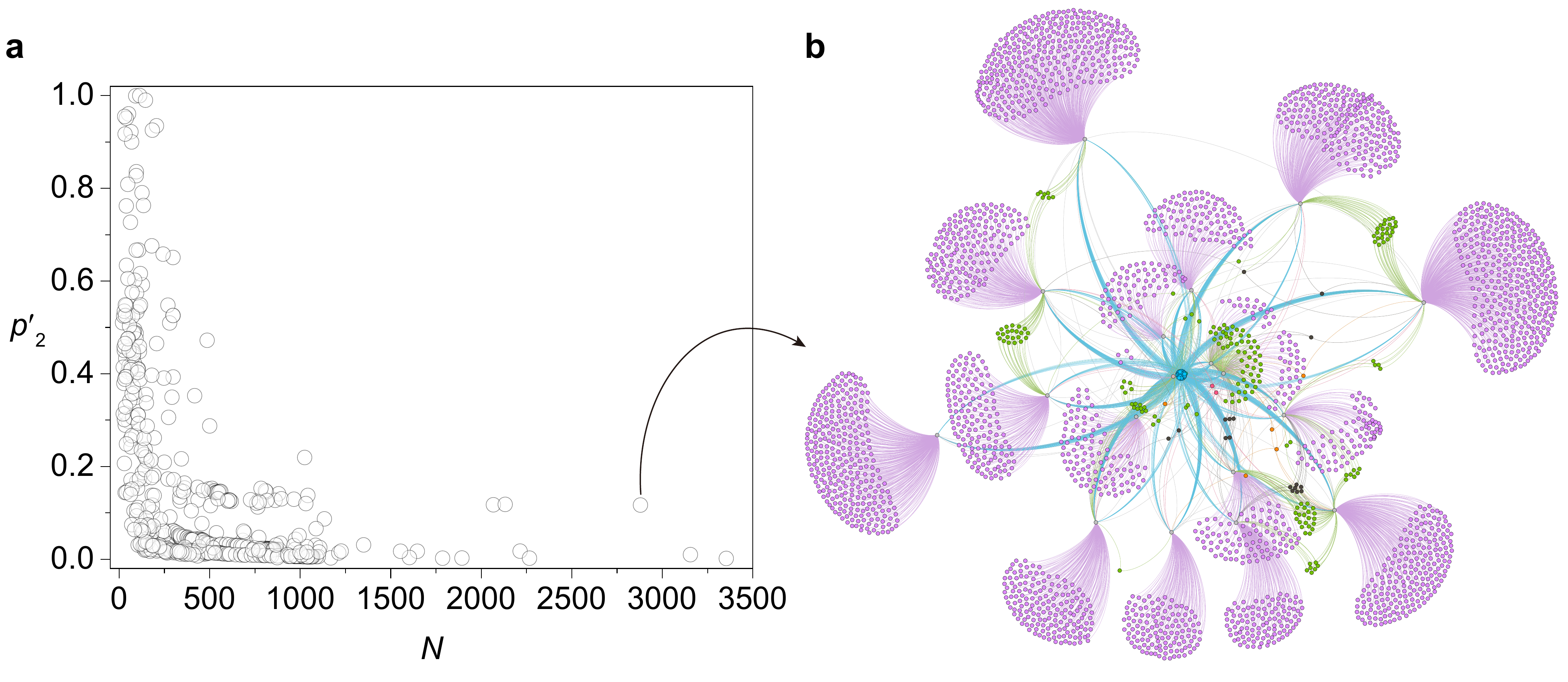}}
\caption{{\bf The structural characteristics of the common neighbor feature.}
$\bf{(a)}$ In the analytical expression of $p^{\prime}_{2}$ (Eq. (\ref{equation:p21})), the denominator is dominated by $N^2$. Hence, $p^{\prime}_{2}$ is expected to decay fast with the network size $N$. For a sufficiently large network, $p^{\prime}_{2}$ is expected to be zero. This is generally held in the empirical analyses of 550 networks. But exceptions are also found. Some large networks have a relatively large $p^{\prime}_{2}$ value. Such networks have many ``leaves'' structures, as illustrated in $\bf{(b)}$. The leaves will make the number of open triangles proportional to $N^2$. Hence, $p^{\prime}_{2}$ will not vanish in such networks. 
The network in $\bf{(b)}$ is ``5886685ba411221d0e7c677e'' in the data set.
}
\label{fig:tp2n}
\end{center}
\end{figure}\noindent 

\clearpage

\begin{table}
\centering
\caption{{\bf The $p_1 \times p_2$ value of a topological feature.}
The $p_1 \times p_2$ (mean$\pm$standard deviation) of a topological feature on the testing set ($L^P$ and $L^N$) applied to 550 empirical networks. For each network, we randomly generate 200 realizations of networks with link removal, as well as 200 pairs of $L^P$ and $L^N$ sets. For the $p_1$ and $p_2$ values of each network, we use the average value of 200 samples.}
\begin{tabular}{cc}
Features & $p_{1} \times p_{2}$  \\
\hline
Common Neighbor & $0.057\pm0.152$ \\
Path of Length Three& $ 0.155\pm0.241$ \\
Path &$ 0.270\pm0.315 $\\
Heterogeneity &$0.695\pm0.126$ \\
\hline
\end{tabular}
\label{table:p1p2}
\end{table}

\clearpage

\section{The 21 indexes used in this study and the classification of these indexes}\label{section:s1}

In this study, we select 21 indexes associated with 4 topological features to validate our quantitative framework presented in the main text. Here, we describe in detail the 21 indexes and how they are classified into 4 families.

{\bf A family of indexes based on common neighbor.} 
We consider 11 indexes that gauge the common neighbor feature.

(1) Common Neighbor Index (CN)

The CN directly counts the number of common neighbors two nodes share \cite{Liben2007The}. It is defined as
\begin{equation}
S_{ab}^{\text{CN}} = \lvert {n(a) \cap n(b)} \rvert, 
\end{equation}
where $n(a)$ denotes the set over all neighbors of node $a$.
 
(2) Adamic-Adar Index (AA)

Adamic and Adar propose the AA index that computes the similarity between two web pages \cite{Adamic2003Friends}. The AA emphasizes less-connected common neighbors. It is defined as 
\begin{equation}
S_{ab}^{\text{AA}} = \sum_{c \in {n(a) \cap n(b)}} \frac{1} {\log k(c)}, 
\end{equation}
where $k(c) = \lvert n(c) \rvert$ is the degree of node $c$.

(3) Resource Allocation Index (RA)

Motivated by the physical process of resource allocation, Zhou \textit{et al.} propose the RA index that puts penalties to large degree nodes \cite{zhou2009predicting}. The RA is defined as
\begin{equation}
S_{ab}^{\text{RA}} = \sum_{c \in {n(a) \cap n(b)}} \frac{1} {k(c)}.
\end{equation}

(4) Salton Index (Salton)

The Salton index is also called cosine similarity \cite{sun2020revealing}, which is defined as 
\begin{equation}
S_{ab}^{\text{Salton}} =  \frac{\lvert {n(a) \cap n(b)} \rvert} {\sqrt {k(a)\times k(b)}}.
\end{equation}

(5) S$\o$rensen Index (SI)

The SI is usually used in ecological science \cite{ghorbanzadeh2021hybrid}, which is defined as 
\begin{equation}
S_{ab}^{\text{SI}} =   \frac{2 \times \lvert {n(a) \cap n(b)} \rvert} {k(a) + k(b)}.
\end{equation}

(6) Hub Promoted Index (HPI)

The HPI aims to measure the degree of topological overlap between two nodes in metabolic networks \cite{Ravasz2002Hierarchical}, which is defined as 
\begin{equation}
S_{ab}^{\text{HPI}} =  \frac{\lvert {n(a) \cap n(b)} \rvert} {\min {(k(a), k(b))}}.
\end{equation}

(7) Hub Depressed Index (HDI)

The HDI is similar to the HPI. The difference is that HDI emphasizes the role of nodes with large degrees \cite{zhou2009predicting}. It is defined as 
\begin{equation}
S_{ab}^{\text{HDI}} =  \frac{\lvert {n(a) \cap n(b)} \rvert} {\max {(k(a), k(b))}}.
\end{equation}

(8) Leicht-Holme-Newman Index (LHN-I)

Leicht \textit{et al.} propose the LHN-I, which assigns a high value to the node pair with many common neighbors \cite{leicht2006vertex}. It is defined as 
\begin{equation}
S_{ab}^{\text{LHN-I}} =  \frac{\lvert {n(a) \cap n(b)} \rvert} {k(a)\times k(b)}.
\end{equation}

(9) Jaccard Index (Jaccard)

The Jaccard directly normalizes the number of common neighbors \cite{Liben2007The}. It is defined as 
\begin{equation}
S_{ab}^{\text{Jaccard}} =  \frac{\lvert {n(a) \cap n(b)} \rvert} {\lvert {n(a) \cup n(b)} \rvert}.
\end{equation}

(10) CH2-L2 Index (CH2-L2)

Muscoloni \textit{et al.} adopt the Cannistraci-Hebb network automaton model to extend the local community paradigm to paths of length 2 \cite{muscoloni2018local,muscoloni2022adaptive,muscoloni2023stealing}. Here, we denote $P_{ab}$ by the set of nodes on all paths of length 2 that connect nodes $a$ and $b$. If node $u$ is the intermediate node on a path of length 2 linking nodes $a$ and $b$, the CH2-L2 is defined as
\begin{equation}
S_{ab}^{\text{CH2-L2}} = \sum_{u}\frac{l_{au}l_{ub}(1+i(u))}{1+e(u)},
\end{equation}
where $i(u)$ is the number of internal links between node $u$ and nodes except nodes $a$ and $b$ in $P_{ab}$, $e(u)$ is the number of external links between node $u$ and nodes not in $P_{ab}$. The $l_{au}=1$ if node $a$ and node $u$ have a link, $l_{au}=0$ otherwise. 

(11) CH3-L2 Index (CH3-L2)

According to the Cannistraci-Hebb rule, Muscoloni \textit{et al.} propose the CH3-L2 index implemented solely based on the minimization of $e(u)$ \cite{muscoloni2018local,muscoloni2022adaptive,muscoloni2023stealing}. Similar to the CH2-L2, the CH3-L2 is defined as
\begin{equation}
S_{ab}^{\text{CH3-L2}} = \sum_{u}\frac{l_{au}l_{ub}}{1+e(u)}.
\end{equation}

{\bf A family of indexes based on path.}
We consider 5 indexes gauging the path feature. To reduce the computational complexity and to unify the path feature, we focus on path lengths less than or equal to 4. 

(1) Local Path Index (LP)

Motivated by the CN index, L{\"u} \textit{et al.} propose the LP index \cite{lu2009similarity}. A large number of studies have shown that the predictive ability of the LP index is stronger than that of the CN on many real networks \cite{cao2019network,martinez2016survey,kumar2020link}. The definition of LP is
\begin{equation}
S_{ab}^{\text{LP}} = A^2_{ab}+ \beta A^3_{ab} + \beta^2 A^4_{ab}, 
\end{equation}
where the $A^i_{ab}$ is the number of paths of length $i$ that links node $a$ and $b$, and $\beta$ controls the weight of paths with different lengths. In this study, we use $\beta = 0.02$. 

(2) Katz Index (Katz)

The Katz can be regarded as an ensemble method that directly sums the number of all paths \cite{ran2021novel}. It is defined as 
\begin{equation}
S_{ab}^{\text{Katz}} =  \beta^2 A^2_{ab}+ \beta^3 A^3_{ab}+\beta^4 A^4_{ab}, 
\end{equation}
where $\beta$ controls the weight of paths with different lengths. Katz is similar to LP. The difference between them is that Katz assigns exponentially decaying weights into long paths whereas the decaying weights in LP are slower. In this study, we use $\beta = 0.02$. 

(3) FriendLink Index (FL)

The basic idea of the FL index is that two people who have more and shorter paths are more likely to become friends on social networks \cite{papadimitriou2011friendlink}. It is defined as 
\begin{equation}
S_{ab}^{\text{FL}} = \sum_{i=2}^{l}\frac{1}{i-1} \cdot \frac{\lvert paths_{ab}^{i} \rvert}{\prod_{j=2}^{i}(N-j)}, 
\end{equation}
where $N$ is the number of nodes in a network, $l$ is the length of the longest path between nodes $a$ and $b$. The $\lvert paths_{ab}^{i} \rvert$ is the number of paths with length $i$ between nodes $a$ and $b$. In this study, we control $l = 4$. 

(4) Relation Strength Similarity Index (RSS)

The RSS proposed by Chen \textit{et al.} is to measure the relative degree of similarity between two nodes \cite{chen2012discovering}. The relation strength is defined as 
\begin{equation*}
R(ab)=
\begin{cases}
 \frac{\alpha_{ab}}{\sum_{\forall x \in n(a)}\alpha_{ax}} &\text{if $a$ and $b$ are adjacent} \\
0&\text{otherwise},
\end{cases}
\end{equation*}
where $\alpha_{ab}$ is the weight between nodes $a$ and $b$, which can be any value. Here, the $R(ab)$ is not symmetric, \emph{i.e.} $R(ab) \neq R(ba)$. To make arbitrary nodes available, Chen \textit{et al.} also propose the generalized relation strength as
\begin{equation*}
R_{p_l}^{*}(ac) = \prod_{k=1}^{K-1}R(b_{k}b_{k+1}), 
\end{equation*}
where the $p_l$ is a set of paths between nodes $a$ and $c$. The $p_l$ is formed by $b_1, b_2, ..., b_K$ in which the $b_1$ represents node $a$ and the $b_K$ represents node $c$. To make the calculation tractable, Chen \textit{et al.} control the path length less than $r$, yielding 
\begin{equation*}
R_{p_l}^{*}(ac)
\begin{cases}
\prod_{k=1}^{K}R(b_{k}b_{k+1}) &\text{if $K \leq r$} \\
0&\text{otherwise}.
\end{cases}
\end{equation*}
Taken together, if there are $P_l$ paths with length shorter than $r$ from node $a$ to node $b$, the RSS is defined as 
\begin{equation}
S_{ab}^{\text{RSS}} = \sum_{l=1}^{P_l}R_{p_l}^{*}(ab).
\end{equation}
In this study, we control $r = 4$. 

(5) Shortest Path Length Index (SPL)

Ran \textit{et al.} show that the shorter the shortest path length between two unconnected nodes is, the more likely they are to form a link in the long-path networks \cite{ran2021novel,ran2022predicting}. The SPL is defined as 
\begin{equation}
S_{ab}^{\text{SPL}} = \frac{1}{d_{ab}-1},
\end{equation}
where the $d_{ab}$ is the length of the shortest path between nodes $a$ and $b$. As the path length is limited to be not greater than 4, $d_{ab}$ in this study falls within the range $[2,4]$. If there is no path within length 4 connecting node $a$ and $b$, $d_{ab} \to \infty$ which gives $S_{ab}^{\text{SPL}} = 0$.

{\bf A family of indexes based on path of length three.} 
Different from the path feature that includes connections with different lengths, the feature path of length three only considers connections with lengths equal to three. Therefore, the path of length three is different from the path feature. We consider 3 indexes that gauge the feature path of length three.

(1) Paths of length three Index (L3)

Kov{\'a}cs \textit{et al.} propose a degree-normalized index based on paths of length three (L3). L3 is found to have a remarkable advantage compared with the indexes based on common neighbors in predicting protein-protein interactions \cite{kovacs2019network}. The L3 index is defined as 
\begin{equation}
S_{ab}^{\text{L3}} = \sum_{uv}\frac{l_{au}l_{uv}l_{vb}}{\sqrt{k(u)k(v)}}, 
\end{equation}
where nodes $u$ and $v$ are the intermediate nodes on a path of length 3 linking nodes $a$ and $b$. The $l_{au}=1$ if node $a$ and node $u$ have a link, $l_{au}=0$ otherwise. 

(2) CH2-L3 Index (CH2-L3)

Motivated by the L3 \cite{kovacs2019network}, Muscoloni \textit{et al.} adopt the Cannistraci-Hebb network automaton model to extend the local community paradigm to paths of length 3 \cite{muscoloni2018local,muscoloni2022adaptive,muscoloni2023stealing}. Denote $P_{ab}$ by the set of nodes on all paths of length 3 that connect nodes $a$ and $b$. The CH2-L3 is defined as
\begin{equation}
S_{ab}^{\text{CH2-L3}} = \sum_{uv}\frac{l_{au}l_{uv}l_{vb}\sqrt{(1+i(u))(1+i(v))}}{\sqrt{(1+e(u))(1+e(v))}}, 
\end{equation}
where $i(u)$ is the number of internal links between node $u$ and nodes except nodes $a$ and $b$ in $P_{ab}$, $e(u)$ is the number of external links between node $u$ and nodes not in $P_{ab}$.

(3) CH3-L3 Index (CH3-L3)

According to the Cannistraci-Hebb rule, Muscoloni \textit{et al.} also propose the CH3-L3 index that puts penalties to the $e(u)$ and $e(v)$ \cite{muscoloni2018local,muscoloni2022adaptive,muscoloni2023stealing}. Similar to the CH2-L3, the CH3-L3 is defined as
\begin{equation}
S_{ab}^{\text{CH3-L3}} = \sum_{uv}\frac{l_{au}l_{uv}l_{vb}}{\sqrt{(1+e(u))(1+e(v))}}.
\end{equation}

{\bf A family of indexes based on heterogeneity.} 
We consider 2 indexes gauging the heterogeneity feature. 

(1) Heterogeneity Index (HEI)

Shang \textit{et al.} find that the indexes based on common neighbors fail to identify missing links in the tree-like networks \cite{shang2019link}. To solve this problem, they take advantage of network heterogeneity and propose the heterogeneity index (HEI). The HEI is defined as 
\begin{equation}
S_{ab}^{\text{HEI}} = \lvert {k(a) - k(b)} \rvert ^ \beta, 
\end{equation}
where $\beta$ is a free heterogeneity exponent. In this work, we set $\beta = 0.02$.

(2) Homogeneity Index (HOI)

Based on the assumption that network homogeneity plays a major role in homogeneous networks \cite{shang2019link}, Shang \textit{et al.} define HOI as
\begin{equation}
S_{ab}^{\text{HOI}} = \frac{1}{\lvert {k(a) - k(b)} \rvert ^ \beta}.
\end{equation}
In this work, we set $\beta = 0.02$.

{\bf Index classification.} 
From the description of the 21 indexes, we can intuitively link the index with the associated topological feature. To further validate the index classification, we consider the following property.

\begin{center}
{\it If two indexes are associated with the same topological feature, they will have the same $p_1$ and $p_2$ values for the given $L^P$ and $L^N$.}
\end{center}

$p_1$ is the percentage of samples in $L^P$ that hold the topological feature, and $p_2$ is the percentage of samples in $L^N$ that hold the topological feature. Technically, $p_1$ and $p_2$ can be measured by counting the number of samples whose index value is 0. From the definition of those indexes, it is obvious that for indexes in the family of the common neighbor feature, they all give $S_{ab} = 0$ for two nodes that do not share any common neighbor. For indexes in the family of the path feature, they all give $S_{ab} = 0$ for two nodes not connected by a path (within length 4). For indexes in the family of the path of length three feature, they all give $S_{ab} = 0$ for two nodes not connected by a path of length 3. For indexes in the family of heterogeneity feature, they all give $S_{ab} = 0$ for two nodes with the same degree. We empirically calculate $p_1$ and $p_2$ values by each index and check if they are the same for indexes classified into the four categories. The numerical $p_1$ and $p_2$ values confirm the classification. 

\clearpage

\section{The results associated with Preferential Attachment}\label{section:s2}
\begin{figure}[ht]
\begin{center}
\resizebox{14cm}{!}{\includegraphics{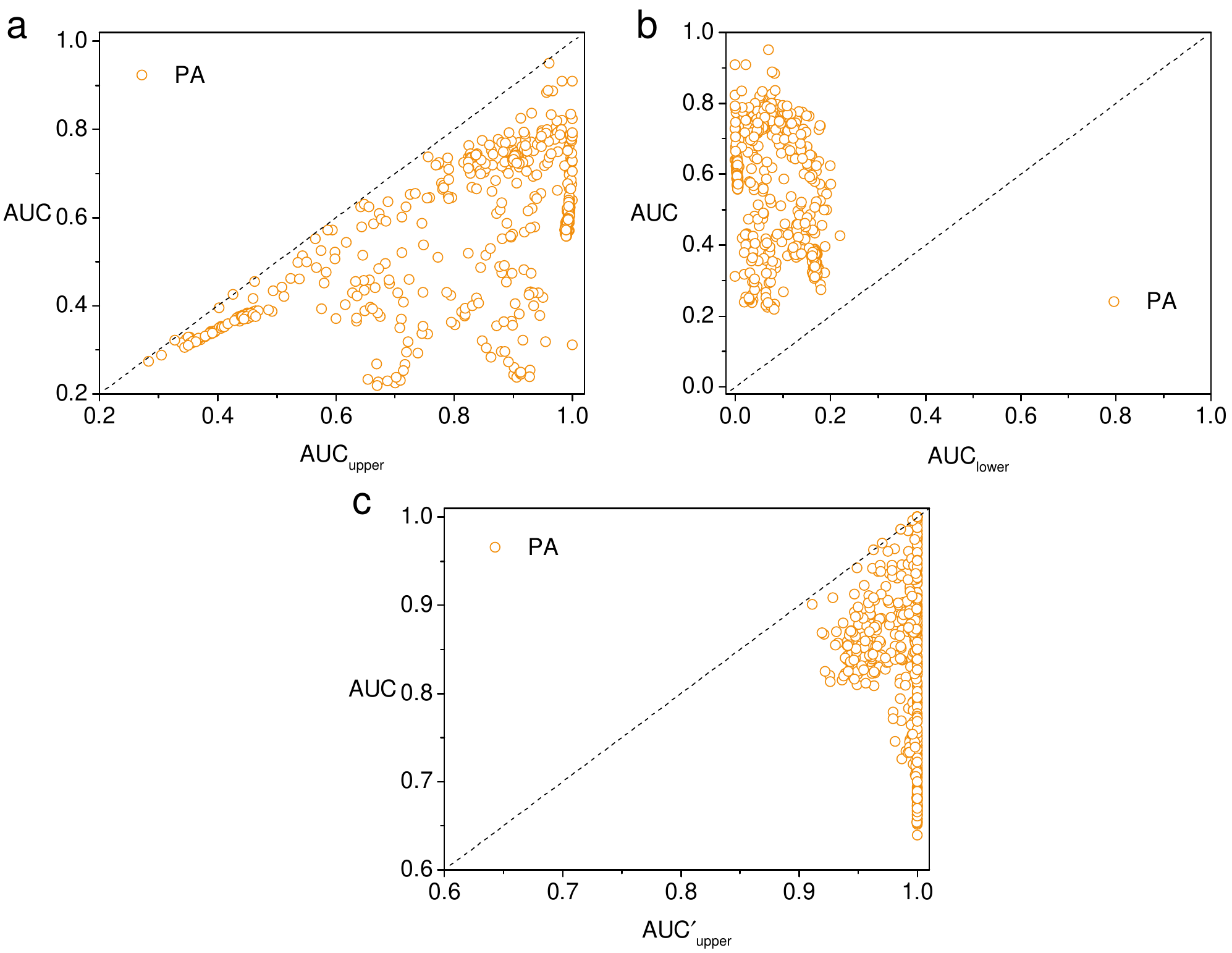}}
\caption{{\bf The capability of the preferential attachment feature measured by AUC.}
Eqs. (\ref{equation:lower}) and (\ref{equation:upper}) in the main text suggest that different indexes have different prediction performances, but all indexes associated with one topological feature share the same $\text{AUC}_\text{upper}$ and $\text{AUC}_\text{lower}$. Here this is confirmed by the PA method related to the preferential attachment feature in $\bf{(a)}$ and $\bf{(b)}$. Eq. (\ref{equation:upper2}) in the main text suggests that $\text{AUC}^{\prime}_\text{upper}$ sets the upper bound of the supervised prediction. This is also confirmed by the PA method related to the preferential attachment feature in $\bf{(c)}$.
}
\label{fig:paauc}
\end{center}
\end{figure}

\begin{figure}[ht]
\begin{center}
\resizebox{16cm}{!}{\includegraphics{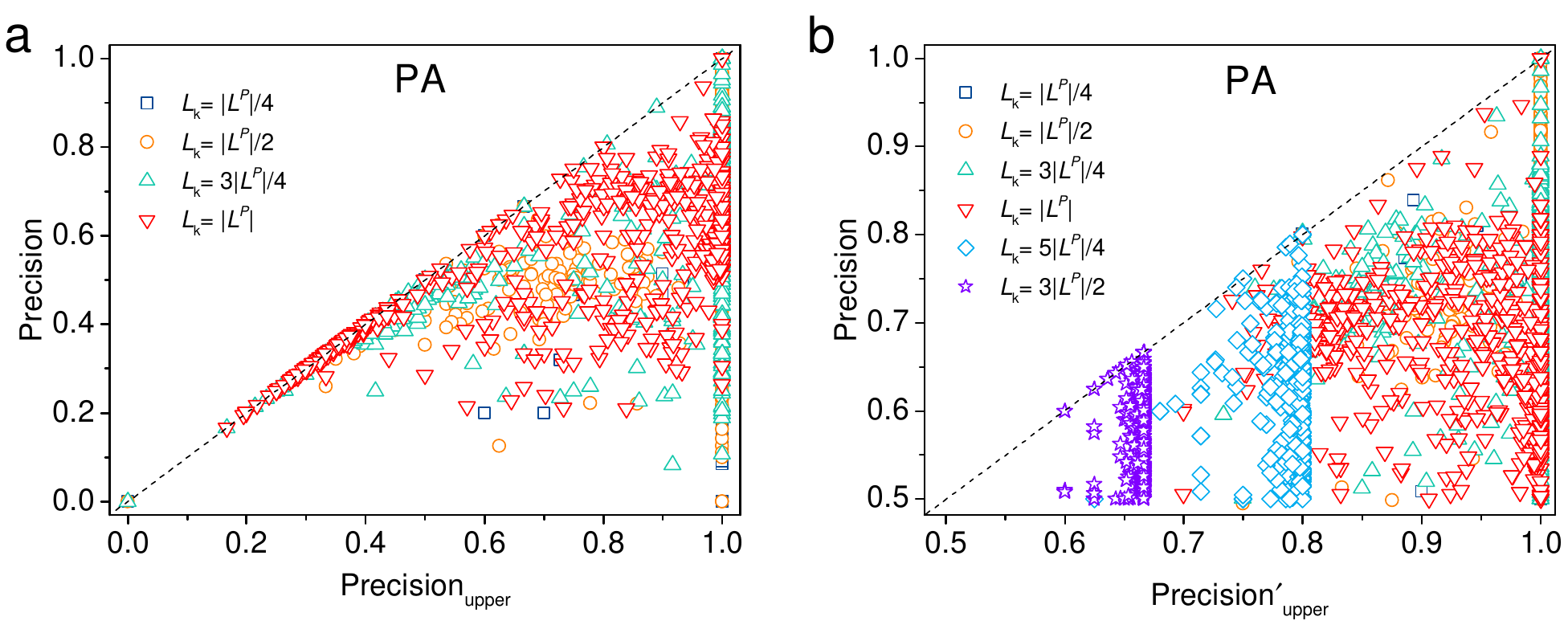}}
\caption{{\bf The maximum capability of the preferential attachment feature measured by precision.}
We choose the PA to measure the performance of the unsupervised and supervised prediction in all 550 networks. $\bf{(a)}$ For different choices of $L_\text{k}$ ($L_\text{k}= |L^P|/4, |L^P|/2, 3|L^P|/4, |L^P|$), the measured precision is equal to or below $\text{Precision}_\text{upper}$, supporting the claim that $\text{Precision}_\text{upper}$ gives the maximum capability of a feature measured by precision in the unsupervised approach. $\bf{(b)}$ For different choices of $L_\text{k}$ ($L_\text{k}= |L^P|/4, |L^P|/2, 3|L^P|/4, |L^P|, 5|L^P|/4, 3|L^P|/2$), the measured precision is equal to or below $\text{Precision}^{\prime}_\text{upper}$, supporting the theoretical results for the maximum capability of a feature measured by precision in the supervised approach.
}
\label{fig:paprecision}
\end{center}
\end{figure}

The index Preferential Attachment (PA) is one of the early indexes proposed in the link prediction problem, which is related to the product of two node’s degree \cite{ran2021novel}. Here we perform prediction results by PA in 550 different real networks. Fig. \ref{fig:paauc} shows that prediction results by PA in the unsupervised and supervised approaches still follow the capability predicted. For all networks, the prediction measured by precision should be below what Eq. (\ref{equation:preupper1}) and Eq. (\ref{equation:preupper2}) yield, respectively. This is also confirmed in Fig. \ref{fig:paprecision}. These show that our finding also works for this feature with one single index.

\noindent

\noindent 

\clearpage

\section{Two alternative experiment setups}\label{section:s3} 
To make the assessment more general, we here evaluate the other two alternative experiment setups. We call the first experiment setup {\it sample\textsl{1}} in the following discussion. {\it Sample\textsl{1}} uses balanced positive and negative samples. It differs from the experiment setup in the main text for the size of the training and testing set. In particular, {\it sample\textsl{1}} randomly removes 20\% of $L$ links as the missing links. 80\% of the removed links (16\% of $L$ links) are used as the positive set in the training and the rest of 20\% of the removed links are used as the positive set in the testing. In both training and testing, randomly selected nonexistent links compose the negative set with the same size as the positive set. The second experiment setup is called {\it sample\textsl{2}} in the following discussion. {\it Sample\textsl{2}} uses imbalanced positive and negative samples. Specifically,  {\it sample\textsl{2}} randomly removes 20\% of $L$ links as the missing links. In the training step, the positive set is composed of 80\% of the removed links (16\% of $L$ links), and the negative set is composed of 80\% of all nonexistent links. In the testing step, the positive set is composed of the rest 20\% of the removed links, and the negative set is composed of the rest 20\% of all nonexistent links. 

\begin{figure}[ht]
\begin{center}
\resizebox{16cm}{!}{\includegraphics{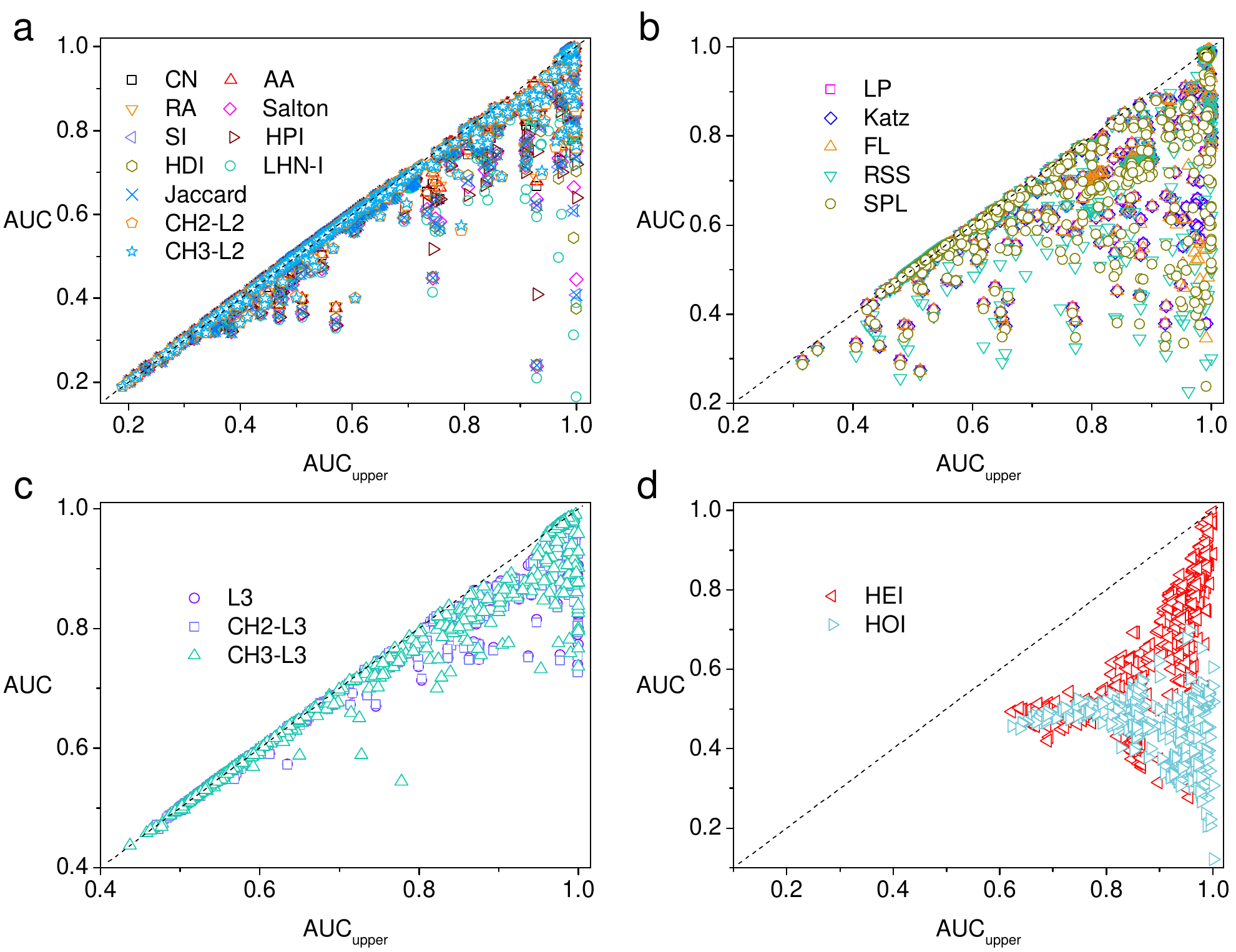}}
\caption{{\bf Unsupervised prediction measured by AUC based on balanced positive and negative samples by {\it sample\textsl{1}}.}
Eqs. (\ref{equation:lower}) and (\ref{equation:upper}) in the main text suggest that different indexes have different prediction performances, but all indexes associated with one topological feature share the same $\text{AUC}_\text{upper}$ and $\text{AUC}_\text{lower}$. This is confirmed by 21 indexes related to 4 topological features: common neighbor $\bf{(a)}$, path $\bf{(b)}$, path of length three $\bf{(c)}$, and heterogeneity $\bf{(d)}$. For each network, we randomly generate 200 realizations of networks with link removal, as well as 200 pairs of $L^P$ and $L^N$ sets. In the figure, we use the average value of 200 samples. The same quantitative analysis in the left panel of Fig. \ref{fig:unsupupper} is repeated.
}
\label{fig:unsupupper08}
\end{center}
\end{figure}

\begin{figure}[ht]
\begin{center}
\resizebox{16cm}{!}{\includegraphics{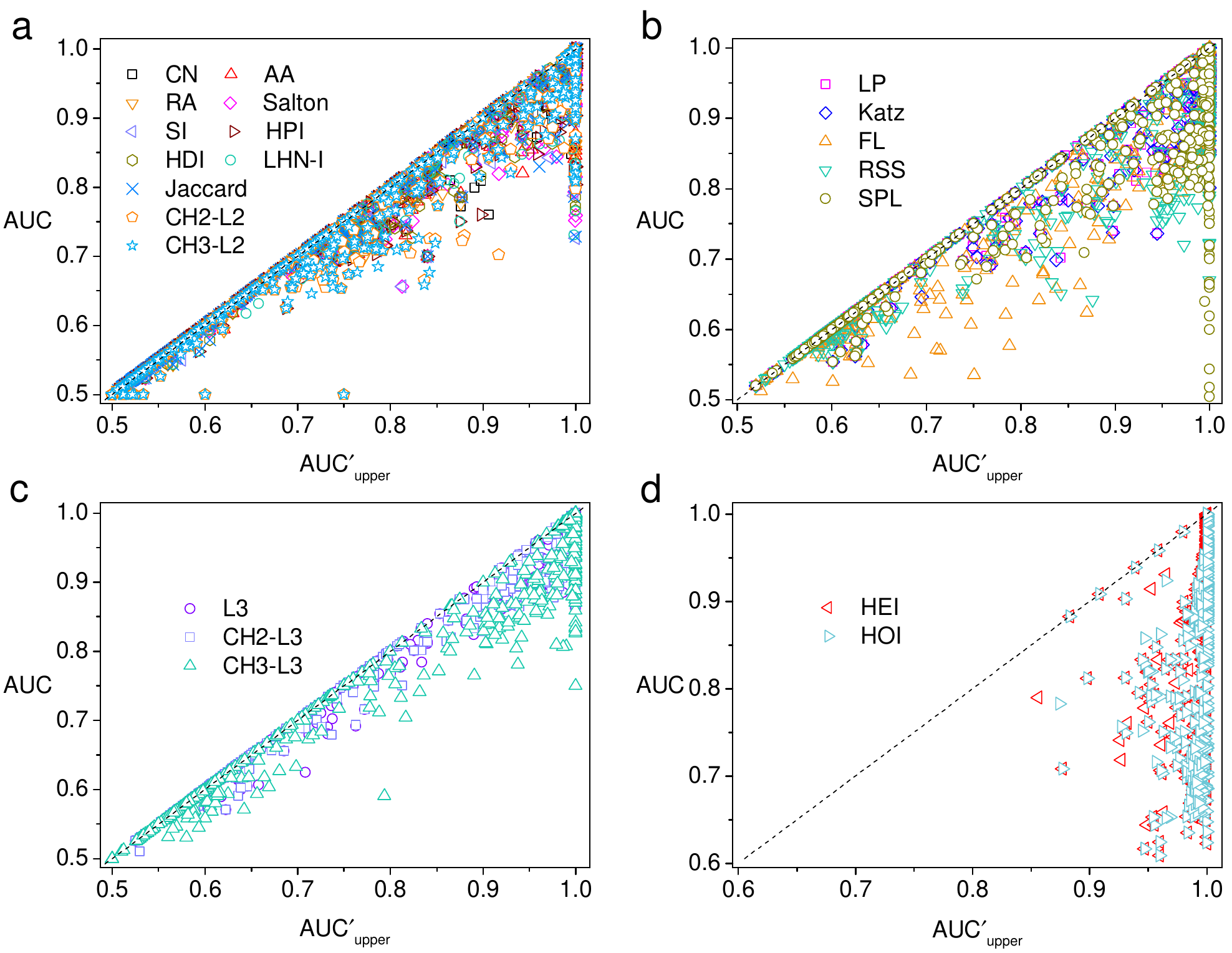}}
\caption{{\bf Supervised prediction measured by AUC based on balanced positive and negative samples by {\it sample\textsl{1}}.}
Eq. (\ref{equation:upper2}) in the main text suggests that $\text{AUC}^{\prime}_\text{upper}$ sets the upper bound of the supervised prediction. This is confirmed by 21 indexes related to 4 topological features: common neighbor $\bf{(a)}$, path $\bf{(b)}$, path of length three $\bf{(c)}$, and heterogeneity $\bf{(d)}$. For each network, we randomly generate 200 realizations of networks with link removal, as well as 200 pairs of $L^P$ and $L^N$ sets. In the figure, we choose the highest AUC from 200 samples as the performance of an index. The same quantitative analysis in the Fig. \ref{fig:supupper} is repeated.
}
\label{fig:supupper08}
\end{center}
\end{figure}

\begin{figure}[ht]
\begin{center}
\resizebox{16cm}{!}{\includegraphics{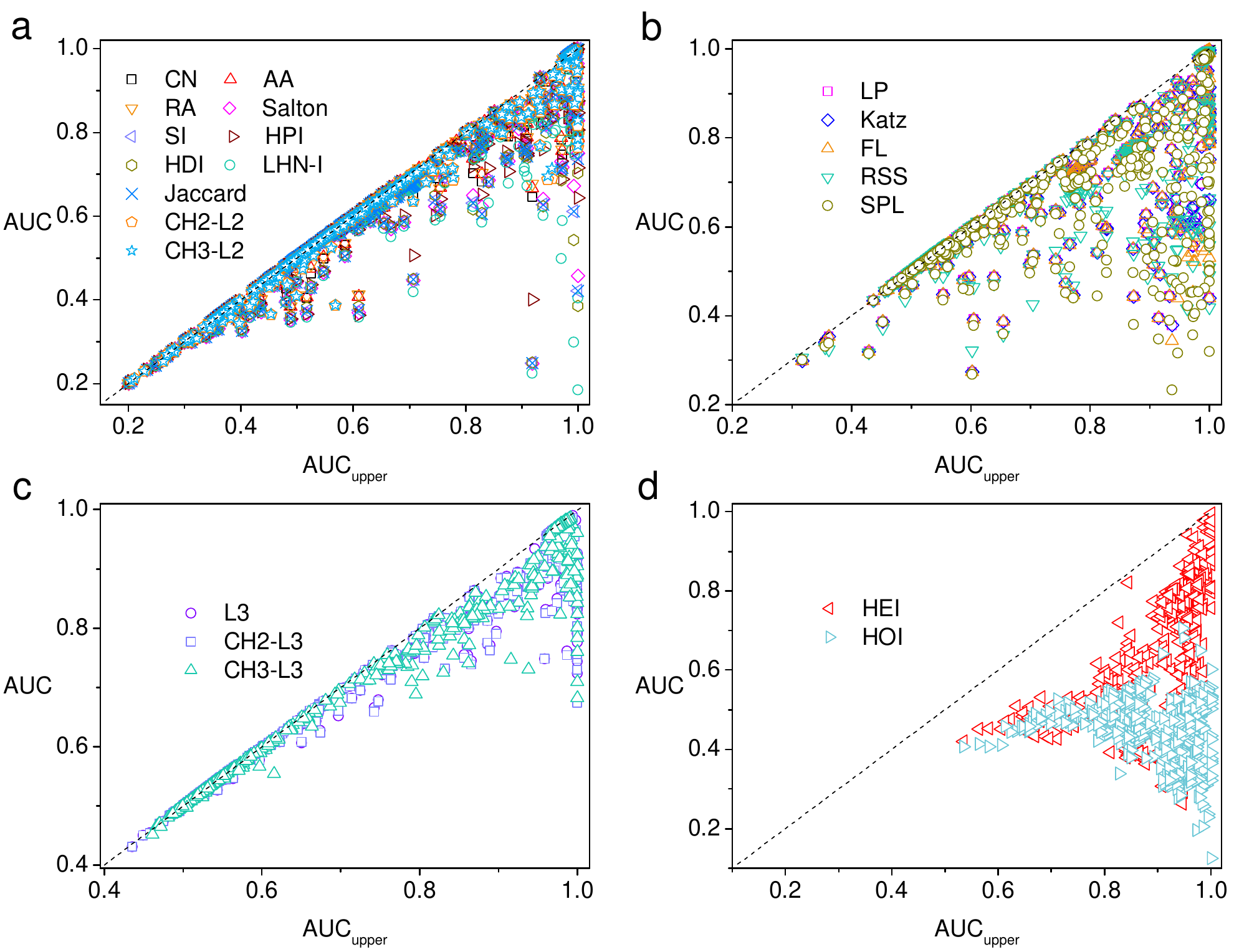}}
\caption{{\bf Unsupervised prediction measured by AUC based on imbalanced positive and negative samples by {\it sample\textsl{2}}.}
Eqs. (\ref{equation:lower}) and (\ref{equation:upper}) in the main text suggest that different indexes have different prediction performances, but all indexes associated with one topological feature share the same $\text{AUC}_\text{upper}$ and $\text{AUC}_\text{lower}$. This is confirmed by 21 indexes related to 4 topological features: common neighbor $\bf{(a)}$, path $\bf{(b)}$, path of length three $\bf{(c)}$, and heterogeneity $\bf{(d)}$. For each network, we randomly generate 200 realizations of networks with link removal, as well as 200 pairs of $L^P$ and $L^N$ sets. In the figure, we use the average value of 200 samples. The same quantitative analysis in the left panel of Fig. \ref{fig:unsupupper} is repeated.
}
\label{fig:unsupuppernon}
\end{center}
\end{figure}

\begin{figure}[ht]
\begin{center}
\resizebox{16cm}{!}{\includegraphics{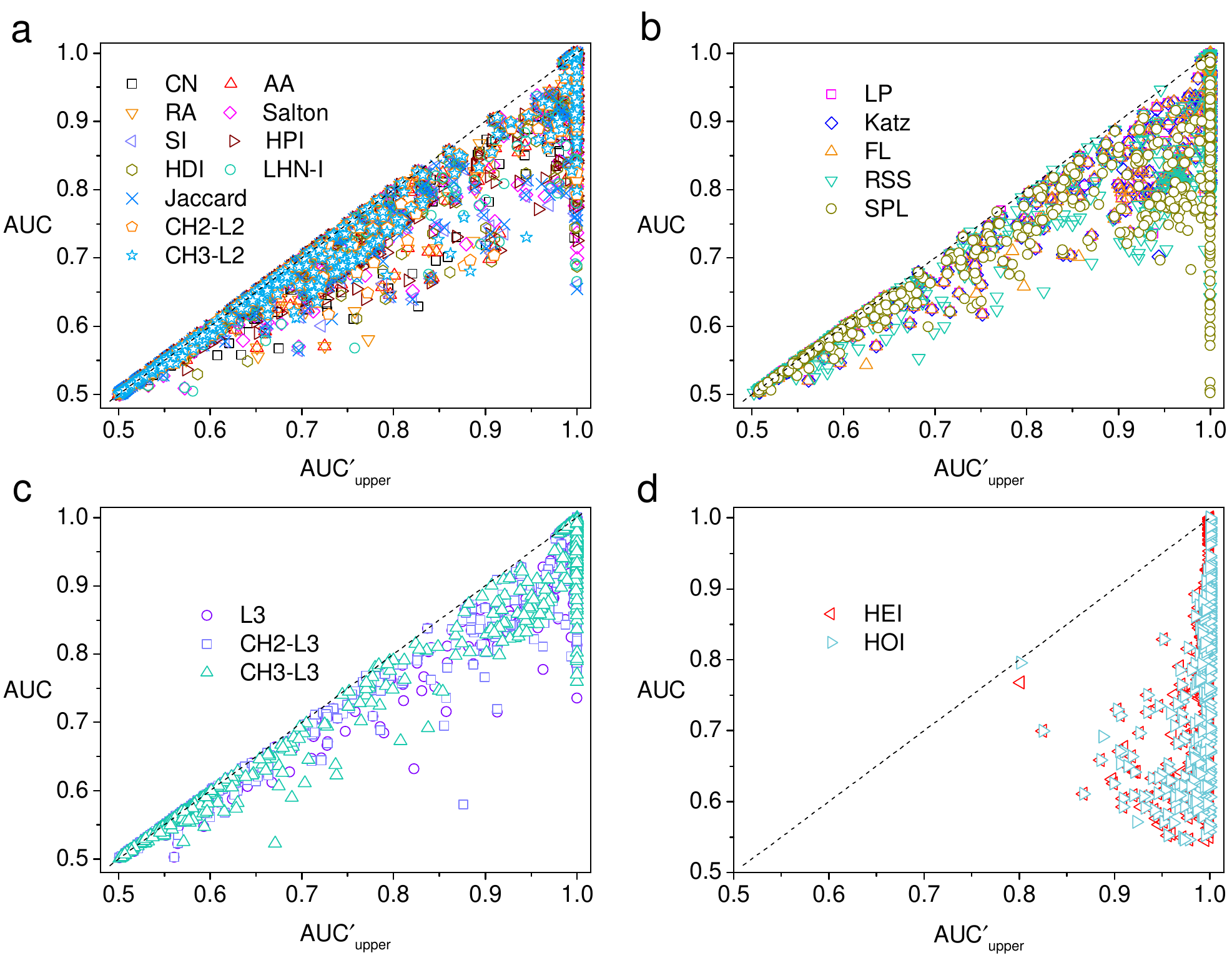}}
\caption{{\bf Supervised prediction measured by AUC based on imbalanced positive and negative samples by {\it sample\textsl{2}}.}
Eq. (\ref{equation:upper2}) in the main text suggests that $\text{AUC}^{\prime}_\text{upper}$ sets the upper bound of the supervised prediction. This is confirmed by 21 indexes related to 4 topological features: common neighbor $\bf{(a)}$, path $\bf{(b)}$, path of length three $\bf{(c)}$, and heterogeneity $\bf{(d)}$. For each network, we randomly generate 200 realizations of networks with link removal, as well as 200 pairs of $L^P$ and $L^N$ sets. In the figure, we choose the highest AUC from 200 samples as the performance of an index. The same quantitative analysis in the Fig. \ref{fig:supupper} is repeated.
}
\label{fig:supuppernon}
\end{center}
\end{figure}

To make sure our results are not affected by different setups, we perform a {\bf robustness check} by repeating the measurement in the main text using {\it sample\textsl{1}} (Fig. \ref{fig:unsupupper08} and Fig. \ref{fig:supupper08}) and {\it sample\textsl{2}} (Fig. \ref{fig:unsupuppernon} and Fig. \ref{fig:supuppernon}). The same capability applies to different setups, supporting the universality of the conclusion drawn.

\noindent 

\noindent 

\noindent 

\noindent 

\clearpage

\section{A topological feature's maximum capability measured by precision} \label{section:s4}
In link prediction, precision can also be used to evaluate the quality of prediction. The precision measures the percentage of the correct prediction (node pairs indeed in $L^P$) among the top-k predicted candidates \cite{lu2015toward,sun2020revealing,garcia2020precision,nasiri2021new,ghorbanzadeh2021hybrid}. After ranking the node pairs in both $L^P$ and $L^N$ according to their scores in descending order, we select $L_\text{k}$ node pairs with the highest score. The precision is given as
\begin{equation}
\text{Precision} = \frac{L_\text{r}}{L_\text{k}},
\label{equation:sprecision}
\end{equation} 
where $L_\text{r}$ is the number of selected node pairs that are included in $L^P$. 

\begin{figure}[ht]
\begin{center}
\resizebox{12cm}{!}{\includegraphics{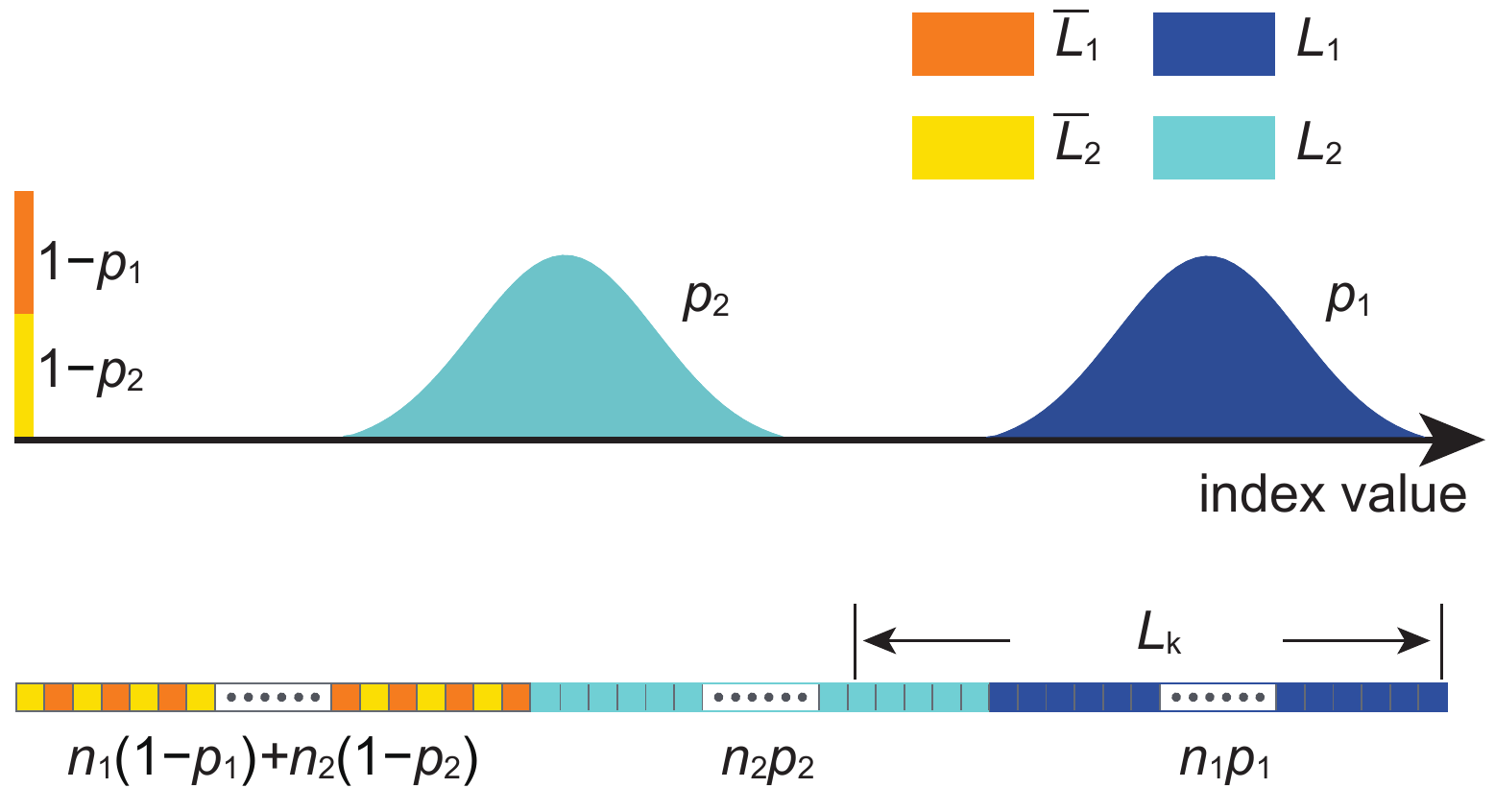}}
\caption{{\bf The best index value ranking for precision in the unsupervised approach.} 
In the best index value ranking illustrated in Fig. \ref{fig:unsupervised}c of the main text, $L_1$ is ranked ahead of $L_2$. Therefore, when ranking node pairs in descending order of their index values, we have three segments in the rank list. The first is $n_1 p_1$ entities of $L_1$, followed by $n_2 p_2$ entities of $L_2$. The $n_1(1-p_1)+n_2(1-p_2)$ entities of $\overline{L}_1 \cup \overline{L}_2$, which all have the same index value, are ranked at the end. Here, $n_1 = |\overline{L}_1 \cup L_1| = |L^P|$ and $n_2 = |\overline{L}_2 \cup L_2|= |L^N|$. Depending on the choice of $L_\text{k}$, the $\text{Precision}_\text{upper}$ for the best index value ranking can be derived.
}
\label{fig:precisiondesc}
\end{center}
\end{figure}\noindent 

Let us first consider the best index value ranking in the unsupervised approach (Fig. \ref{fig:unsupervised}c presented in the main text and Fig. \ref{fig:precisiondesc}), in which the lowest index value of $L_1$ is greater than the highest index value of $L_2$. Assume $n_1 = |\overline{L}_1 \cup L_1| = |L^P|$ and $n_2 = |\overline{L}_2 \cup L_2|= |L^N|$ for the number of node pairs in $L^P$ and $L^N$. When ranking node pairs in descending order of their index values, we have three segments in the ranking list. The first is $n_1 p_1$ entities of $L_1$, followed by $n_2 p_2$ entities of $L_2$. The $n_1(1-p_1)+n_2(1-p_2)$ entities of $\overline{L}_1 \cup \overline{L}_2$, which all have the same index value, are ranked at the end. Once the rank list is known, the $\text{Precision}_\text{upper}$ for the best index value ranking only depends on how $L_\text{k}$ cuts the top entities of the list, which can be formulated as 
\begin{equation}
\text{Precision}_\text{upper} =
\begin{dcases}
1, & \text{ $n_1p_1 \geq L_\text{k}$},\\
\frac{n_1p_1}{L_\text{k}}, & \text{ $n_1p_1 < L_\text{k} \leq n_1p_1+n_2p_2$},\\
\frac{n_1p_1+\frac{n_1(1-p_1)(L_\text{k}-n_1p_1-n_2p_2)}{n_1(1-p_1)+n_2(1-p_2)}}{L_\text{k}}, & \text{ $n_1p_1+n_2p_2 < L_\text{k} \leq n_1+n_2$}.
\label{equation:preupper1}
\end{dcases}
\end{equation}
Note that node pairs in $\overline{L}_1 \cup \overline{L}_2$ have the same index value. The relative position of one node pair among all $n_1(1-p_1)+n_2(1-p_2)$ entities is random. In Eq. (\ref{equation:preupper1}), $\frac{n_1(1-p_1)}{n_1(1-p_1)+n_2(1-p_2)}$ corresponds to the probability of finding a missing link in $\overline{L}_1 \cup \overline{L}_2$. Therefore, $\frac{n_1(1-p_1)(L_\text{k}-n_1p_1-n_2p_2)}{n_1(1-p_1)+n_2(1-p_2)}$ is the expected number of missing links for a given $L_\text{k}$ value.

\begin{figure}[ht]
\begin{center}
\resizebox{16cm}{!}{\includegraphics{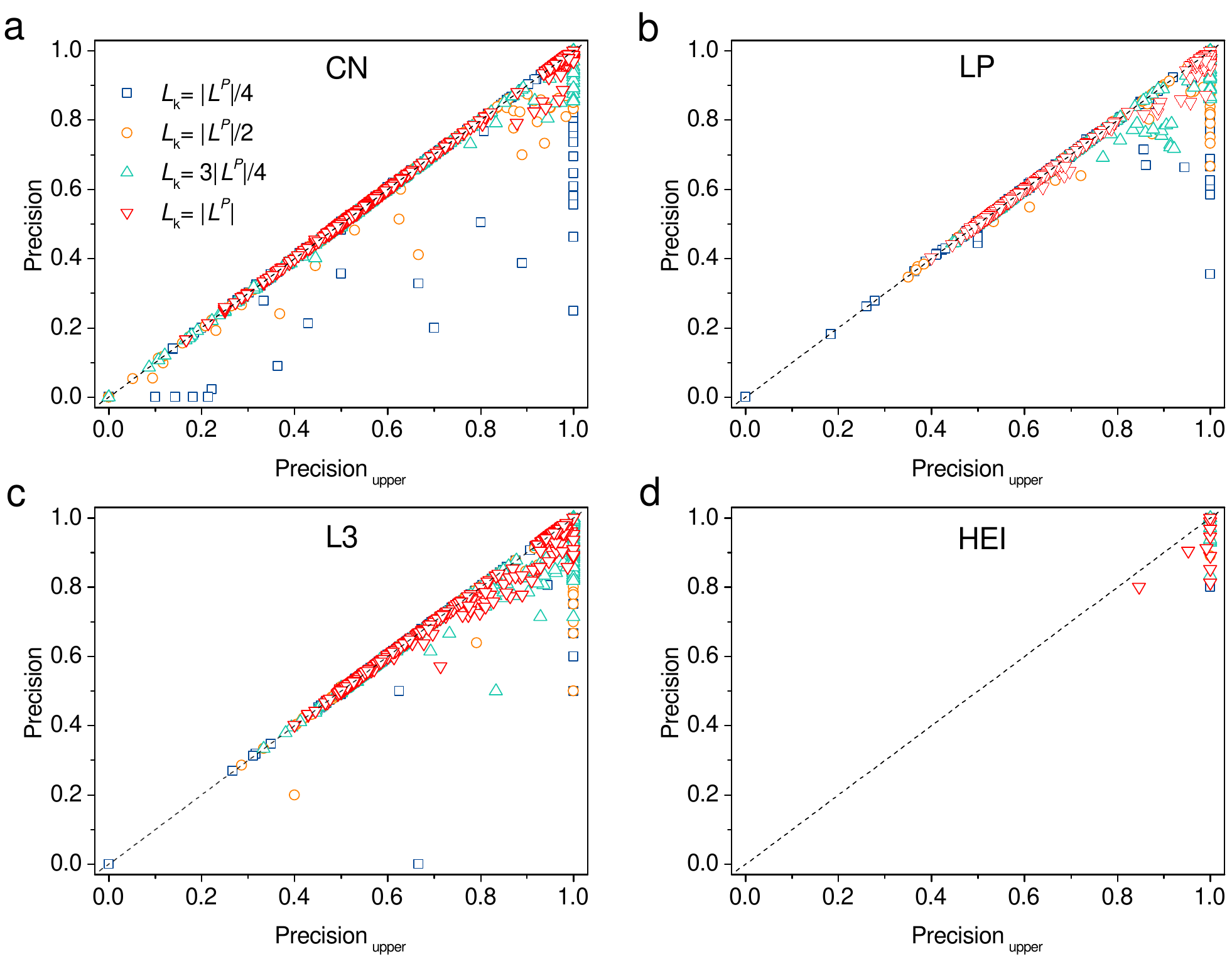}}
\caption{{\bf Networks whose AUC is close to the upper bound in the unsupervised approach.}
We choose the index CN, LP, L3, and HEI from each of the four families. We select the networks in which the unsupervised prediction by one index is already close to the upper bound measured by AUC (measured AUC is more than 95\% of $\text{AUC}_\text{upper}$). For such networks, it is expected that the performance of this index measured by precision should follow Eq. (\ref{equation:preupper1}). Indeed, for different choices of $L_\text{k}$ ($L_\text{k}= |L^P|/4, |L^P|/2, 3|L^P|/4, |L^P|$), the precision measured is almost on the line $y=x$, supporting the theoretical prediction.
}
\label{fig:precisionv1}
\end{center}
\end{figure}

\begin{figure}[ht]
\begin{center}
\resizebox{16cm}{!}{\includegraphics{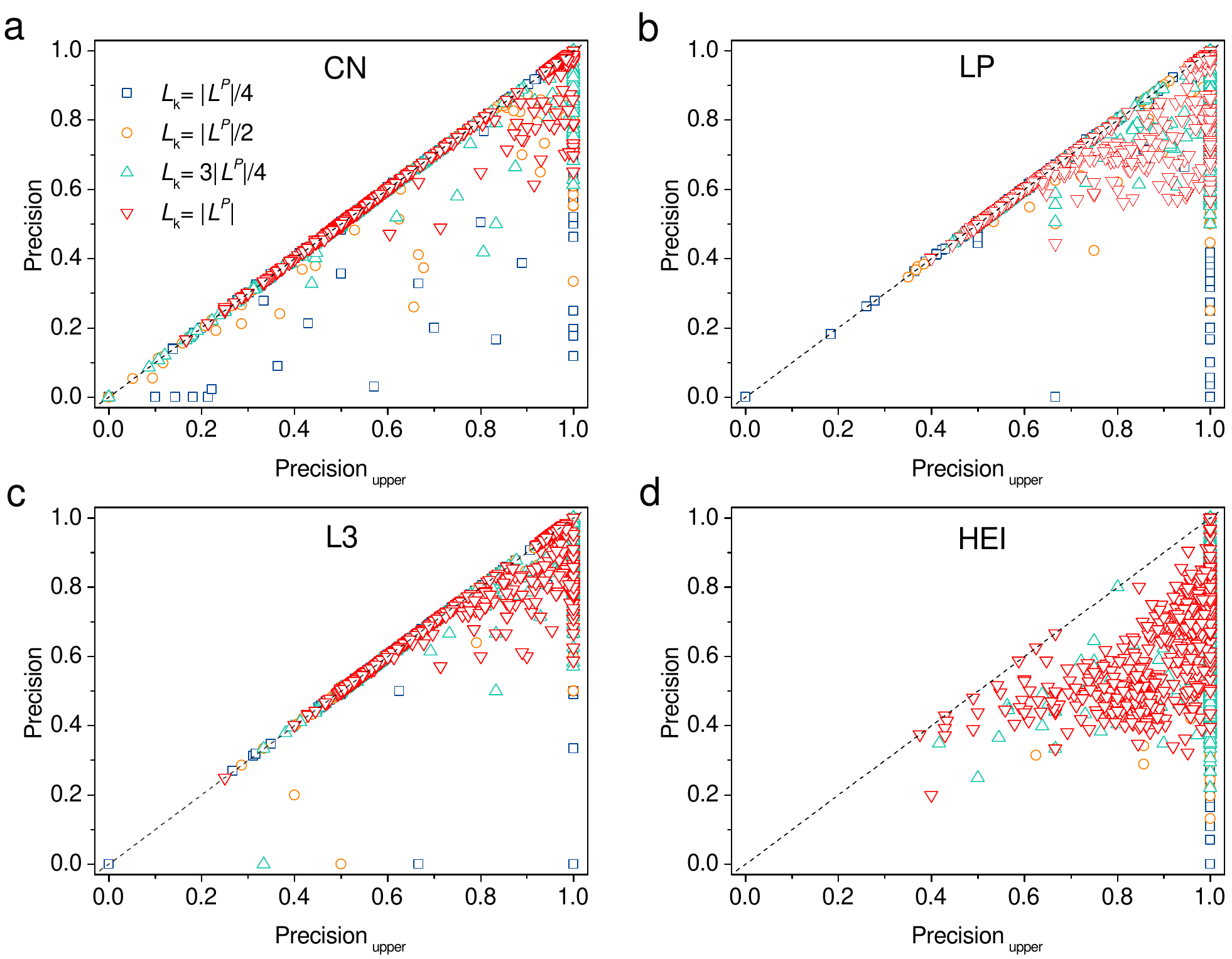}}
\caption{{\bf The maximum capability of the topological features measured by precision in the unsupervised approach.}
We choose the index CN, LP, L3, and HEI from each of the four families and measure the performance of the unsupervised prediction by these indexes in all 550 networks. For different choices of $L_\text{k}$ ($L_\text{k}= |L^P|/4, |L^P|/2, 3|L^P|/4, |L^P|$), the measured precision is equal to or below $\text{Precision}_\text{upper}$, supporting the claim that $\text{Precision}_\text{upper}$ gives the maximum capability of a topological feature measured by precision. For each network, we randomly generate 200 realizations of networks with link removal, as well as 200 pairs of $L^P$ and $L^N$ sets. In the figure, we use the average value of 200 samples.
}
\label{fig:precisionupper1}
\end{center}
\end{figure}

The above deduction suggests that under the best index value ranking, the measured precision should exactly follow $\text{Precision}_\text{upper}$ given by Eq. (\ref{equation:preupper1}). To test it, we select one index from each family and identify networks in which the unsupervised prediction by this index has an AUC value greater than 95\% of $\text{AUC}_\text{upper}$. For such networks, we could approximate that the index value ranking is close to the best scenario. We then quantify the performance of the unsupervised prediction using precision. In these networks, the measured precision well follows what Eq. (\ref{equation:preupper1}) depicts (Fig. \ref{fig:precisionv1}). Moreover, when measuring the prediction performance of this index in all 550 networks, the measured precision is below $\text{Precision}_\text{upper}$ (Fig. \ref{fig:precisionupper1}), supporting the claim that Eq. (\ref{equation:preupper1}) captures the maximum capability of a topological feature measured by precision in the unsupervised approach. 

\noindent 

\clearpage

\noindent 

\clearpage

\begin{figure}[ht]
\begin{center}
\resizebox{12cm}{!}{\includegraphics{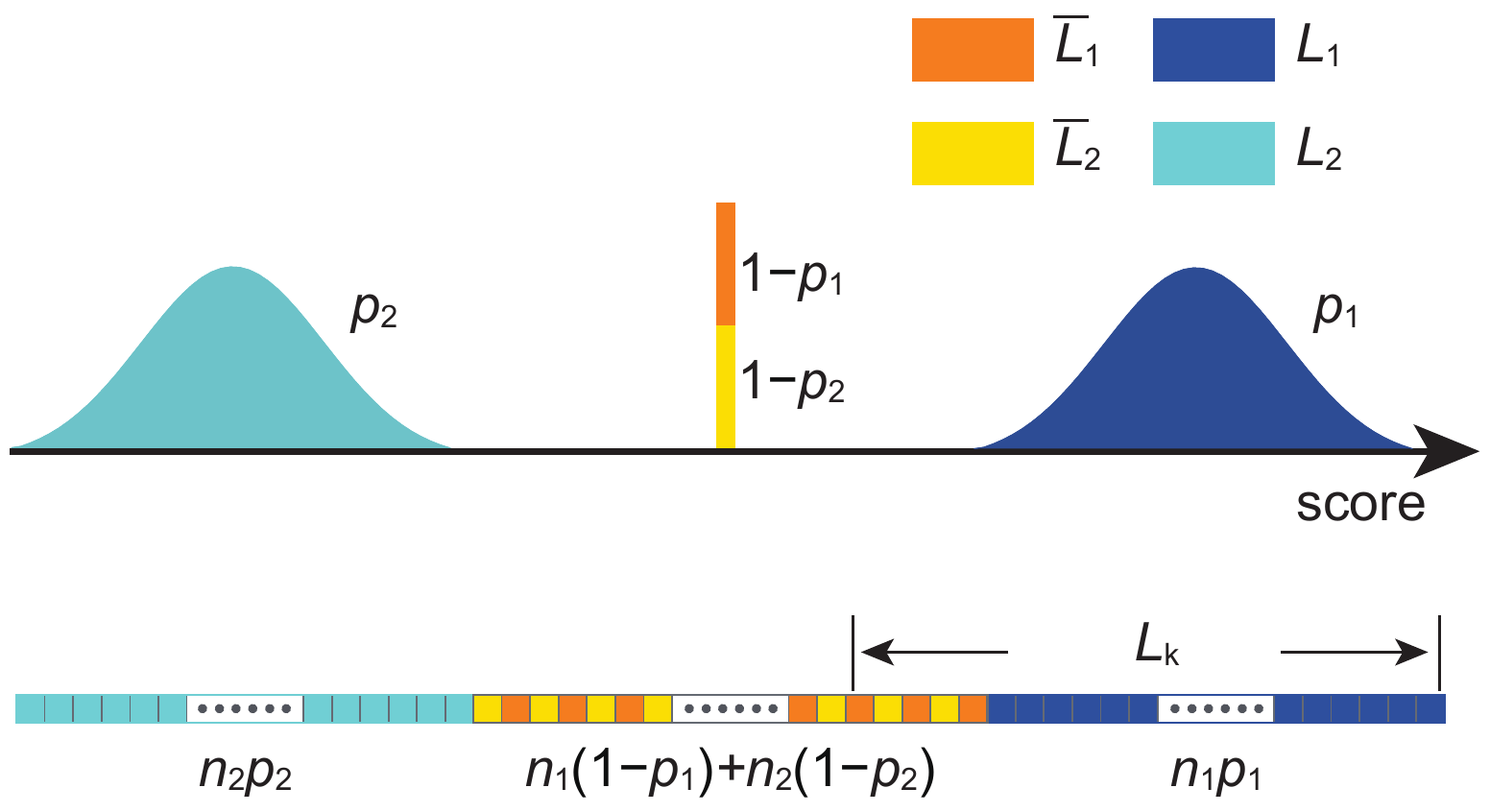}}
\caption{{\bf The optimal score ranking for precision in the supervised approach.}
The optimal score ranking after the mapping function is to have $\overline{L}_1 \cup \overline{L}_2$ ranked in between $L_1$ and $L_2$, as illustrated in Fig. \ref{fig:unsupervised}d of the main text. When ranking node pairs in descending order of their scores, we have three segments in the rank list. The first is $n_1 p_1$ entities of $L_1$, followed by $n_1(1-p_1)+n_2(1-p_2)$ entities of $\overline{L}_1 \cup \overline{L}_2$, which all have the same score. $n_2 p_2$ entities of $L_2$ are ranked at the end. Here, $n_1 = |\overline{L}_1 \cup L_1| = |L^P|$ and $n_2 = |\overline{L}_2 \cup L_2|= |L^N|$. Depending on the choice of $L_\text{k}$, the $\text{Precision}^{\prime}_\text{upper}$ in the supervised approach can be derived.
}
\label{fig:precisiondescml}
\end{center}
\end{figure}\noindent

In the supervised approach, the mapping function can further optimize the score ranking from the index value ranking. As discussed in the main text, the optimal score ranking is such that $L_1$ ranks ahead, followed by $\overline{L}_1 \cup \overline{L}_2$, and $L_2$ ranks at the end (Fig. \ref{fig:precisiondescml}). Similar to Eq. (\ref{equation:preupper1}), we can derive the upper bound of supervised prediction measured by precision as
\begin{equation}
\text{Precision}^{\prime}_\text{upper} =
\begin{dcases}
1, & \text{ $n_1p_1 \geq L_\text{k}$},\\
\frac{n_1p_1+\frac{n_1(1-p_1)(L_\text{k}-n_1p_1)}{n_1(1-p_1)+n_2(1-p_2)}}{L_\text{k}}, & \text{ $n_1p_1 < L_\text{k} \leq n_1+n_2-n_2p_2$},\\
\frac{n_1}{L_\text{k}}, & \text{ $n_1+n_2-n_2p_2 < L_\text{k} \leq n_1+n_2$},
\label{equation:preupper2}
\end{dcases}
\end{equation}
where $\frac{n_1(1-p_1)}{n_1(1-p_1)+n_2(1-p_2)}$ corresponds to the probability of finding a missing link in $\overline{L}_1 \cup \overline{L}_2$. It is noteworthy that the Eq. (\ref{equation:preupper1}) and Eq. (\ref{equation:preupper2}) still hold when the $L_\text{k}=|L^P|$ (where the precision is equivalent to the recall \cite{zhou2021progresses}).

\begin{figure}[ht]
\begin{center}
\resizebox{16cm}{!}{\includegraphics{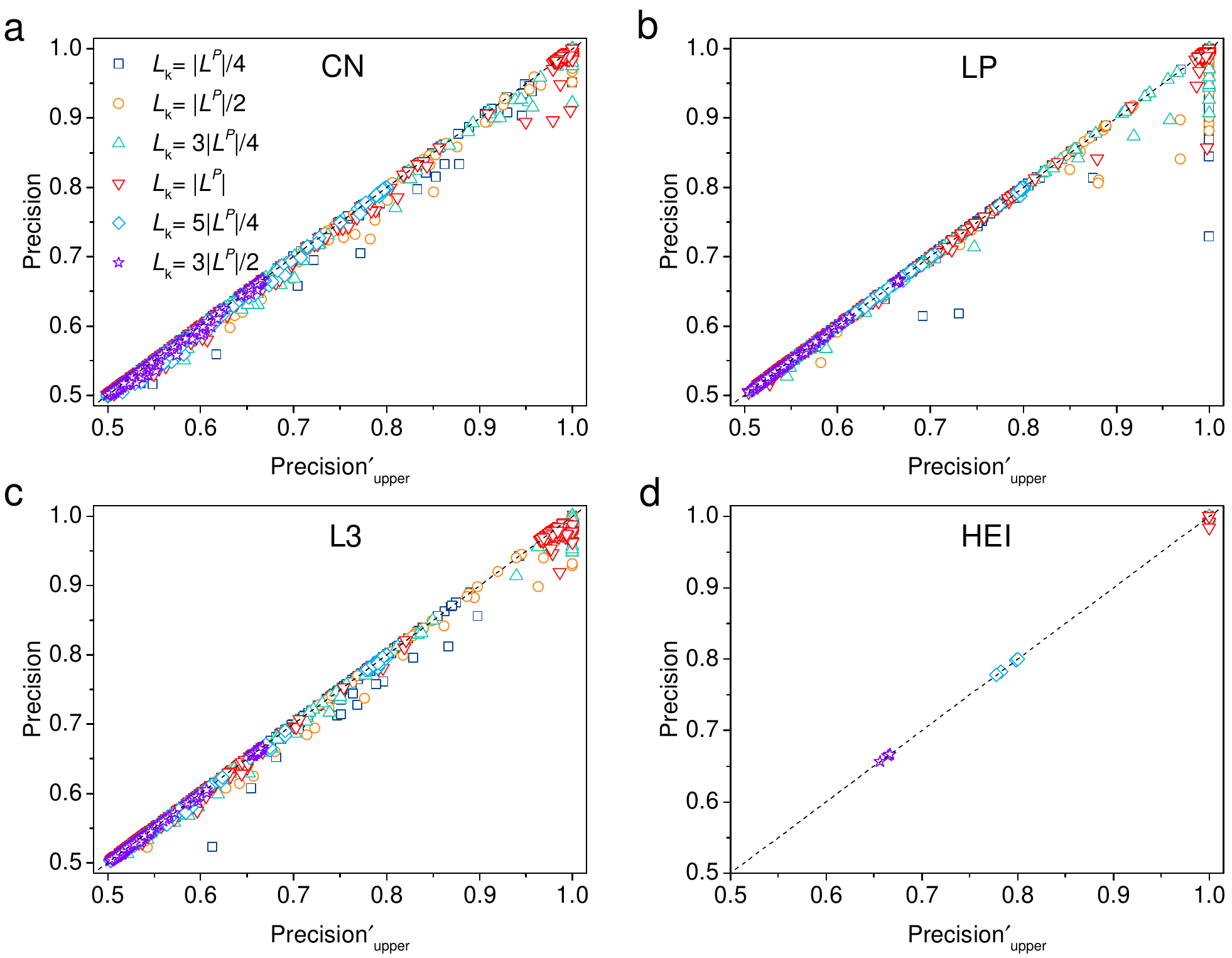}}
\caption{{\bf Networks whose AUC is close to the upper bound in the supervised approach.}
We choose the index CN, LP, L3, and HEI from each of the four families. After we select the networks in which the supervised prediction by one index is already close to the upper bound measured by AUC (measured AUC is more than 95\% of $\text{AUC}^{\prime}_\text{upper}$). For such networks, it is expected that the performance of this index measured by precision should follow Eq. (\ref{equation:preupper2}). Indeed, for different choices of $L_\text{k}$ ($L_\text{k}= |L^P|/4, |L^P|/2, 3|L^P|/4, |L^P|, 5|L^P|/4, 3|L^P|/2$), the precision measured is almost on the line $y=x$, supporting the theoretical prediction.
}
\label{fig:precisionv2}
\end{center}
\end{figure}

\begin{figure}[ht]
\begin{center}
\resizebox{16cm}{!}{\includegraphics{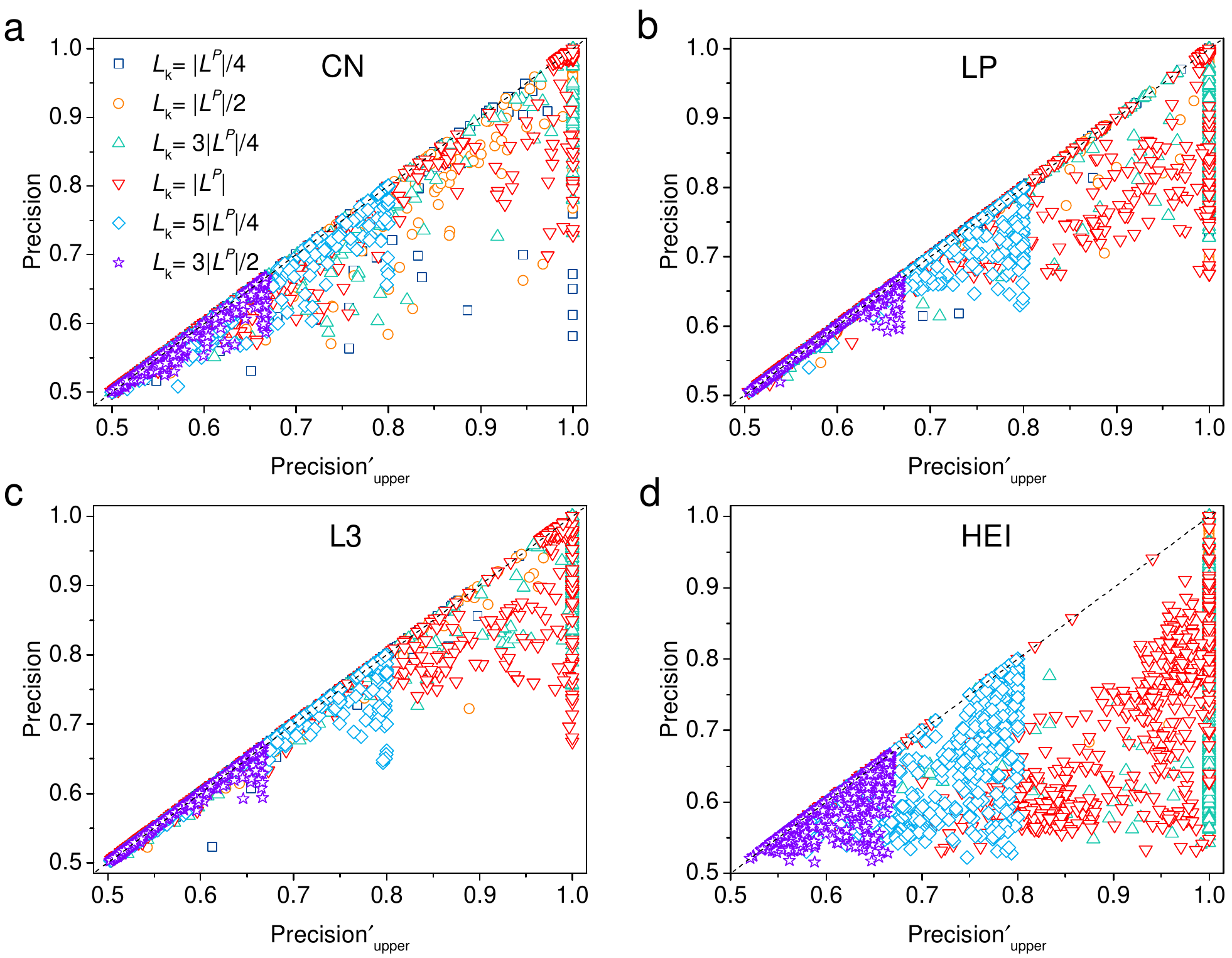}}
\caption{{\bf The maximum capability of the topological features measured by precision in the supervised approach.}
We choose the index CN, LP, L3, and HEI from each of the four families and measure the performance of the supervised prediction by these indexes in all 550 networks. For different choices of $L_\text{k}$ ($L_\text{k}= |L^P|/4, |L^P|/2, 3|L^P|/4, |L^P|, 5|L^P|/4, 3|L^P|/2$), the measured precision is equal to or below $\text{Precision}^{\prime}_\text{upper}$, supporting the theoretical results for the maximum capability of a topological feature measured by precision. For each network, we randomly generate 200 realizations of networks with link removal, as well as 200 pairs of $L^P$ and $L^N$ sets. In the figure, we choose the highest precision from 200 samples as the performance of an index.
}
\label{fig:precisionupper2}
\end{center}
\end{figure}

We perform the same test for Eq. (\ref{equation:preupper2}) as for Eq. (\ref{equation:preupper1}). For networks in which an index gives a supervised prediction already close to the upper bound $\text{AUC}^{\prime}_\text{upper}$, the prediction measured by precision is expected to follow Eq. (\ref{equation:preupper2}). This is confirmed in Fig. (\ref{fig:precisionv2}). For all networks, the prediction measured by precision should be below what Eq. (\ref{equation:preupper2}) yields. This is also confirmed in Fig. (\ref{fig:precisionupper2}).

\noindent 

\clearpage

\noindent 

\clearpage

\begin{figure}[ht]
\begin{center}
\resizebox{16cm}{!}{\includegraphics{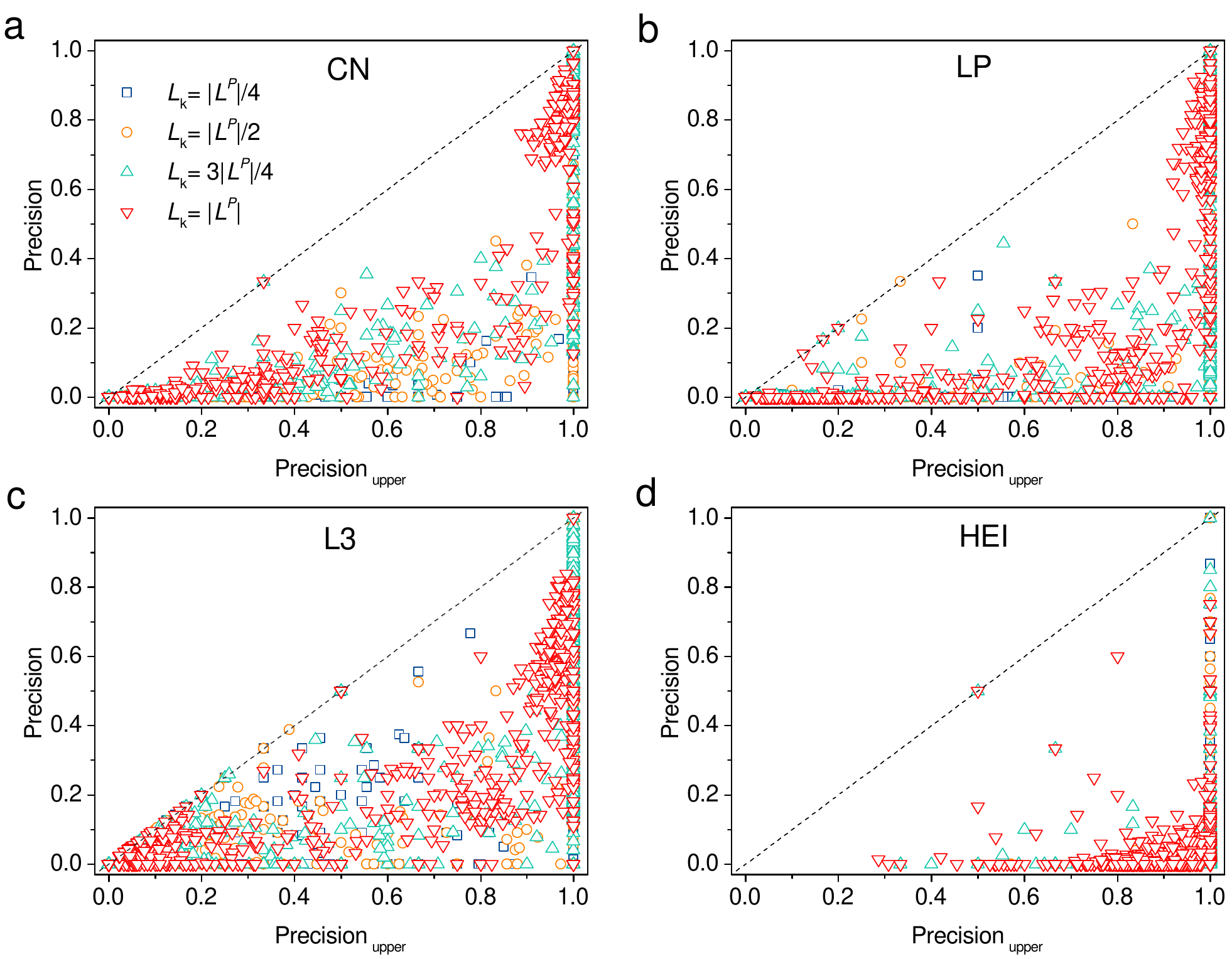}}
\caption{{\bf Unsupervised prediction measured by precision based on imbalanced positive and negative samples by {\it sample\textsl{2}}.}
We choose the index CN, LP, L3, and HEI from each of the four families and measure the performance of the unsupervised prediction by these indexes in all 550 networks. For different choices of $L_\text{k}$ ($L_\text{k}= |L^P|/4, |L^P|/2, 3|L^P|/4, |L^P|$), the measured precision is equal to or below $\text{Precision}_\text{upper}$, supporting the claim that $\text{Precision}_\text{upper}$ gives the maximum capability of a topological feature measured by precision. The same quantitative analysis in the Fig. \ref{fig:precisionupper1} is repeated.
}
\label{fig:precisionunnon}
\end{center}
\end{figure}

\begin{figure}[ht]
\begin{center}
\resizebox{16cm}{!}{\includegraphics{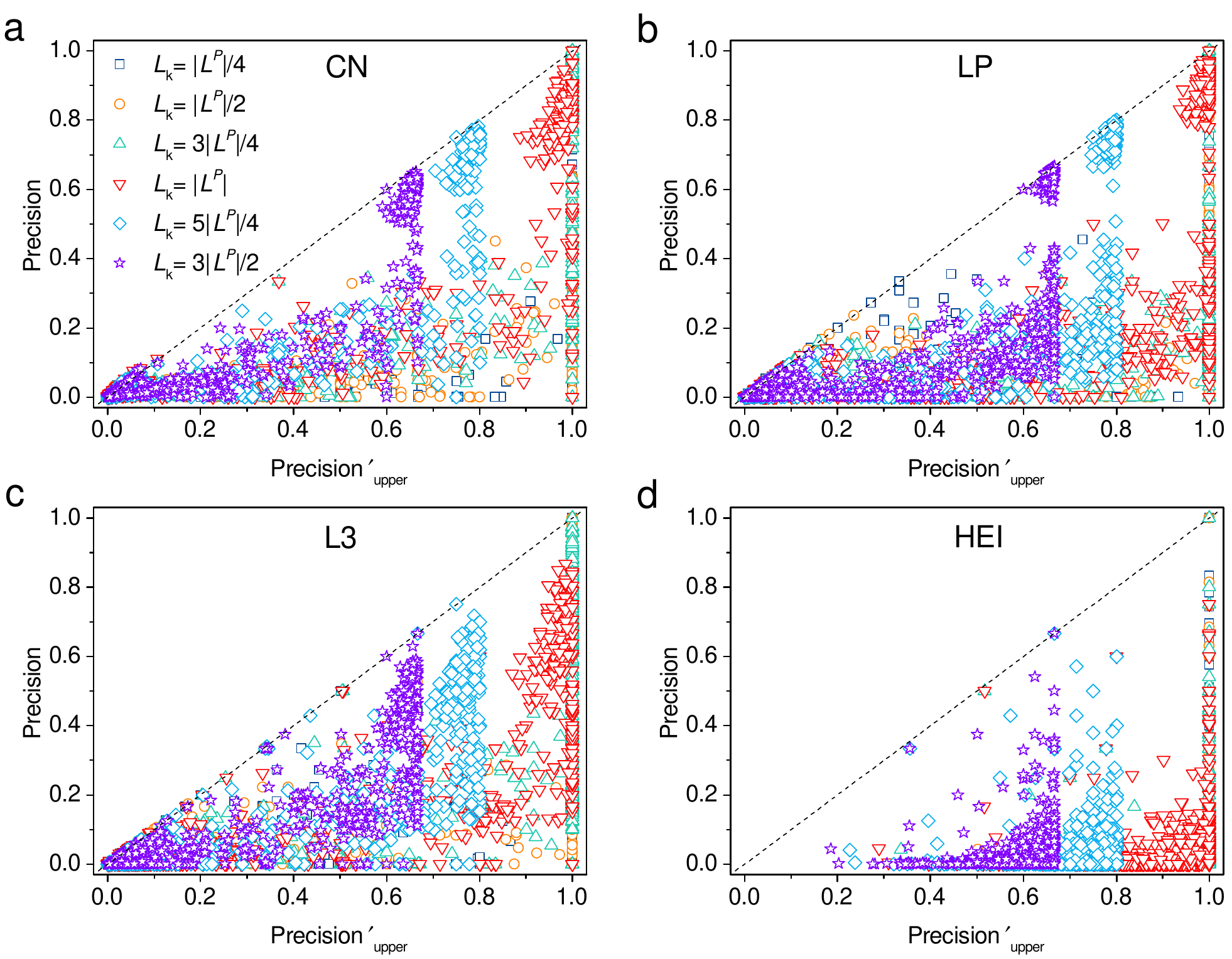}}
\caption{{\bf Supervised prediction measured by precision based on imbalanced positive and negative samples by {\it sample\textsl{2}}.}
We choose the index CN, LP, L3, and HEI from each of the four families and measure the performance of the supervised prediction by these indexes in all 550 networks. For different choices of $L_\text{k}$ ($L_\text{k}= |L^P|/4, |L^P|/2, 3|L^P|/4, |L^P|, 5|L^P|/4, 3|L^P|/2$), the measured precision is equal to or below $\text{Precision}^{\prime}_\text{upper}$, supporting the theoretical results for the maximum capability of a topological feature measured by precision. The same quantitative analysis in the Fig. \ref{fig:precisionupper2} is repeated.
}
\label{fig:precisionsunon}
\end{center}
\end{figure}

Finally, some studies argue that the imbalanced samples extremely affect the results measured by precision \cite{lichtnwalter2012link,yang2015evaluating,fawcett2006introduction}. For this reason, we test Eq. (\ref{equation:preupper1}) and Eq. (\ref{equation:preupper2}) under the imbalanced sampling {\it sample\textsl{2}}. The imbalanced sample indeed reduces the performance by precision (Fig. \ref{fig:precisionunnon} and Fig.  \ref{fig:precisionsunon}). But Eq. (\ref{equation:preupper1}) and Eq. (\ref{equation:preupper2}) still capture the upper bound of the performance, which correctly gives the maximum capability of a topological feature. It is worth noting that the $\text{Precision}^{\prime}_\text{upper}$ for supervised prediction will be lower than 0.5 when the positive testing set $L^P$ and the negative testing set $L^N$ are imbalanced (Fig. \ref{fig:precisionsunon}).

\noindent 

\clearpage

\noindent 

\clearpage

\section{A topological feature's maximum capability measured by AUC-mROC}\label{section:s5}

To deal with the imbalanced positive and negative samples, Muscoloni \textit{et al.} propose the area under the magnified ROC (AUC-mROC) to measure the prediction performance of one index \cite{muscoloni2022early}. The mROC can make the performance of a random predictor always equal to 0.5. Denote $n_1 = |L^P|$ and $n_2 = |L^N|$ by the size of the positive and negative testing set. Given the index value of a predictor, we here define the true positive (\text{TP}@r) and false positive (\text{FP}@r) as the number of samples in $L^P$ and $L^N$ at a varying ranking threshold $r \in [1, n_1+n_2]$. Hence, we can denote the non-normalized magnified TPR (\text{nmTPR}@r) and non-normalized magnified FPR (\text{nmFPR}@r) at a varying ranking threshold $r \in [1, n_1+n_2]$ as 
\begin{equation}
\text{nmTPR}@r = \frac{\ln(1+\text{TP}@r)}{\ln(1+n_1)},
\label{equation:nmtpr}
\end{equation} 
\begin{equation}
\text{nmFPR}@r = \frac{\ln(1+\text{FP}@r)}{\ln(1+n_2)}.
\label{equation:nmfpr}
\end{equation} 
To make the AUC-mROC of the random predictor 0.5, the \text{nmTPR}@r is normalized to \text{mTPR}@r. The \text{mTPR}@r is defined as 
\begin{equation}
\text{mTPR}@r = \text{nmFPR}@r+\frac{\text{nmTPR}@r-\ln_{(1+n_1)}(1+\text{FP}@r \cdot \frac{n_1}{n_2})}{1-\ln_{(1+n_1)}(1+\text{FP}@r \cdot \frac{n_1}{n_2})} \cdot (1-\text{nmFPR}@r).
\label{equation:mtpr}
\end{equation} 
Therefore, the mROC curve is composed of the points at coordinates (\text{nmFPR}@r, \text{mTPR}@r) for each $r \in [1, n_1+n_2]$. Finally, the AUC-mROC can be obtained by computing the area under the mROC curve using the trapezoidal rule.

\begin{figure}[ht]
\begin{center}
\resizebox{12cm}{!}{\includegraphics{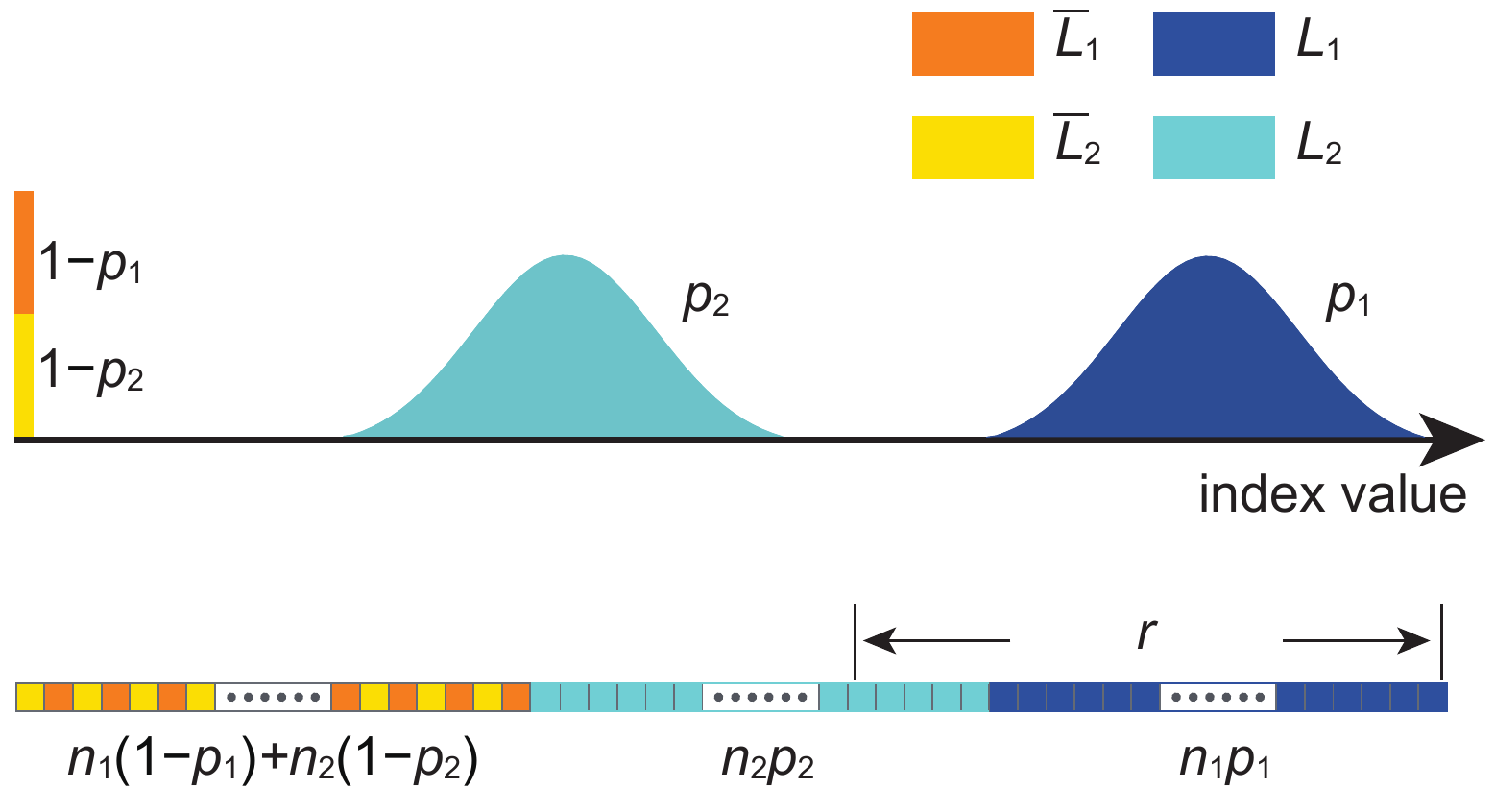}}
\caption{{\bf The best index value ranking for AUC-mROC in the unsupervised approach.} 
In the best index value ranking illustrated in Fig. \ref{fig:unsupervised}c of the main text, $L_1$ is ranked ahead of $L_2$. Therefore, when ranking node pairs in descending order of their index values, we have three segments in the rank list. The first is $n_1 p_1$ entities of $L_1$, followed by $n_2 p_2$ entities of $L_2$. The $n_1(1-p_1)+n_2(1-p_2)$ entities of $\overline{L}_1 \cup \overline{L}_2$, which all have the same index value, are ranked at the end. Here, $n_1 = |\overline{L}_1 \cup L_1| = |L^P|$ and $n_2 = |\overline{L}_2 \cup L_2|= |L^N|$. Depending on the different ranking threshold $r$, the \text{TP}@r and \text{FP}@r for the best index value ranking can be derived.
}
\label{fig:aucmrocdesc}
\end{center}
\end{figure}\noindent 

Let us first consider the best index value ranking in the unsupervised approach (Fig. \ref{fig:unsupervised}c presented in the main text and Fig. \ref{fig:aucmrocdesc}), in which the lowest index value of $L_1$ is greater than the highest index value of $L_2$. When ranking node pairs in descending order of their index values, we have three segments in the ranking list. The first is $n_1 p_1$ entities of $L_1$, followed by $n_2 p_2$ entities of $L_2$. The $n_1(1-p_1)+n_2(1-p_2)$ entities of $\overline{L}_1 \cup \overline{L}_2$, which all have the same index value, are ranked at the end. Once the rank list is known, the \text{TP}@r and \text{FP}@r for the best index value ranking only depends on how $r$ cuts the top entities of the list, which can be formulated as 
\begin{equation}
\text{TP}@r =
\begin{dcases}
r, & \text{ $n_1p_1 \geq r$},\\
n_1p_1, & \text{ $n_1p_1 < r \leq n_1p_1+n_2p_2$},\\
n_1p_1+\frac{n_1(1-p_1)(r-n_1p_1-n_2p_2)}{n_1(1-p_1)+n_2(1-p_2)}, & \text{ $n_1p_1+n_2p_2 < r \leq n_1+n_2$}.
\label{equation:tpupper1}
\end{dcases}
\end{equation}
\begin{equation}
\text{FP}@r =
\begin{dcases}
0, & \text{ $n_1p_1 \geq r$},\\
r-n_1p_1, & \text{ $n_1p_1 < r \leq n_1p_1+n_2p_2$},\\
n_2p_2+\frac{n_2(1-p_2)(r-n_1p_1-n_2p_2)}{n_1(1-p_1)+n_2(1-p_2)}, & \text{ $n_1p_1+n_2p_2 < r \leq n_1+n_2$}.
\label{equation:fpupper1}
\end{dcases}
\end{equation}
Note that node pairs in $\overline{L}_1 \cup \overline{L}_2$ have the same index value. The relative position of one node pair among all $n_1(1-p_1)+n_2(1-p_2)$ entities is random. In Eqs. (\ref{equation:tpupper1}) and (\ref{equation:fpupper1}), $\frac{n_1(1-p_1)}{n_1(1-p_1)+n_2(1-p_2)}$ and $\frac{n_2(1-p_2)}{n_1(1-p_1)+n_2(1-p_2)}$ correspond to the probability of finding a missing link and nonexistent link in $\overline{L}_1 \cup \overline{L}_2$, respectively. Therefore, $\frac{n_1(1-p_1)(r-n_1p_1-n_2p_2)}{n_1(1-p_1)+n_2(1-p_2)}$ and $\frac{n_2(1-p_2)(r-n_1p_1-n_2p_2)}{n_1(1-p_1)+n_2(1-p_2)}$ are the expected number of missing links and nonexistent links for a given $r$ value, respectively.

\begin{figure}[ht]
\begin{center}
\resizebox{16cm}{!}{\includegraphics{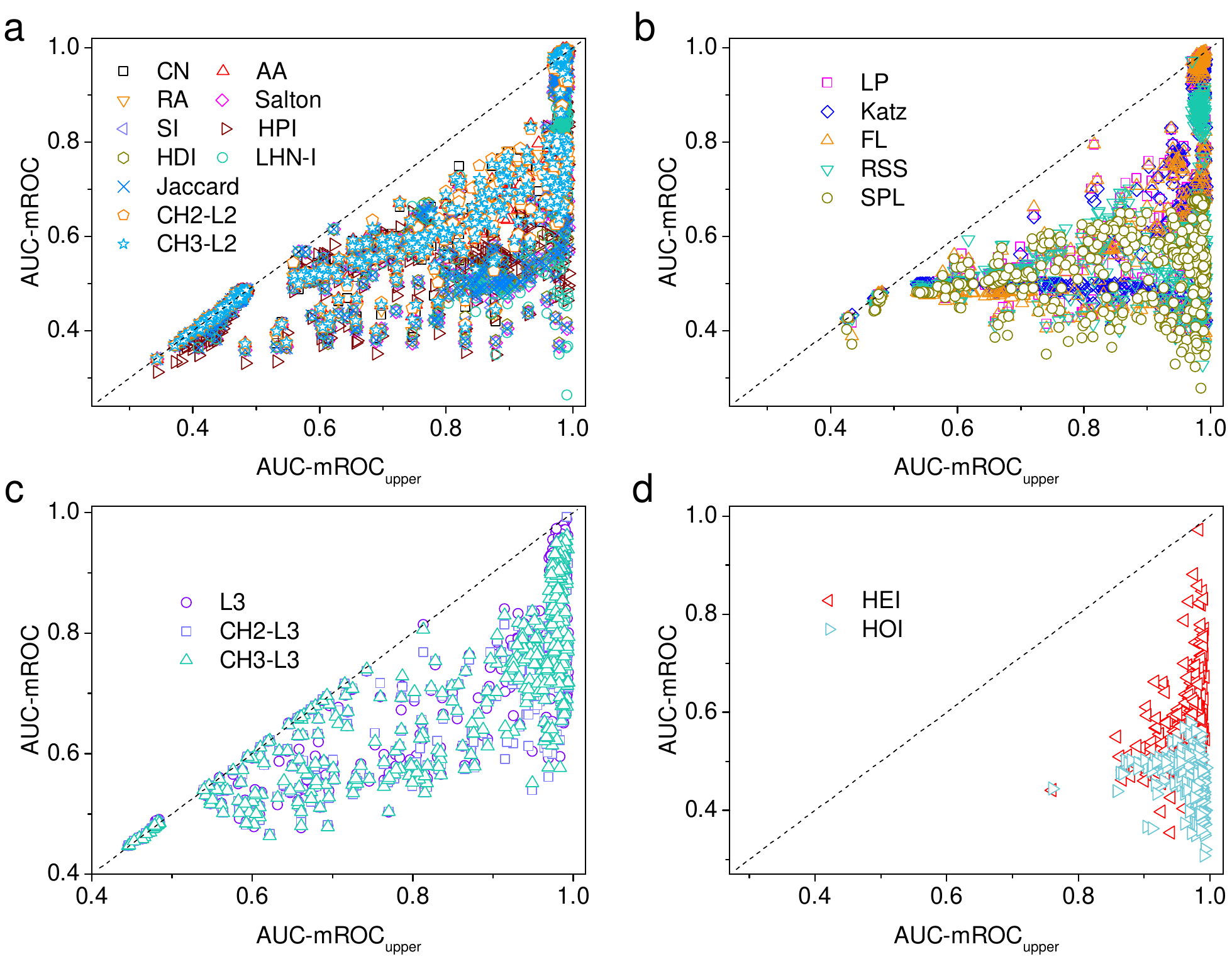}}
\caption{{\bf The maximum capability of the topological features measured by AUC-mROC in the unsupervised approach.}
The imbalanced positive and negative samples are generated by {\it sample\textsl{2}}. We use 21 indexes from four families and measure the performance of the unsupervised prediction by these indexes in all 550 networks. The prediction measured by AUC-mROC is equal to or below $\text{AUC-mROC}_\text{upper}$, supporting the claim that $\text{AUC-mROC}_\text{upper}$ gives the maximum capability of a topological feature measured by AUC-mROC. For each network, we randomly generate 200 realizations of networks with link removal, as well as 200 pairs of $L^P$ and $L^N$ sets. In the figure, we use the average value of 200 samples. 
}
\label{fig:aucmrocupper1}
\end{center}
\end{figure}

By using Eqs. (\ref{equation:nmtpr})-(\ref{equation:fpupper1}), we can derive a topological feature's maximum capability ($\text{AUC-mROC}_\text{upper}$) measured by AUC-mROC in the unsupervised approach. The above deduction suggests that under the best index value ranking, the measured AUC-mROC should exactly follow $\text{AUC-mROC}_\text{upper}$. To test it, we measure the prediction performance of 21 indexes in all 550 networks. And we use the experiment setup {\it sample\textsl{2}} to make the prediction.  The {\it sample\textsl{2}} randomly removes 20\% of $L$ links as the missing links. In the training step, the positive set is composed of 80\% of the removed links (16\% of $L$ links), and the negative set is composed of 80\% of all nonexistent links. In the testing step, the positive set is composed of the rest 20\% of the removed links, and the negative set is composed of the rest 20\% of all nonexistent links. The measured AUC-mROC is below $\text{AUC-mROC}_\text{upper}$ (Fig. \ref{fig:aucmrocupper1}), supporting the claim that $\text{AUC-mROC}_\text{upper}$ captures the maximum capability of a topological feature measured by AUC-mROC in the unsupervised approach.

\noindent 

\clearpage

\begin{figure}[ht]
\begin{center}
\resizebox{12cm}{!}{\includegraphics{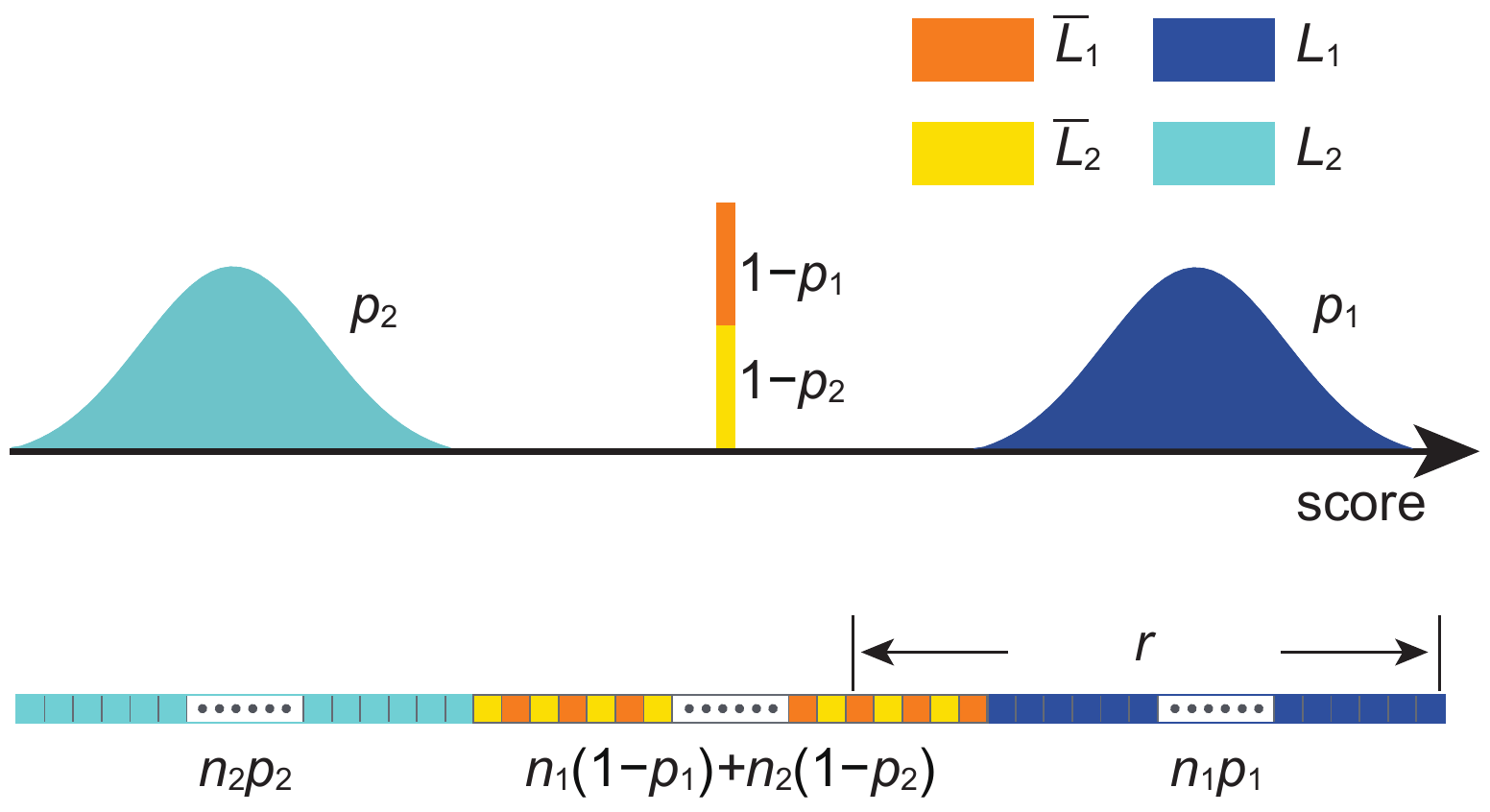}}
\caption{{\bf The optimal score ranking for AUC-mROC in the supervised approach.}   
The optimal score ranking after the mapping function is to have $\overline{L}_1 \cup \overline{L}_2$ ranked in between $L_1$ and $L_2$, as illustrated in Fig. \ref{fig:unsupervised}d of the main text. When ranking node pairs in descending order of their scores, we have three segments in the rank list. The first is $n_1 p_1$ entities of $L_1$, followed by $n_1(1-p_1)+n_2(1-p_2)$ entities of $\overline{L}_1 \cup \overline{L}_2$, which all have the same score. $n_2 p_2$ entities of $L_2$ are ranked at the end. Here, $n_1 = |\overline{L}_1 \cup L_1| = |L^P|$ and $n_2 = |\overline{L}_2 \cup L_2|= |L^N|$. Depending on the different ranking threshold $r$, the \text{TP}@r and \text{FP}@r for the best index value ranking can be derived.
}
\label{fig:aucmrocdescml}
\end{center}
\end{figure}\noindent

In the supervised approach, the mapping function can further optimize the score ranking from the index value ranking. As discussed in the main text, the optimal score ranking is such that $L_1$ ranks ahead, followed by $\overline{L}_1 \cup \overline{L}_2$, and $L_2$ ranks at the end (Fig. \ref{fig:aucmrocdescml}). Here we can derive the \text{TP}@r and \text{FP}@r for the best index value ranking in the supervised approach as
\begin{equation}
\text{TP}@r =
\begin{dcases}
r, & \text{ $n_1p_1 \geq r$},\\
n_1p_1+\frac{n_1(1-p_1)(r-n_1p_1)}{n_1(1-p_1)+n_2(1-p_2)}, & \text{ $n_1p_1 < r \leq n_1+n_2-n_2p_2$},\\
n_1, & \text{ $n_1+n_2-n_2p_2 < r \leq n_1+n_2$},
\label{equation:tpupper2}
\end{dcases}
\end{equation}
\begin{equation}
\text{FP}@r =
\begin{dcases}
0, & \text{ $n_1p_1 \geq r$},\\
\frac{n_2(1-p_2)(r-n_1p_1)}{n_1(1-p_1)+n_2(1-p_2)}, & \text{ $n_1p_1 < r \leq n_1+n_2-n_2p_2$},\\
r-n_1, & \text{ $n_1+n_2-n_2p_2 < r \leq n_1+n_2$},
\label{equation:fpupper2}
\end{dcases}
\end{equation}
where $\frac{n_1(1-p_1)}{n_1(1-p_1)+n_2(1-p_2)}$ and $\frac{n_2(1-p_2)}{n_1(1-p_1)+n_2(1-p_2)}$ correspond to the probability of finding a missing link and nonexistent link in $\overline{L}_1 \cup \overline{L}_2$, respectively.

\begin{figure}[ht]
\begin{center}
\resizebox{16cm}{!}{\includegraphics{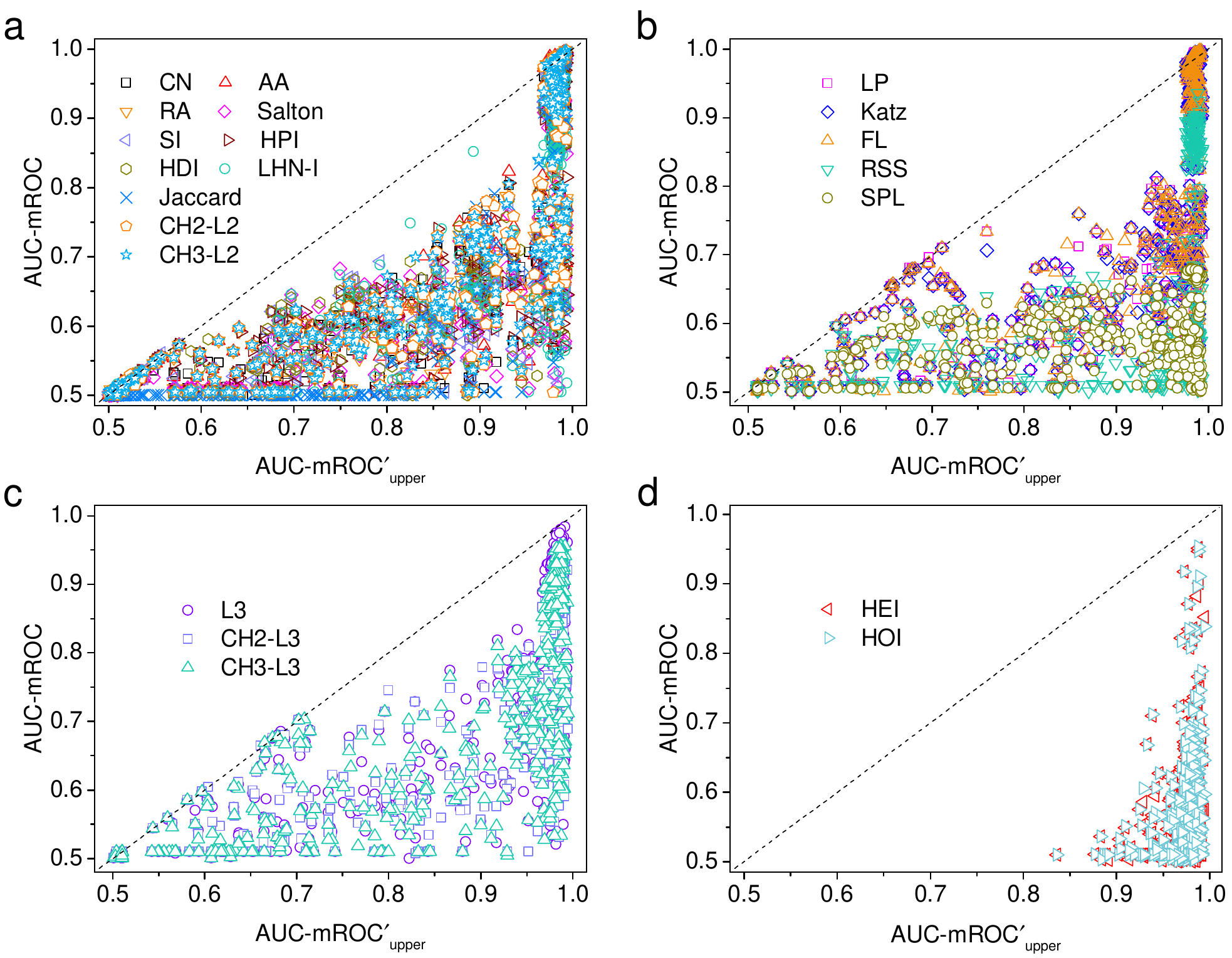}}
\caption{{\bf The maximum capability of the topological features measured by AUC-mROC in the supervised approach.}
The imbalanced positive and negative samples are generated by {\it sample\textsl{2}}. We use 21 indexes from four families and measure the performance of the supervised prediction by these indexes in all 550 networks. The prediction measured by AUC-mROC is equal to or below $\text{AUC-mROC}^{\prime}_\text{upper}$, supporting the claim that $\text{AUC-mROC}^{\prime}_\text{upper}$ gives the maximum capability of a topological feature measured by AUC-mROC. For each network, we randomly generate 200 realizations of networks with link removal, as well as 200 pairs of $L^P$ and $L^N$ sets. In the figure, we choose the highest AUC-mROC from 200 samples as the performance of an index.
}
\label{fig:aucmrocupper2}
\end{center}
\end{figure}

By putting Eqs. (\ref{equation:tpupper2}) and (\ref{equation:fpupper2}) into Eqs. (\ref{equation:nmtpr})-(\ref{equation:mtpr}), we can derive a topological feature's maximum capability ($\text{AUC-mROC}^{\prime}_\text{upper}$) measured by AUC-mROC in the supervised approach. We perform the same test for $\text{AUC-mROC}^{\prime}_\text{upper}$ as for $\text{AUC-mROC}_\text{upper}$. For all networks, the prediction measured by AUC-mROC should be below what $\text{AUC-mROC}^{\prime}_\text{upper}$ yields. This is also confirmed in Fig. (\ref{fig:aucmrocupper2}).

\noindent 

\clearpage

\section{Extended discussion on the lowest index value}\label{section:s6}

An index for a topological feature is designed to quantify the expression of this feature. Hence, the logic behind the index is that its value increases monotonically with the extent to which the feature is expressed. For entities that do not hold the feature at all, they should have the same and the lowest value. If an entity that does not hold the feature has a higher index value than that does hold the feature, the index must be incorrectly designed. Likewise, if two entities that do not hold the feature have different index values, the index is also incorrectly designed.

Currently, there is no unified rule on what value should be the lowest. In the main text, we consider the lowest value to be zero. Indeed, all 21 indexes in this study assign the value 0 to entities that do not hold the feature they are designed to quantify. There could be exceptions. For example, one can design an index equal to the CN value plus 0.5. This is a valid index. However, it is easy to see that our theoretical framework still holds when the lowest value is not zero. Indeed, the validity of based on the fact that entities that do not hold the feature are assigned the same and the lowest value, which is guaranteed by the design of the index.

\clearpage

\section{Extended discussion on the scaling of the unsupervised prediction performance} \label{section:s7}

\begin{figure}[ht]
\begin{center}
\resizebox{16cm}{!}{\includegraphics{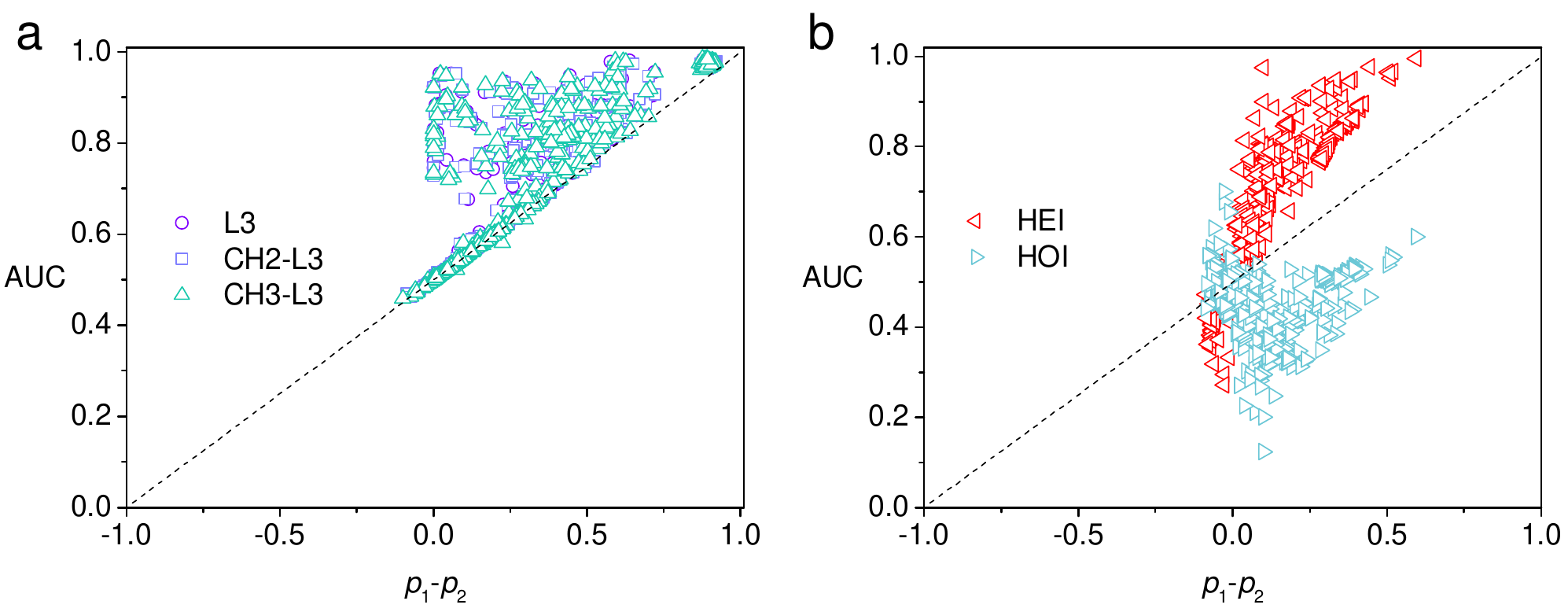}}
\caption{{\bf The scaling of the AUC values from the path of length three feature and the heterogeneity feature.}
Eqs. (\ref{equation:lower}) and (\ref{equation:upper}) in the main text suggest that the actual prediction by an index fluctuates within $p_1 \times p_2$. ($\bf{a}$) The path of length three feature has a relatively small average $p_1 \times p_2$ value. Hence the data points roughly scale as $p_{1}-p_{2}$. ($\bf{b}$) For the heterogeneity feature, as the average $p_1 \times p_2$ value is large, the data points do not collapse to the line $y=x$. For each network, we randomly generate 200 realizations of networks with link removal, as well as 200 pairs of $L^P$ and $L^N$ sets. In the figure, we use the average value of 200 samples. 
}
\label{fig:featurep1p2}
\end{center}
\end{figure}

As illustrated in Eqs. (\ref{equation:lower}) and (\ref{equation:upper}) of the main text, the gap between the upper and lower bound of the unsupervised prediction is $p_1 \times p_2$. Therefore, if the average $p_1 \times p_2$ is relatively small for a given feature, the performance of different indexes associated with it should roughly scale as $p_{1}-p_{2}$. However, the final scaling depends on how even the measured AUC data points distribute within the gap. In the main text, we show the scaling behavior of the common neighbor feature and path feature. For the path of length three feature, the average $p_1 \times p_2$ is smaller than that of the path feature (Table \ref{table:p1p2}). The scaling also holds (Fig. \ref{fig:featurep1p2}a). But because the AUC for the path of length three feature is closer to the upper bound compared with the path feature (Figs. \ref{fig:unsupupper}b, c), the data points are systematically located in the upper corner. For the heterogeneity feature, the average $p_1 \times p_2$ is much larger (Table \ref{table:p1p2}). Hence, the AUC does not follow $p_{1}-p_{2}$.

\noindent 

\clearpage

\section{Test on other machine learning algorithms}\label{section:s8}

\begin{figure}[ht]
\begin{center}
\resizebox{16cm}{!}{\includegraphics{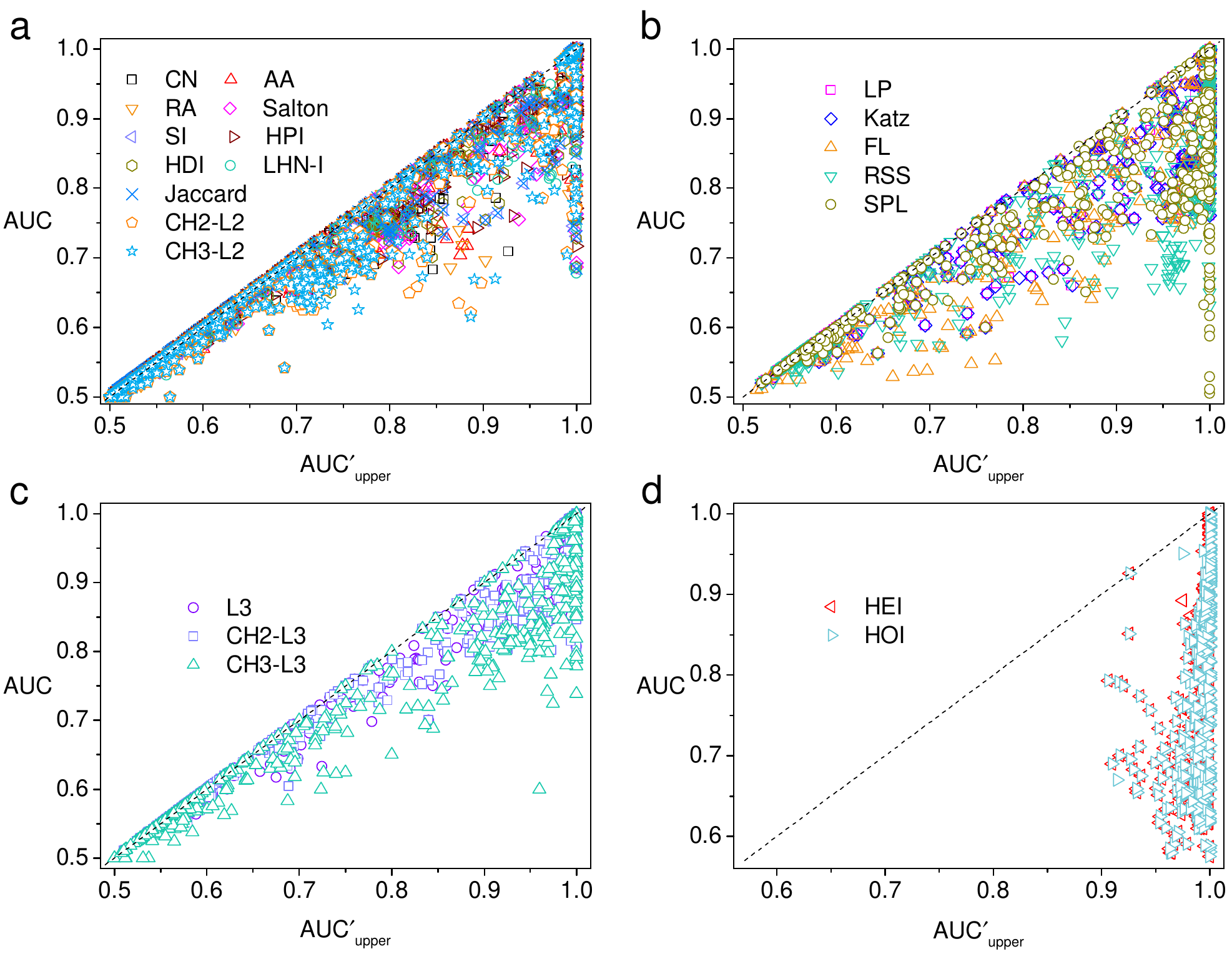}}
\caption{{\bf Supervised prediction measured by AUC based on the Gradient Boosting classifier.}
Eq. (\ref{equation:upper2}) in the main text suggests that $\text{AUC}^{\prime}_\text{upper}$ sets the upper bound of the supervised prediction. This is confirmed by 21 indexes related to 4 topological features: common neighbor $\bf{(a)}$, path $\bf{(b)}$, path of length three $\bf{(c)}$, and heterogeneity $\bf{(d)}$. For each network, we randomly generate 200 realizations of networks with link removal, as well as 200 pairs of $L^P$ and $L^N$ sets. In the figure, we choose the highest AUC from 200 samples as the performance of an index. The same quantitative analysis in the Fig. \ref{fig:supupper} is repeated.
}
\label{fig:supuppergb}
\end{center}
\end{figure}

\begin{figure}[ht]
\begin{center}
\resizebox{16cm}{!}{\includegraphics{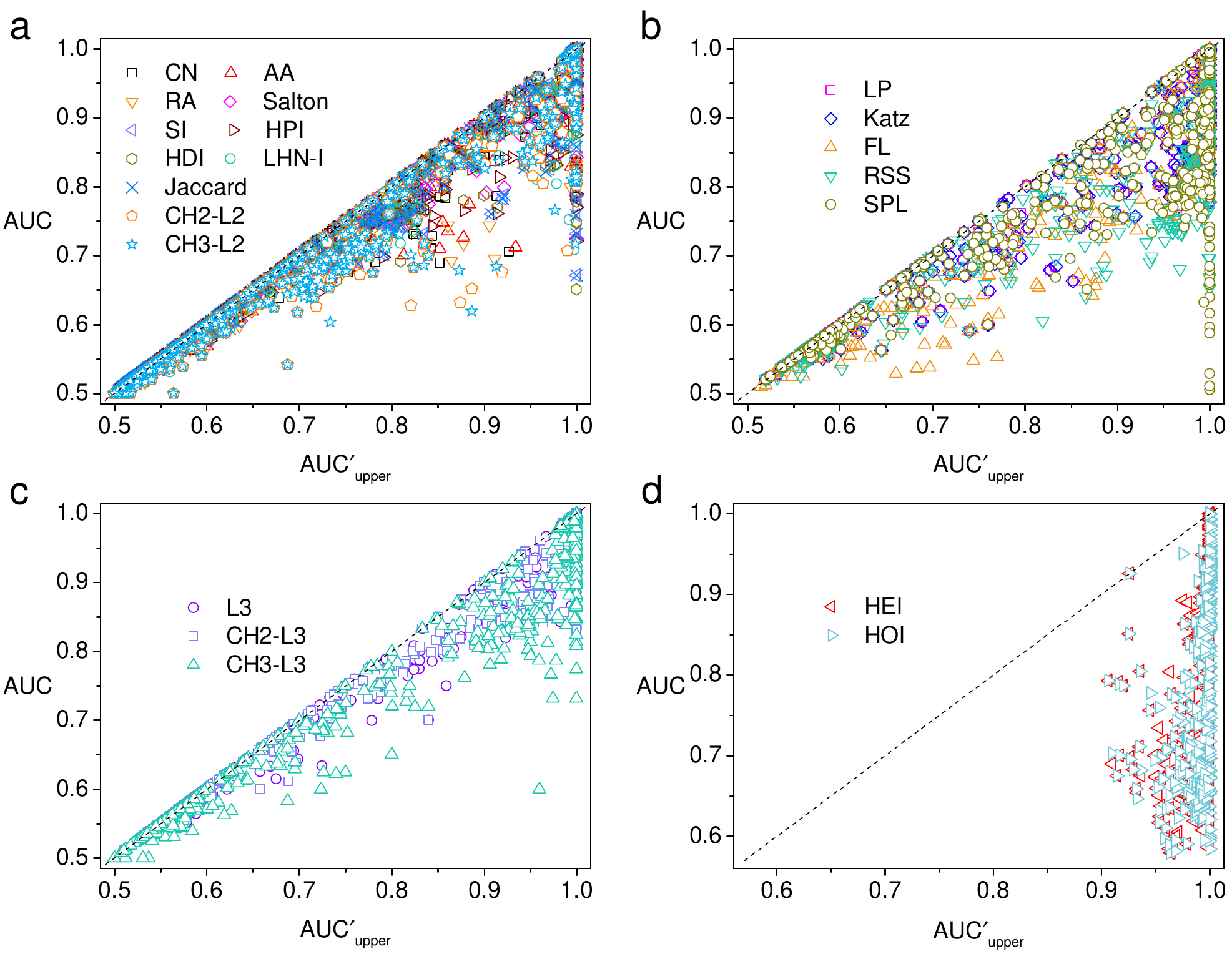}}
\caption{{\bf Supervised prediction measured by AUC based on the AdaBoost classifier.}
Eq. (\ref{equation:upper2}) in the main text suggests that $\text{AUC}^{\prime}_\text{upper}$ sets the upper bound of the supervised prediction. This is confirmed by 21 indexes related to 4 topological features: common neighbor $\bf{(a)}$, path $\bf{(b)}$, path of length three $\bf{(c)}$, and heterogeneity $\bf{(d)}$. For each network, we randomly generate 200 realizations of networks with link removal, as well as 200 pairs of $L^P$ and $L^N$ sets. In the figure, we choose the highest AUC from 200 samples as the performance of an index. The same quantitative analysis in the Fig. \ref{fig:supupper} is repeated.
}
\label{fig:supupperab}
\end{center}
\end{figure}

In the main text, we use the Random Forest classifier to analyze the capability of a topological feature. Here, we also consider Gradient Boosting and AdaBoost classifiers to display similar tests. The sampling method is the same as that of the main text. The results from Figs. \ref{fig:supuppergb} and \ref{fig:supupperab} show that our quantitative framework is not affected by the machine learning algorithm. Hence, this further indicates that the machine learning algorithm can find the optimal mapping function and improve the prediction compared with that of the unsupervised prediction.

\noindent 

\clearpage

\noindent 

\clearpage

\section{The optimal score ranking}\label{section:s9}

\begin{figure}[ht]
\begin{center}
\resizebox{10cm}{!}{\includegraphics{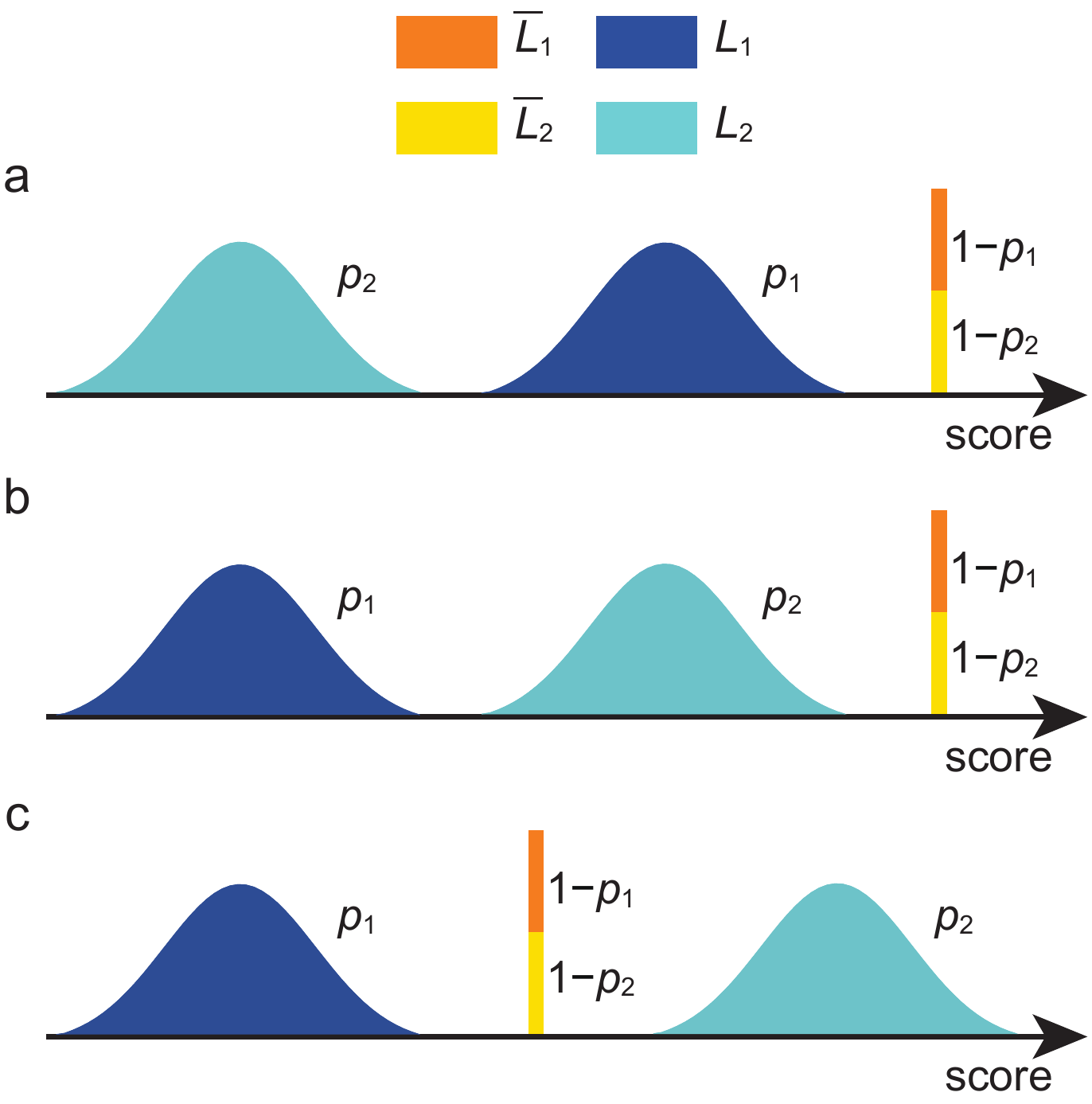}}
\caption{{\bf The supplement to prove the optimal score ranking.}
The link prediction performance relies on the relative rank of the three sets: $L_1$, $L_2$, and $\overline{L}_1 \cup \overline{L}_2$. It is a total of 6 different rankings. This is the other 3 different permutations excluding ones of the main text. 
({\bf a}) The $\overline{L}_1 \cup \overline{L}_2$ is systematically ranked ahead of $L_1$, and $L_2$ is ranked at the end.
({\bf b}) The $\overline{L}_1 \cup \overline{L}_2$ is systematically ranked ahead of $L_2$, and $L_1$ is ranked at the end.
({\bf c}) The $L_2$ is systematically ranked ahead of $\overline{L}_1 \cup \overline{L}_2$, and $L_1$ is ranked at the end.
The variables applied here are the same as those in Fig. \ref{fig:unsupervised} of the main text: $p_1$ and $p_2$ represent the fraction of samples in the positive set $L^P$ and the negative set $L^N$ in which each node pair has an index value greater than 0, respectively.
}
\label{fig:optimaldesc}
\end{center}
\end{figure}\noindent

The link prediction performance relies on the relative rank of the three sets: $L_1$, $L_2$, and $\overline{L}_1 \cup \overline{L}_2$. Totally there are 6 different rankings. In Figs. \ref{fig:unsupervised}b, \ref{fig:unsupervised}c and \ref{fig:unsupervised}d of the main text, we show 3 scenarios based on different rankings (in descending order of the scores/values): $L_2 > L_1 > \overline{L}_1 \cup \overline{L}_2$, $L_1 > L_2 > \overline{L}_1 \cup \overline{L}_2$, and $L_1 >  \overline{L}_1 \cup \overline{L}_2 > L_2$. Here we show the other 3 different permutations in Fig. \ref{fig:optimaldesc} and derive the AUC value in these ranking scenarios.

For the ranking in Fig. \ref{fig:optimaldesc}a, we have $\overline{L}_1 \cup \overline{L}_2 > L_1 > L_2$. A positive sample outscores a negative sample only when one node pair is from $L_1 \cup \overline{L}_1$ (the whole set $L^P$), and the other is from $L_2$. Hence, this gives
\begin{equation}
\text{AUC}_\text{s1} = \frac{n'}{n} + \frac{1}{2}\frac{n''}{n} = p_{2}+\frac{1}{2}(1-p_{1})(1-p_{2}).
\label{equation:s1}
\end{equation}

For the ranking in Fig. \ref{fig:optimaldesc}b, we have $\overline{L}_1 \cup \overline{L}_2 > L_2 > L_1$. A positive sample outscores a negative sample only when one node pair is from $\overline{L}_1$, and the other is from $L_2$ (Fig. \ref{fig:optimaldesc}b). Correspondingly, we have
\begin{equation}
\text{AUC}_\text{s2} = \frac{n'}{n} + \frac{1}{2}\frac{n''}{n} = (1-p_{1})p_{2}+\frac{1}{2}(1-p_{1})(1-p_{2}).
\label{equation:s2}
\end{equation}

For the ranking in Fig. \ref{fig:optimaldesc}c, we have $L_2 > \overline{L}_1 \cup \overline{L}_2 > L_1$. Node pairs in $L_2$ outscore all positive samples (the whole set $L^P$). Hence, this gives
\begin{equation}
\text{AUC}_\text{s3} = \frac{n'}{n} + \frac{1}{2}\frac{n''}{n} = \frac{1}{2}(1-p_{1})(1-p_{2}).
\label{equation:s3}
\end{equation}

By comparing the 6 equations (Eq. (\ref{equation:s1}), Eq. (\ref{equation:s2}), Eq. (\ref{equation:s3}), $\text{AUC}_\text{lower} = p_{1}(1-p_{2})+\frac{1}{2}(1-p_{1})(1-p_{2})$, $\text{AUC}_\text{upper} = p_{1}+\frac{1}{2}(1-p_{1})(1-p_{2})$, and $\text{AUC}^{\prime}_\text{upper} = p_{1} + (1-p_{1})p_{2} +\frac{1}{2}(1-p_{1})(1-p_{2})$), we can find that the $\text{AUC}^{\prime}_\text{upper}$ gives the highest AUC value. Hence, Fig \ref{fig:unsupervised}d of the main text gives the optimal rankings of the three sets. Indeed, this is in line with the theory of machine learning, as the score of the positive sample should be greater than that of the negative one if we assume the positive one is expected to be predicted during the learning process.

\begin{figure}[ht]
\begin{center}
\resizebox{16cm}{!}{\includegraphics{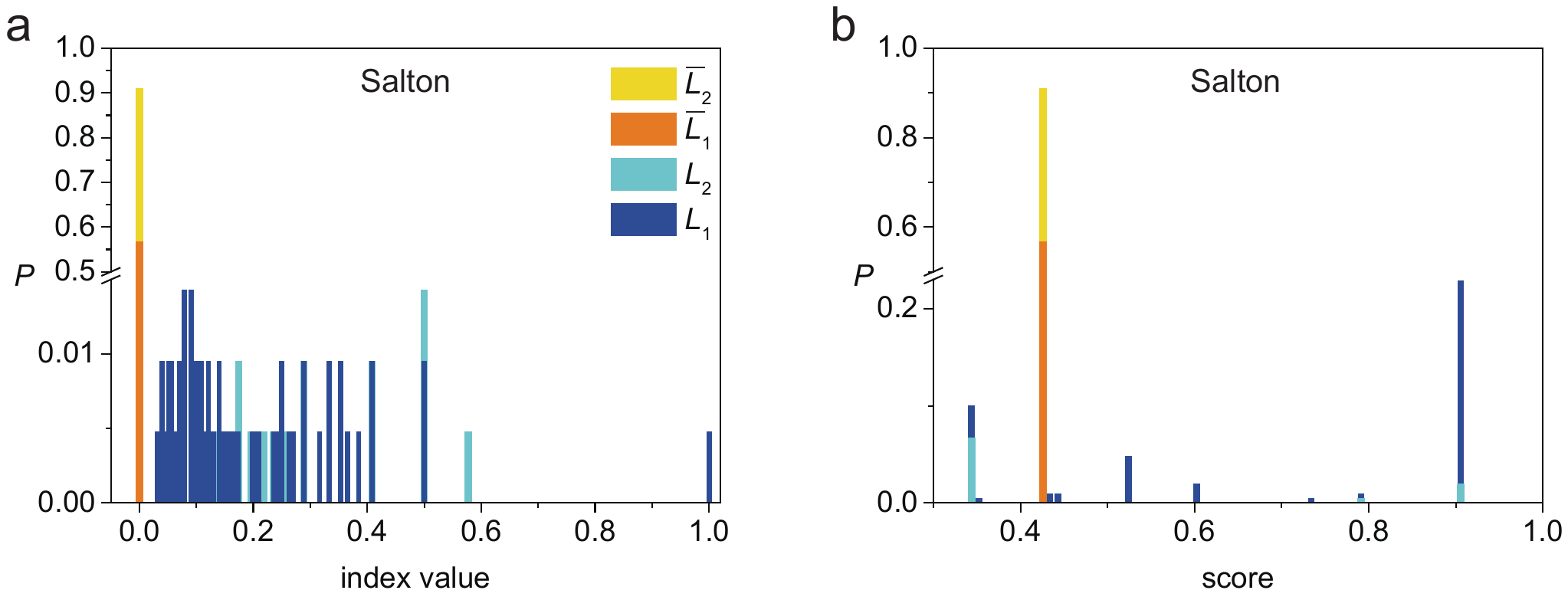}}
\caption{{\bf A real case that has achieved the mapping from the index value to the score.}
$\bf{(a)}$ The distribution of index values obtained. $\bf{(b)}$ The distribution of scores obtained by using the Random Forest classifier in the supervised approach. This network is ``56e9e0d7a6d70217090cdffa'' in the data set.
}
\label{fig:cndistribution}
\end{center}
\end{figure}

To further test if a machine learning algorithm practically re-arranges the ranking, we show an example below. The Salton index is used. The distribution of index values in different sets is shown in Fig. \ref{fig:cndistribution}a. After applying the Random Forest algorithm, the classifier finds the mapping function to transform the index value into the score. The corresponding distribution of the score is presented in Fig. \ref{fig:cndistribution}b. The set $\overline{L}_1 \cup \overline{L}_2$ indeed moves to the middle, giving rise to a ranking similar to the optimal ranking derived theoretically.

\noindent 

\clearpage

\section{Another example of feature and index selection in link prediction}\label{section:s10}

In the main text, taking the common neighbor feature as an example, we show the theoretical finding can help us to determine the feature and index selection. To further validate the preceding analysis, we here conduct a second analysis using the path feature (Table \ref{table:unsupervised}). When using the index SPL to make unsupervised predictions in two real networks, we obtain $\text{AUC}=0.567$ in both networks. The prediction results are relatively low, so it is difficult for us to decide whether the path feature is suitable for these two real networks. Under this case, we have to try one by one for the other 4 indexes on the two networks when the $p_1$ and $p_2$ values are unknown. However, calculating the maximum capability of the path feature ($\text{AUC}_\text{upper}$ presented in the main text) shows that we should switch to a new feature for prediction in the network E and should try the other indexes based on the path feature in the network F (Table \ref{table:unsupervised}). This analysis further shows that the theoretical finding presented in the main text can be applied to optimize the feature and index selection.

Moreover, as the theoretical expression can confirm whether an index is superior to the other indexes, our theoretical finding can provide strong support for experimental validation such as the interaction between genes, and protein-protein interactions. More importantly, the theoretical expression can also help to estimate how close to the upper bound for observed performance in the same topological feature.

\begin{table}[htbp]
	\centering
	\caption{{\bf Another example that utilizes the maximum capability for feature selection.} The AUC performance of the unsupervised prediction using SPL is the same for both networks E and F. However, using the $p_1$ and $p_2$ values, the maximum capability of the path feature can be estimated. The path feature is not suitable for network E but has potential in network F. The network E is ``56e98770a6d70217090cde08'', and the network F is ``Cat\_cerebral\_hemisphere\_cortex\_only'' in the dataset.}
	\begin{tabular}{l|rrrrrrrrrrrrrrrr}
		  \multicolumn{1}{c}{}& \multicolumn{1}{c}{SPL} & \multicolumn{1}{c}{$p_{1}$} & \multicolumn{1}{c}{$p_{2}$}& \multicolumn{1}{c}{$\text{AUC}_\text{upper}$}& \multicolumn{1}{c}{LP} & \multicolumn{1}{c}{Katz} & \multicolumn{1}{c}{FL}& \multicolumn{1}{c}{RSS}\\
		\hline
		\multicolumn{1}{c}{Network E} &\multicolumn{1}{c}{0.567} & \multicolumn{1}{c}{$\textsl{0.158}$}& \multicolumn{1}{c}{$\textsl{0.026}$}& \multicolumn{1}{c}{${\bf 0.568}$} &\multicolumn{1}{c}{0.567} & \multicolumn{1}{c}{0.567}& \multicolumn{1}{c}{0.567} & \multicolumn{1}{c}{0.567} \\
		\multicolumn{1}{c}{Network F} & \multicolumn{1}{c}{0.567} & \multicolumn{1}{c}{$\textsl{1.0}$}& \multicolumn{1}{c}{$\textsl{1.0}$}& \multicolumn{1}{c}{${\bf 1.0}$} & \multicolumn{1}{c}{0.867} & \multicolumn{1}{c}{0.867}& \multicolumn{1}{c}{0.869} & \multicolumn{1}{c}{0.841} \\
		\hline
	\end{tabular}%
\label{table:unsupervised}
\end{table}%

\clearpage

\section{The theoretical expression of $p_{1}$ and $p_{2}$}\label{section:s11}

\begin{figure}[ht]
\begin{center}
\resizebox{6cm}{!}{\includegraphics{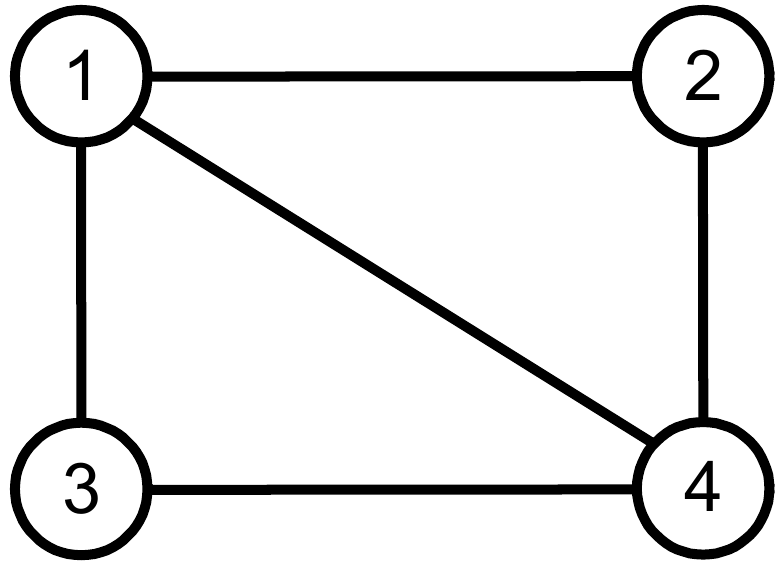}}
\caption{{\bf An illustration of explaining parameters in Eqs. (\ref{equation:p11}) and (\ref{equation:p21}).}
Since the two closed triangles ($\triangle_{134}$ and $\triangle_{124}$) share the same link (1-4), $N_{\triangle} = 2$ and $S_{\triangle} = 1$.
Similarly, the two open triangles ($\wedge_{123}$ and $\wedge_{243}$) share the unconnected node pair (2-3), hence $N_{\wedge} = 2$ and $S_{\wedge} = 1$.
}
\label{fig:equation_theory}
\end{center}
\end{figure}

To give more insights into the structural characteristics that make a topological feature effective in link prediction, we deduct the theoretical expression of $p_{1}$ and $p_{2}$. Here, we take the common neighbor feature as an example. The idea behind the feature is that two unconnected nodes that share the same neighborhood nodes are likely to become a link. Hence, the indexes based on the common neighbor assign a value greater than 0 to a node pair only if this node pair belongs to a link of the closed triangle. Since $p_{1}$ is the fraction of samples in the positive set $L^P$ with an index value greater than 0, $p_{1}$ can be quantified as the probability that a randomly picked node pair from all existing links of a network is exactly from one link in a closed triangle. Therefore, the theoretical expression of $p_{1}$ is defined as
\begin{equation}
p^{\prime}_{1} =  \frac{3*N_{\triangle}-S_{\triangle}}{L},
\label{equation:p11}
\end{equation}
where the $N_{\triangle}$ is the number of closed triangles in a network. The $S_{\triangle}$ is the number of times that a link is shared by multiple triangles (Fig. \ref{fig:equation_theory}). Because a link can belong to multiple triangles, the $3*N_{\triangle}$ would over-count the number of links belonging to a triangle and has to subtract the number of times a link appears in other triangles. 

Likewise, as $p_{2}$ is the fraction of samples in the negative set $L^N$ with an index value greater than 0, $p_{2}$ can be quantified as the probability that a randomly picked node pair from all nonexistent links of a network shares a common neighbor (this node pair and an open triangle constitute a closed triangle). Hence, the theoretical expression of $p_{2}$ is formulated as
\begin{equation}
p^{\prime}_{2} =  \frac{N_{\wedge}-S_{\wedge}}{\frac{N(N-1)}{2}-L},
\label{equation:p21}
\end{equation}
where the $N_{\wedge}$ is the number of open triangles in a network (the triad that two nodes that are not directly connected but both connect to the third node). The $S_{\wedge}$ is the number of times that an unconnected node pair is shared by other open triangles, which can cause an overcount if it is not subtracted (Fig. \ref{fig:equation_theory}). The $\frac{N(N-1)}{2}-L$ corresponds to the total number of unconnected node pairs. 

\noindent 

\begin{figure}[ht]
\begin{center}
\resizebox{16cm}{!}{\includegraphics{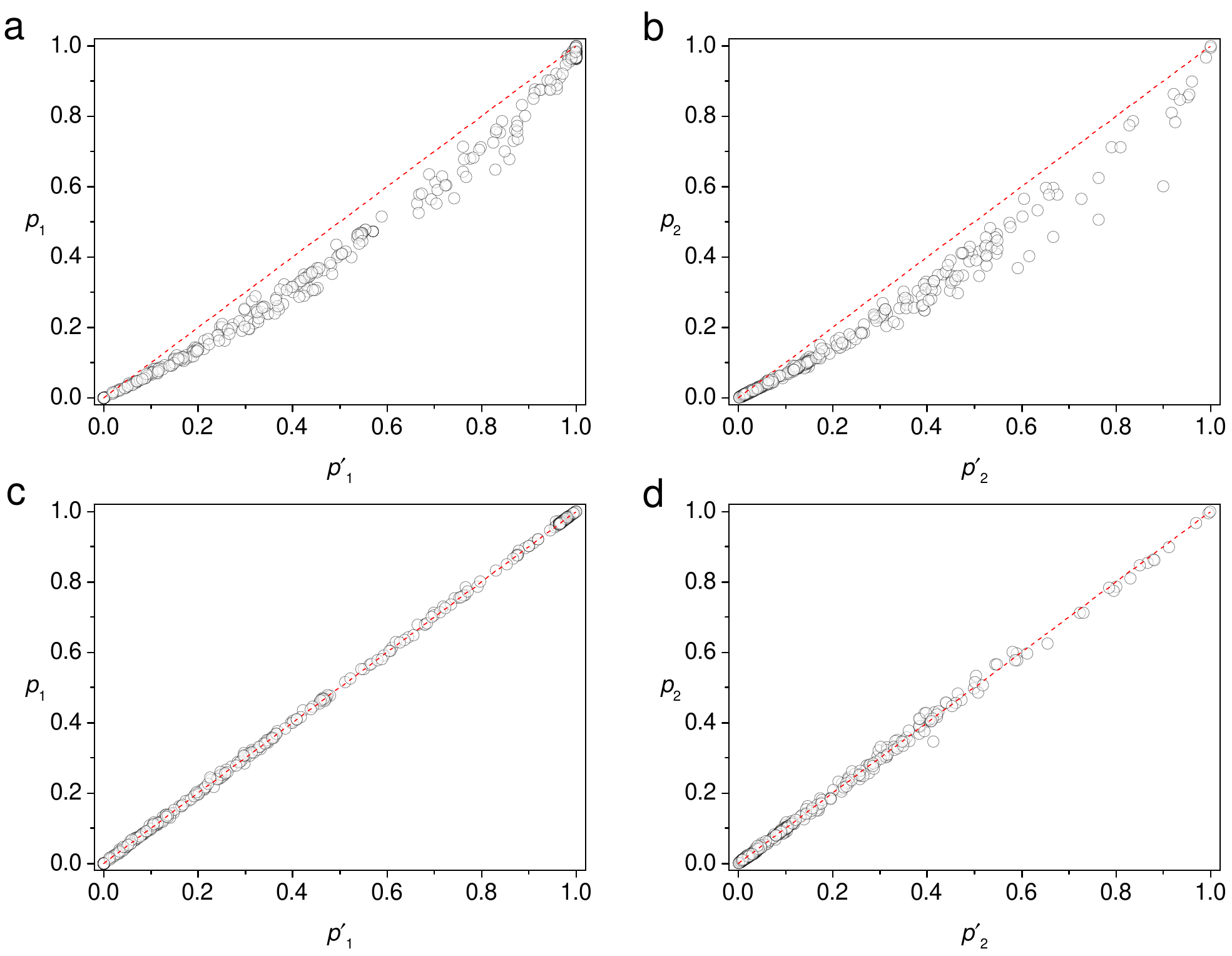}} 
\caption{{\bf Evidence of the theoretical values $p^{\prime}_{1}$ and $p^{\prime}_{2}$ applied to 550 real networks.}
$\bf{(a, b)}$ The theoretical values $p^{\prime}_{1}$ and $p^{\prime}_{2}$ are given by Eq. (\ref{equation:p11}) and Eq. (\ref{equation:p21}) from the rest 80\% of links, respectively.
$\bf{(c, d)}$ The theoretical values $p^{\prime}_{1}$ and $p^{\prime}_{2}$ are given by Eq. (\ref{equation:p11}) and Eq. (\ref{equation:p21}) from the original network, respectively. 
For $p_{1}$ and $p_{2}$, we generate 200 independent pairs of $L^P$ and $L^N$ sets based on the random sampling (removed 20\% of links) in each network, and we use the average value of them. 
}
\label{fig:stp1p2}
\end{center}
\end{figure}

To validate Eqs. (\ref{equation:p11}) and (\ref{equation:p21}), we take 550 empirical networks to directly calculate the theoretical values $p^{\prime}_{1}$ and $p^{\prime}_{2}$. Shading multiple links will change the existing topology of the original network which makes the estimate of $N_{\triangle}$ or $N_{\wedge}$ incorrect. Still, we observe an overall nice agreement between the theoretical and empirical values even when 20\% of links are temporally removed (Figs. \ref{fig:stp1p2}a, b). Moreover, to give more reasonable evidence of Eqs. (\ref{equation:p11}) and (\ref{equation:p21}), we here also consider the situation that computing the $p^{\prime}_{1}$ and $p^{\prime}_{2}$ from the original network. The results from Fig. \ref{fig:stp1p2}c and Fig. \ref{fig:stp1p2}d show that the theoretical values ($p^{\prime}_{1}$ and $p^{\prime}_{2}$) are line with the empirical values ($p_{1}$ and $p_{2}$), demonstrating a perfect agreement. The theoretical expressions explain why $C$ itself is insufficient to characterize the capability of the common neighbor feature. More importantly, the Eqs. (\ref{equation:p11}) and (\ref{equation:p21}) help us understand network characteristics associated with the utilization of a topological feature in link prediction.

\noindent 

\clearpage

\section{Extended discussion on the prediction performance measured by precision}\label{section:s12}

The precision measure involves a hyper-parameter $L_\text{k}$ for the cutoff of the top-k node pairs. Therefore, we might not well measure the performance of an index when only using precision. Here, we show the link prediction performance in two networks (Table \ref{table:unsupervisedprecision}). When using the CN index to make unsupervised predictions in network G, we obtain $\text{AUC}=0.533$, suggesting that the CN index overall has limited potential ($\text{AUC}_\text{upper}=0.533$). But when measuring the performance by precision, the obtained measure can vary significantly with different choices of $L_\text{k}$. When choosing $L_\text{k} < p_1|L^P|$, the predicted upper bound is 1. In this case, we also obtain a high precision value $\text{Precision}=0.973$. But because $p_1$ is very small for network G, a high precision value only suggests that the prediction is correct for the very top candidates. When $L_\text{k}$ becomes larger, the precision drops drastically ($\text{Precision}=0.550$ for $L_\text{k} = 121$).

Likewise, in network H, the performance of unsupervised prediction by LP index yields $\text{AUC}=0.826$. The LP index also has a high potential in terms of the AUC measure ($\text{AUC}_\text{upper}=0.977$). But if measured by precision, we obtain a low value ($\text{Precision}=0.381$ for $L_\text{k} < p_1|L^P|$). Hence, the prediction accuracy is low for the top candidates. But if the number of candidates increases, the precision goes up again. When $L_\text{k} > p_1|L^P|$, we obtain $\text{Precision}=0.711$.

The two networks in Table \ref{table:unsupervisedprecision} demonstrate a vivid example of the complexity of interpreting the precision measure. One has to take the $p_1$ into consideration in order to explain the number of populations the precision is measured from.

\begin{table}[htbp]
	\centering
	\caption{{\bf The interpretation of the performance by precision measure.} The precision gives different performance when choosing different $L_\text{k}$. The network G is ``Water\_Distribution\_Network\_EXNET'', and the network H is ``Freshwater\_stream\_webs\_Stony'' in the data set.} 
	\begin{tabular}{l|rrrrrrrrrrrrrrrr}
		 \multicolumn{1}{c}{}&  \multicolumn{1}{c}{$\text{AUC}$}& \multicolumn{1}{c}{$p_{1}$} &\multicolumn{1}{c}{$|L^P|$} &\multicolumn{1}{c}{$p_{1}|L^P|$} &\multicolumn{1}{c}{$\text{Precision}$} & \multicolumn{1}{c}{$\text{Precision}$} \\
		\hline
		\multicolumn{1}{c}{Network G} & \multicolumn{1}{c}{${\bf 0.533}$} &  \multicolumn{1}{c}{$\textsl{0.067}$}&\multicolumn{1}{c}{241} & \multicolumn{1}{c}{16} & \multicolumn{1}{c}{${\bf 0.973}$ $(L_\text{k}=10)$} & \multicolumn{1}{c}{${\bf 0.550}$ $(L_\text{k}=121)$} \\
		\multicolumn{1}{c}{Network H} & \multicolumn{1}{c}{${\bf 0.826}$} &  \multicolumn{1}{c}{$\textsl{0.976}$}&\multicolumn{1}{c}{83} & \multicolumn{1}{c}{81} & \multicolumn{1}{c}{${\bf 0.381}$ $(L_\text{k}=21)$} & \multicolumn{1}{c}{${\bf 0.711}$ $(L_\text{k}=83)$} \\
		\hline
	\end{tabular} %
\label{table:unsupervisedprecision}
\end{table} %

\clearpage

\section{The analysis for 10\% random removal links}\label{section:s13}
\begin{figure}[ht]
\begin{center}
\resizebox{16cm}{!}{\includegraphics{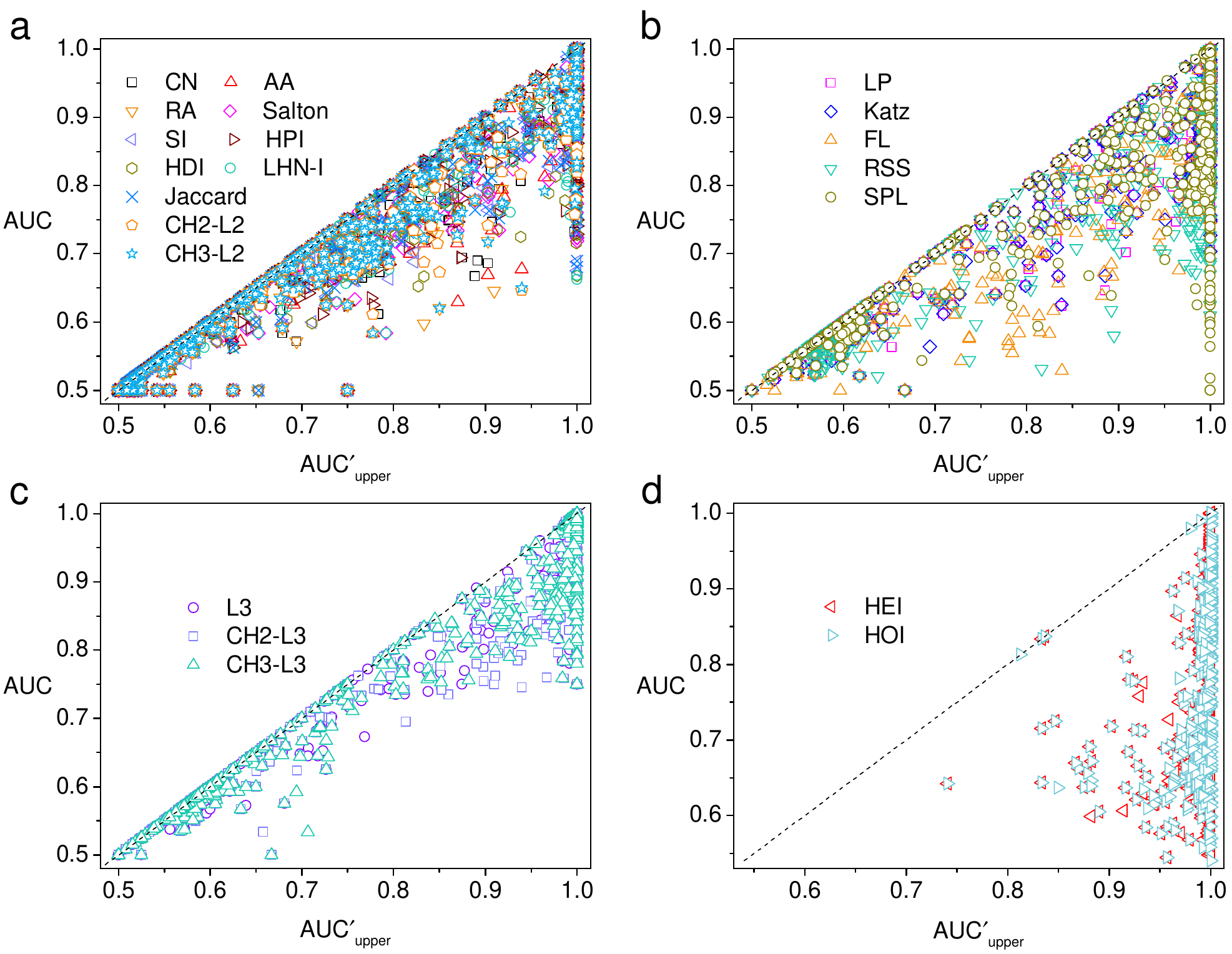}}
\caption{{\bf The supervised prediction measured by AUC based on 10\% random removal.}
Eq. (\ref{equation:upper2}) in the main text suggests that $\text{AUC}^{\prime}_\text{upper}$ sets the upper bound of the supervised prediction. This is confirmed by 21 indexes related to 4 topological features: common neighbor $\bf{(a)}$, path $\bf{(b)}$, path of length three $\bf{(c)}$, and heterogeneity $\bf{(d)}$. For each network, we randomly generate 200 realizations of networks with link removal, as well as 200 pairs of $L^P$ and $L^N$ sets. In the figure, we choose the highest AUC from 200 samples as the performance of an index. The same quantitative analysis in the Fig. \ref{fig:supupper} is repeated.
}
\label{fig:supupper01}
\end{center}
\end{figure}

\begin{figure}[ht]
\begin{center}
\resizebox{16cm}{!}{\includegraphics{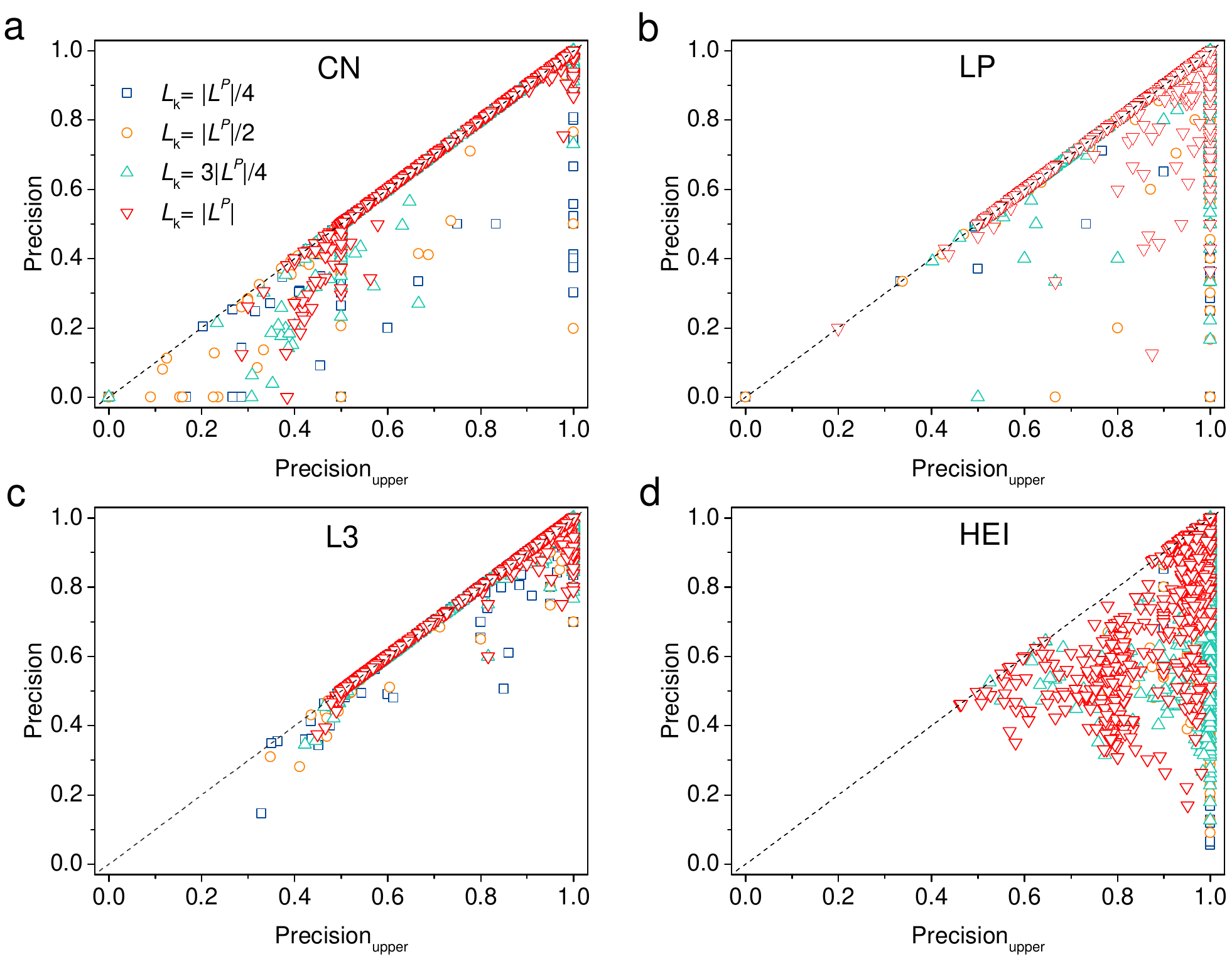}}
\caption{{\bf The unsupervised prediction measured by precision based on 10\% random removal.}
We choose the index CN, LP, L3, and HEI from each of the four families and measure the performance of the unsupervised prediction by these indexes in all 550 networks. For different choices of $L_\text{k}$ ($L_\text{k}= |L^P|/4, |L^P|/2, 3|L^P|/4, |L^P|$), the measured precision is equal to or below $\text{Precision}_\text{upper}$, supporting the claim that $\text{Precision}_\text{upper}$ gives the maximum capability of a topological feature measured by precision. For each network, we randomly generate 200 realizations of networks with link removal, as well as 200 pairs of $L^P$ and $L^N$ sets. In the figure, we use the average value of 200 samples.
}
\label{fig:precisionupper01}
\end{center}
\end{figure}

\begin{figure}[ht]
\begin{center}
\resizebox{16cm}{!}{\includegraphics{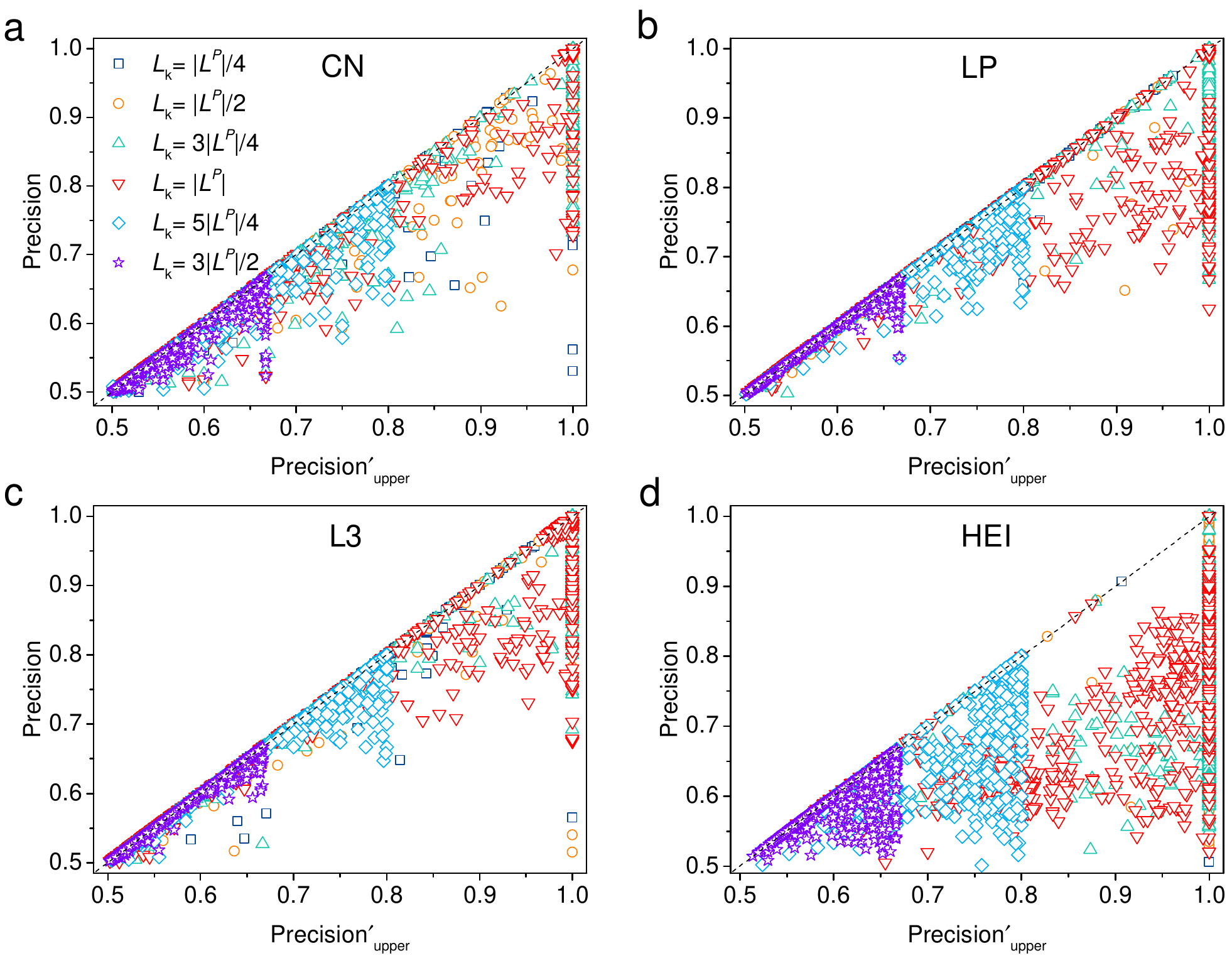}}
\caption{{\bf The supervised prediction measured by precision based on 10\% random removal.}
We choose the index CN, LP, L3, and HEI from each of the four families and measure the performance of the supervised prediction by these indexes in all 550 networks. For different choices of $L_\text{k}$ ($L_\text{k}= |L^P|/4, |L^P|/2, 3|L^P|/4, |L^P|, 5|L^P|/4, 3|L^P|/2$), the measured precision is equal to or below $\text{Precision}^{\prime}_\text{upper}$, supporting the theoretical results for the maximum capability of a topological feature measured by precision. For each network, we randomly generate 200 realizations of networks with link removal, as well as 200 pairs of $L^P$ and $L^N$ sets. In the figure, we choose the highest precision from 200 samples as the performance of an index.
}
\label{fig:precisionupper02}
\end{center}
\end{figure}

In the main text, we show the analysis for 20\% random removal links. For robustness check, we here repeat the analysis for 10\% random removal. The similar AUC results (the Fig. \ref{fig:unsupupper01} is similar to Fig. \ref{fig:unsupupper}, the Fig. \ref{fig:supupper01} is similar to Fig. \ref{fig:supupper}) are obtained. Analogously, the similar precision results (the Fig. \ref{fig:precisionupper01} is similar to Fig. \ref{fig:precisionupper1}, the Fig. \ref{fig:precisionupper02} is similar to Fig. \ref{fig:precisionupper2}) are obtained. These show that our quantitative framework is not affected by the random removal percentage.


\noindent

\begin{figure}[ht]
\begin{center}
\resizebox{14cm}{!}{\includegraphics{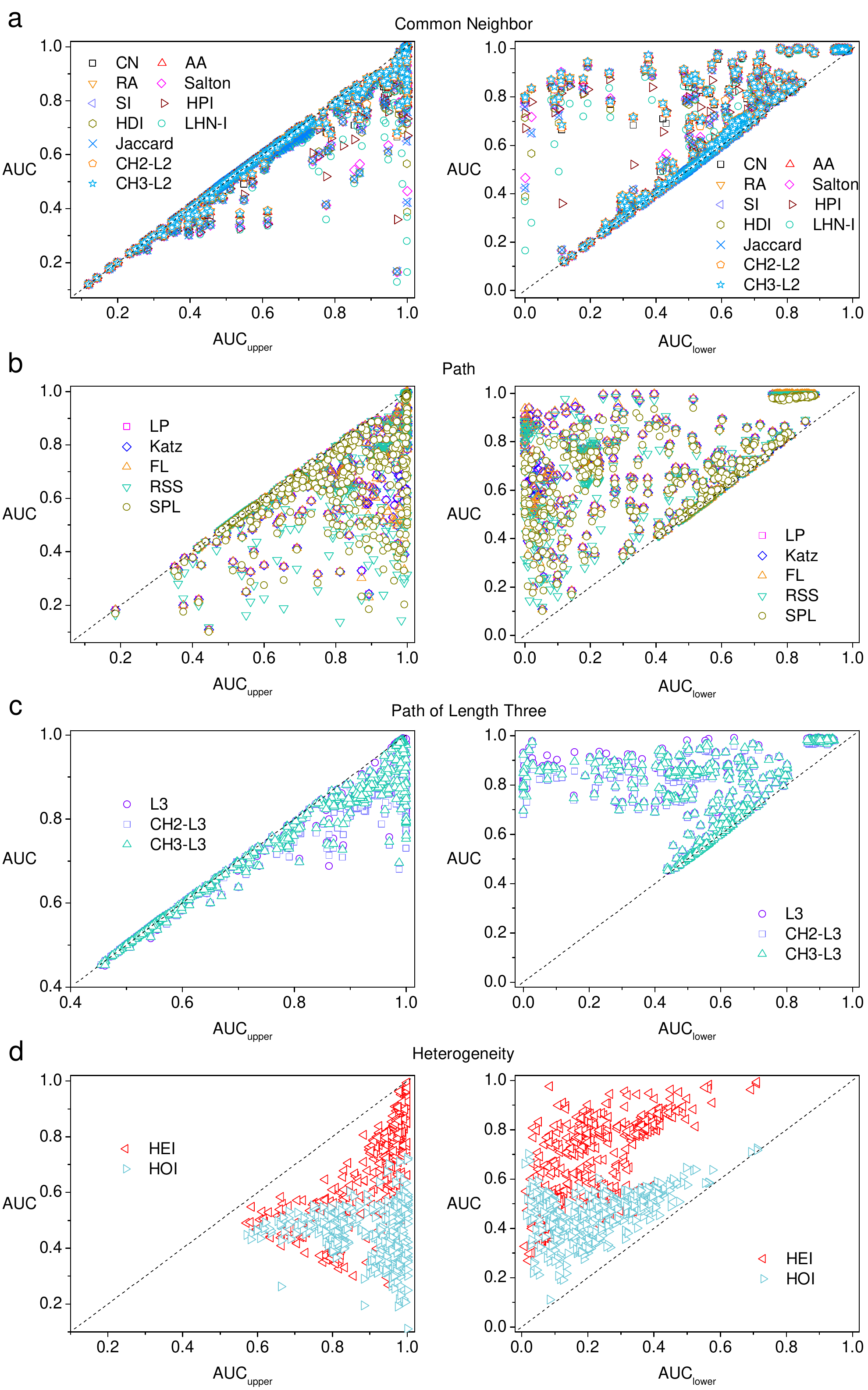}}
\end{center}
\end{figure}\noindent 

\begin{figure}[ht]
\begin{center}
\caption{{\bf The unsupervised prediction measured by AUC based on 10\% random removal.}
Eqs. (\ref{equation:lower}) and (\ref{equation:upper}) in the main text suggest that different indexes have different prediction performances, but all indexes associated with one topological feature share the same $\text{AUC}_\text{upper}$ and $\text{AUC}_\text{lower}$. This is confirmed by 21 indexes related to 4 topological features: common neighbor $\bf{(a)}$, path $\bf{(b)}$, path of length three $\bf{(c)}$, and heterogeneity $\bf{(d)}$. For each network, we randomly generate 200 realizations of networks with link removal, as well as 200 pairs of $L^P$ and $L^N$ sets. In the figure, we use the average value of 200 samples. The same quantitative analysis in the left panel of Fig. \ref{fig:unsupupper} is repeated.
}
\label{fig:unsupupper01}
\end{center}
\end{figure}\noindent

\noindent

\noindent 

\clearpage

\section{The detailed information of empirical dataset}\label{section:s14}
To verify the universality of the pattern uncovered about the maximum capability of the topological features, we use the dataset from ``CommunityFitNet corpus'' \cite{ghasemian2020stacking}. The basic statistics of different networks in each domain are shown in Table \ref{table:statistics}.

\begin{table}[htbp]
	\centering
	\caption{{\bf The basic statistics of 6 different-domain networks.} We here show the average statistics of networks in 6 different domains. $\langle N \rangle$ and $\langle L \rangle$ are the average number of nodes and links, respectively. $\langle C \rangle$ is the average number of clustering coefficients. $\langle k \rangle$ is the average degree, $\langle d \rangle$ is the average shortest path length.} 
	\begin{tabular}{l|rrrrrrrrrrrrrrrr}
		 \multicolumn{1}{c}{Network domains}&  \multicolumn{1}{c}{$\langle N \rangle$}& \multicolumn{1}{c}{$\langle L \rangle$} &\multicolumn{1}{c}{$\langle k \rangle$} &\multicolumn{1}{c}{$\langle C \rangle$} &\multicolumn{1}{c}{$\langle d \rangle$} \\
		\hline
		\multicolumn{1}{c}{biological} & \multicolumn{1}{c}{$294.235$} &  \multicolumn{1}{c}{$780.151$}&\multicolumn{1}{c}{6.302} & \multicolumn{1}{c}{0.135} & \multicolumn{1}{c}{$4.577$}  \\
		\multicolumn{1}{c}{social} & \multicolumn{1}{c}{$558.581$} &  \multicolumn{1}{c}{$1988.331$}&\multicolumn{1}{c}{7.592} & \multicolumn{1}{c}{0.840} & \multicolumn{1}{c}{$6.017$}  \\
		\multicolumn{1}{c}{economic} & \multicolumn{1}{c}{$701.677$} &  \multicolumn{1}{c}{$865.685$}&\multicolumn{1}{c}{3.338} & \multicolumn{1}{c}{0.040} & \multicolumn{1}{c}{$11.697$}  \\
		\multicolumn{1}{c}{technological} & \multicolumn{1}{c}{$532.543$} &  \multicolumn{1}{c}{$1061.029$}&\multicolumn{1}{c}{4.034} & \multicolumn{1}{c}{0.118} & \multicolumn{1}{c}{$6.076$}  \\
		\multicolumn{1}{c}{transportation} & \multicolumn{1}{c}{$721.343$} &  \multicolumn{1}{c}{$1274.143$}&\multicolumn{1}{c}{3.493} & \multicolumn{1}{c}{0.101} & \multicolumn{1}{c}{$12.555$}  \\
		\multicolumn{1}{c}{information} & \multicolumn{1}{c}{$494.167$} &  \multicolumn{1}{c}{$1266.222$}&\multicolumn{1}{c}{5.315} & \multicolumn{1}{c}{0.223} & \multicolumn{1}{c}{$3.419$}  \\
		\hline
	\end{tabular} %
\label{table:statistics}
\end{table} %

\clearpage

\end{document}